\documentclass[twocolumn,aps,floatfix,superscriptaddress]{revtex4-1}
\usepackage{amsmath,amssymb}
\usepackage{graphicx,float}
\usepackage{times,txfonts}
\usepackage[makeroom]{cancel}
\usepackage{color}
\usepackage{multirow}
\usepackage{bbold}
\usepackage{color}
\usepackage{soul}
\usepackage{hyperref}
\usepackage{times,txfonts}
\usepackage{natbib}
\usepackage{blindtext}
\usepackage[export]{adjustbox}
\usepackage{mathrsfs}
\usepackage{textcomp}
\usepackage{blkarray}
\usepackage{multirow}
\usepackage{qcircuit}

\newcommand{\abs}[1]{\left| #1 \right|}
\newcommand{\bra}[1]{\left\langle #1 \right|}
\newcommand{\ket}[1]{\left| #1 \right\rangle}
\newcommand{\braket}[2]{\left\langle {#1{\left| \vphantom{#1 #2} \right.} #2} \right\rangle}

\renewcommand{\epsilon}{\varepsilon}

\def\VR{\kern-\arraycolsep\strut\vrule &\kern-\arraycolsep}
\def\vr{\kern-\arraycolsep & \kern-\arraycolsep}
\usepackage{sistyle}
\usepackage{soul}
\definecolor{lightblue}{RGB}{185,210,248}
\sethlcolor{lightblue}

\begin{document}
\title{Two-photon interference: the Hong-Ou-Mandel effect}
\author{Fr\'ed\'eric Bouchard}
\affiliation{Department of Physics, University of Ottawa, Advanced Research Complex, 25 Templeton Street, Ottawa ON Canada, K1N 6N5}
\affiliation{Current address: National Research Council of Canada, 100 Sussex Drive, Ottawa, Ontario K1A 0R6, Canada}
\author{Alicia Sit}
\affiliation{Department of Physics, University of Ottawa, Advanced Research Complex, 25 Templeton Street, Ottawa ON Canada, K1N 6N5}
\author{Yingwen Zhang}
\affiliation{National Research Council of Canada, 100 Sussex Drive, Ottawa, Ontario K1A 0R6, Canada}
\author{Robert Fickler}
\affiliation{Department of Physics, University of Ottawa, Advanced Research Complex, 25 Templeton Street, Ottawa ON Canada, K1N 6N5}
\affiliation{Current address: Photonics Laboratory, Physics Unit, Tampere University, Tampere, FI-33720, Finland}
\author{Filippo M. Miatto}
\affiliation{T\'el\'ecom Paris, LTCI, Institut Polytechnique de Paris, 19 Place Marguerite Peray, 91120 Palaiseau, France}
\author{Yuan Yao}
\affiliation{T\'el\'ecom Paris, LTCI, Institut Polytechnique de Paris, 19 Place Marguerite Peray, 91120 Palaiseau, France}
\author{Fabio Sciarrino}
\affiliation{Dipartimento di Fisica, Sapienza Universit\`a di Roma, Piazzale Aldo Moro 5, I-00185 Roma, Italy}
\author{Ebrahim Karimi}
\affiliation{Department of Physics, University of Ottawa, Advanced Research Complex, 25 Templeton Street, Ottawa ON Canada, K1N 6N5}
\affiliation{National Research Council of Canada, 100 Sussex Drive, Ottawa, Ontario K1A 0R6, Canada}
\email{ekarimi@uottawa.ca}
%



\begin{abstract}
Nearly 30 years ago, two-photon interference was observed, marking the beginning of a new quantum era. Indeed, two-photon interference has no classical analogue, giving it a distinct advantage for a range of applications. The peculiarities of quantum physics may now be used to our advantage to outperform classical computations, securely communicate information, simulate highly complex physical systems and increase the sensitivity of precise measurements. This separation from classical to quantum physics has motivated physicists to study two-particle interference for both fermionic and bosonic quantum objects. So far, two-particle interference has been observed with massive particles, among others, such as electrons and atoms, in addition to plasmons, demonstrating the extent of this effect to larger and more complex quantum systems. A wide array of novel applications to this quantum effect is to be expected in the future. This review will thus cover the progress and applications of two-photon (two-particle) interference over the last three decades.
\end{abstract}

\maketitle

\section{Introduction}

At the inception of quantum mechanics, a series of seminal experiments were performed showing the superposition principle. For instance, the double-slit experiment, with photons or electrons, demonstrates the interference of a single particle with itself, revealing the ``blurring" of the quantum wavefunction prior to measurement. Such sort of experiments open up fundamental and philosophical questions regarding the non-local (and contextual) nature of quantum mechanics. Although, for the case of particles, superposition and interference remain counterintuitive and surprising, when applied to classical waves, they become instinctive and common~\cite{dilip:20}. Thus, quantum phenomena arising from the wave-particle duality does not encapsulate the whole essence of quantum weirdness. In the search for the ``most'' quantum phenomenon, three groups---Hong-Ou-Mandel, Fearn-Loudon, and Rarity-Tapster---independently investigated the interference of ``quantum paths", all within a couple of years during the late 1980s. As to the precise ordering of who ``discovered''---or rather more appropriately ``clarified''---this new effect first is perhaps only known to a few and the original authors themselves; however, in the interest of being unbiased, we recount here only the story of \textit{two-photon interference} as historically reported by the literature.

Beginning in early 1987, several theoretical papers were published in quick succession which formulated the action of the humble beam splitter in terms of quantum mechanics. To start, Prasad, Scully, and Martienssen derived the unitary transformation that couples the input mode set to the outgoing transmitted and reflected mode set~\cite{prasad:87}. A brief comment is made on the consequence of having a non-zero number state in each input mode, stating that a highly correlated superposition of states is created, given by the different ways the total number of input photons can be distributed between the two output modes. Of interest is that a general formulation for the transformed output state is given (in which two-photon interference is embedded, but not mentioned), with a footnote pertaining to private communication with Loudon, whose own work on this subject was forthcoming as we will see.

Published two months later in the same journal, we find a paper by Ou, Hong, and Mandel outlining how to express the output of a beam splitter using the diagonal coherent state representation~\cite{ou:87}; in particular, they give the explicit example of what happens when two photons, one horizontally and one vertically polarized, are incident from different input ports. Indeed, they note that, for a balanced beam splitter, the simultaneous detections at the two output ports behaves like the singlet state for two orthogonally polarized photons, i.e., a Bell state measurement, as we know it today. In tandem with their theoretical work, we come to the (seminal) experimental work by Hong, Ou, and Mandel demonstrating that this interference of two identical photons at a beam splitter, shown as the hallmark ``dip'' in coincidence detections at the output, can be used to measure very short time intervals~\cite{hong:87}. Barely a month later, the work of Fearn and Loudon formulating the action of a lossless beam splitter is published, presumably that which was hinted at by Prasad, Scully, and Martienssen, looking more rigorously into the physics through quantization in terms of both the input and output mode contributions~\cite{fearn:87}. Finally, our last set of players who investigated this two-photon interference effect around the same time as Hong-Ou-Mandel and Fearn-Loudon are Rarity and Tapster. A conference paper of theirs appears in a workshop book, outlining the nonclassical effect in parametric downconversion~\cite{rarity:88}, and later published as a separate paper~\cite{rarity:89}. Much like Hong-Ou-Mandel, Rarity and Tapster experimentally demonstrate two-photon interference from parametric down conversion for the use of subpicosecond measurements.
Curiously, a two-photon interferometer was demonstrated in 1986 by Alley and Shih, as noted in~\cite{shih:88}; however, their results were published in 1988. They performed correlation measurements on two identically polarized photons that were superposed on a beam splitter, observing maximum and minimum coincidences for measurements with parallel- and perpendicularly-aligned polarizers, respectively.

No matter the historical details, two-photon interference - popularly termed now as the Hong-Ou-Mandel (HOM) effect - has been referred to as ``the heart of quantum mechanics'' for being exquisitely quantum in nature, with absolutely no analogue in classical physics. It is precisely this separation from classical to quantum physics that gives two-photon interference a distinct advantage to outperform classical computations, securely communicate information, simulate highly complex physical systems and increase the sensitivity of precise measurements. It has also motivated physicists to study more generally two-particle interference for fermionic and bosonic quantum objects, with successful experiments using massive particles, among others, such as electrons and atoms, in addition to plasmons, demonstrating the extent of this effect to larger and more complex quantum systems. A wide array of novel applications to this quantum effect is to be expected in the future. 

This review attempts to aggregate the most prevelant uses of two-photon interference over the past three decades into one coherent text. To be consistent with the present-day literature, we will simply refer to two-photon (particle) interference as the HOM effect hereafter, or simply HOM for brevity. We begin with the fundamental theory behind the HOM effect in an intuitive manner, along with the experimental considerations. After this, the majority of the review details the many uses of HOM in the optical domain: precise measurements (as per the intention of the original experimental paper), quantum state analysis/engineering, quantum communication/computation, and a generalization to multipartite and multimode systems. Of course, the HOM effect is not limited to photons, and thus we close the review with implementations in non-photonic systems, such as plasmons, phonons, atoms and electrons.

\subsection{Action of the Beam Splitter}
The action of the beam splitter (BS) is at the heart of the two-photon interference effect, providing the mixing between the two input modes. Formally, it can be represented by a unitary transformation, thus conserving energy and preserving orthogonality between input states. Although simple, the BS is an essential building block of any linear optical quantum information processing system. As a starting point for the theory of the HOM effect, we establish the theoretical framework describing the operational action of the BS.  

\begin{figure}[t]
	\centering
		\includegraphics[width=0.47\textwidth]{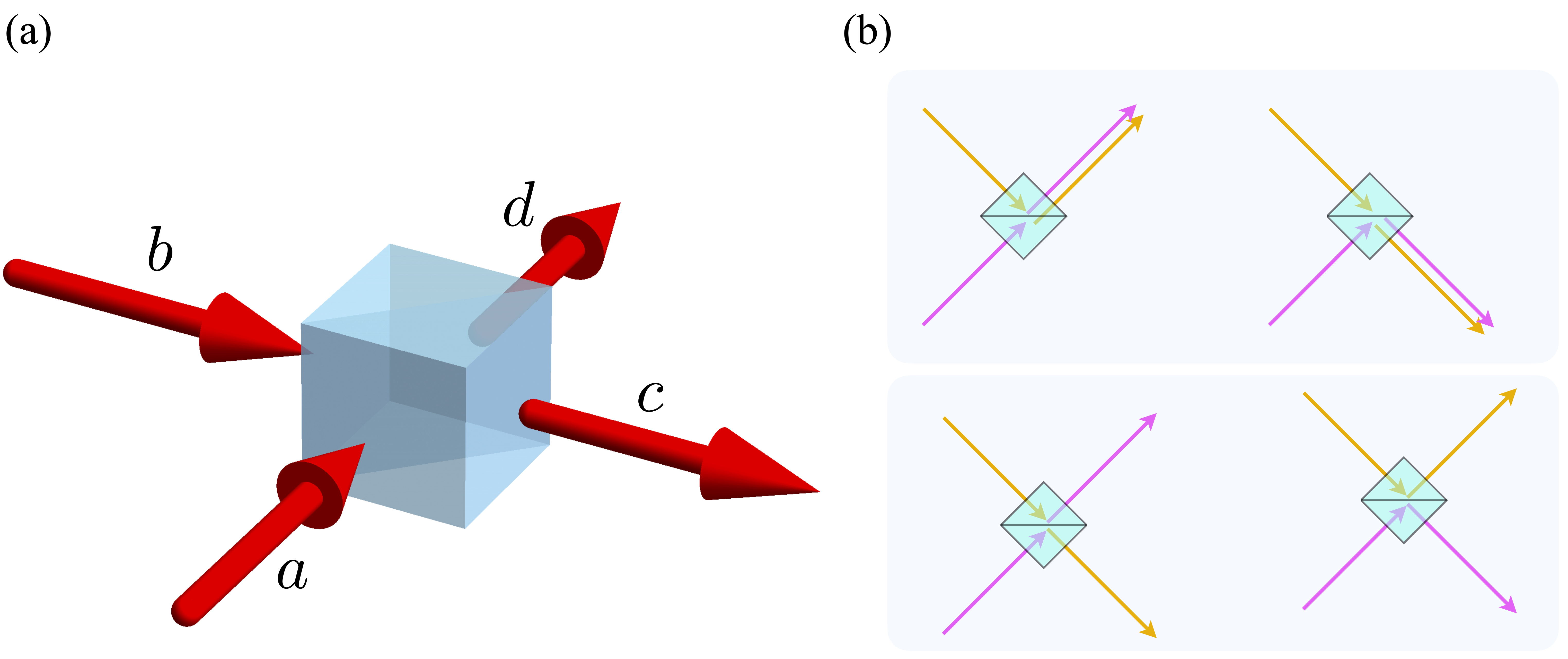}
	\caption{\textbf{Action of a beam splitter.} a) Beam splitter with input ports labelled $a$ and $b$, and output ports labelled $c$ and $d$. Arrows indicate the field propagation directions. b) The four ways the two photons can exit from the beam splitter---through the same port (top row) or different ports (bottom row).}
	\label{fig:BS}
\end{figure}

A BS is an optical device with two input ports, labelled $a$ and $b$, and two output ports, labelled $c$ and $d$, see Fig.~\ref{fig:BS}. A beam incident on a BS at the input port $a$, or similarly for $b$, is split between output ports $c$ and $d$ in proportions depending on complex parameters $r$ and $t$, known as the reflectance and transmittance of the BS, here taken to be lossless. From this point forward, we will consider the prominent case of the balanced BS where $|r|=|t|=1/\sqrt{2}$, also known as the 50:50 BS. Classically, the electric fields in output modes $c$ and $d$ are given in terms of the electric fields in the input modes according to $E_c = \left(E_a + E_b\right)/\sqrt{2}$ and $E_d = \left(E_a - E_b\right)/\sqrt{2}$, where we have chosen a specific phase relation between the reflected and transmitted beams. In particular, this phase relation depends on technical design of the BS, e.g. number of dielectric layers and the coating design~\cite{hamilton:00}. Nevertheless, the physical aspects of the BS discussed hereafter are not affected by this phase relation, as long as energy conservation and unitarity are fulfilled.

We are now ready to move to the quantum description of the BS according to the \emph{second quantization} formalism. This is done by employing a set of bosonic annihilation and creation operators ($\hat{a}_i$ and ${\hat{a}_i}^\dagger$, respectively) to represent electromagnetic fields in mode $i$. The annihilation and creation operators must satisfy the standard bosonic commutation relation, i.e. $[ \hat{a}_i , {\hat{a}_j}^\dagger ] = \delta_{ij}$, where $\delta_{ij}$ is the Kronecker delta symbol. Explicitly, the effect of the creation operator acting on the vacuum is given by,
\begin{eqnarray}\label{eq:photonnumberop}
\left( \hat{a}_i^\dagger \right)^n |0\rangle = \sqrt{n!} \ |n\rangle_i,
\end{eqnarray}
where $|0\rangle$ is the vacuum state and $|n\rangle_i$ is an $n$-photon Fock state in mode $i$. For the input and output modes shown in Fig.~\ref{fig:BS}, we use the notation $\hat{a}$, $\hat{b}$, $\hat{c}$ and $\hat{d}$ to represent annihilation operators in modes $a$, $b$, $c$ and $d$, respectively.  Hence, the operation of a 50:50 BS is given in terms of field operators by,
\begin{eqnarray} \label{eq:transf}
\left\{\begin{array}{c}
\hat{a} =\frac{1}{\sqrt{2}}\left(\hat{c} + \hat{d}\right) \cr
\hat{b} =\frac{1}{\sqrt{2}}\left(\hat{c} - \hat{d}\right)
\end{array}\right.,
\end{eqnarray}
where we have inverted the transformation by representing the input annihilation operators in terms of the output annihilation operators. We have now laid out the theoretical framework necessary to describe the two-photon interference effect.

\subsection{Two-Photon Interference}
The two-photon interference effect is traditionally introduced as follows: assume two photons are incident at each input port of a 50:50 BS. Let us additionally consider that the two photons may be distinguished due to an additional degree of freedom, for example polarization, timing, frequency or spatial modes. Using the notation introduced above, we consider the initial state as ${|\psi_\mathrm{in} \rangle = \ket{1}_{a,H}\ket{1}_{b,V} =  \ket{H}_a\ket{V}_b = \hat{a}_H^\dagger \, \hat{b}_V^\dagger \, |0\rangle}$, where $\hat{a}_H^\dagger$ and $\hat{b}_V^\dagger$ are the creation operators corresponding to the modes of \textit{horizontally} polarized light in input port $a$, $\ket{H}_a$, and \textit{vertically} polarized light in input port $b$, $\ket{V}_b$, respectively. Using the BS transformation relations from Eq.~(\ref{eq:transf}), the output state is given by,
\begin{eqnarray}\label{eq:eq3}
\hat{a}_H^\dagger \, \hat{b}_V^\dagger \, |0\rangle &\stackrel{\mathrm{BS\,\,}}{\longrightarrow}& \frac{1}{2} \left( \hat{c}_H^\dagger + \hat{d}_H^\dagger \right) \left(  \hat{c}_V^\dagger - \hat{d}_V^\dagger \right) \, |0\rangle \\
&=& \frac{1}{2} \left( \hat{c}_H^\dagger \,  \hat{c}_V^\dagger - \hat{c}_H^\dagger \,  \hat{d}_V^\dagger  + \hat{c}_V^\dagger \,  \hat{d}_H^\dagger - \hat{d}_H^\dagger \,  \hat{d}_V^\dagger  \right) \, |0\rangle. \nonumber
\end{eqnarray}

From the output state, we can infer four distinct possibilities for the two photons: the two photons exit the BS together through the same output port, corresponding to the first and last terms above, i.e. $\hat{c}_H \,  \hat{c}_V$ and $\hat{d}_H \,  \hat{d}_V$, or the photons exit the BS separately through different output ports, corresponding to the second and third terms above, i.e. $\hat{c}_H \,  \hat{d}_V$ and $\hat{c}_V \,  \hat{d}_H$. With no surprises, we obtain all the classically expected outcomes.

The two-photon interference now occurs when considering indistinguishable photons. Starting from Eq.~(\ref{eq:eq3}), let the two input photons have the same polarization state, such that they are identical. For simplicity, we can then remove the polarization subscripts such that the output state now reads,
\begin{eqnarray}
\hat{a}^\dagger \, \hat{b}^\dagger \, |0\rangle &\stackrel{\mathrm{BS\,\,}}{\longrightarrow}& \frac{1}{2} \left( \hat{c}^\dagger \,  \hat{c}^\dagger - \cancel{\hat{c}^\dagger \,  \hat{d}^\dagger}  + \cancel{\hat{c}^\dagger \,  \hat{d}^\dagger}- \hat{d}^\dagger \,  \hat{d}^\dagger  \right) \, |0\rangle \\
&=& \frac{1}{2} \left( \Big( \hat{c}^\dagger \Big)^2 - \Big( \hat{d}^\dagger \Big)^2  \right) \, |0\rangle \nonumber \\
&=& \frac{1}{\sqrt{2}} \left(\ket{2}_c - \ket{2}_d  \right). \nonumber
\label{eq:dist}
\end{eqnarray}
Astonishingly, we find a new type of interference effect, physically distinct from the interference of a single photon or classical fields, where it is the overall two-photon states that interfere. In this case, we observe destructive interference of the two-photon states corresponding to photons exiting opposite output ports, and constructive interference of the two-photon states corresponding to photons exiting through the same output ports. Nevertheless, destructive and constructive interference can be adjusted by tailoring the symmetry of the overall input two-photon state, as will be discussed in the following section, noting again that the relative phase is inconsequential.

In the case of fermionic particles, say electrons, the creation operators will anti-commute, implying that $\hat{c}^\dagger\hat{d}^\dagger=-\hat{d}^\dagger\hat{c}^\dagger$. Therefore, $\ket{1}_a\ket{1}_b\stackrel{\mathrm{BS\,\,}}{\longrightarrow}\ket{1}_c\ket{1}_d$ and the fermions will exit through different output ports. This will be discuss in more details in Sec.~\ref{section8}.

\subsection{Entangled States}
In addition to being one of the most peculiar phenomenon in quantum optics, two-photon interference is one of the primary tools used in quantum information processing. At the heart of quantum information is the notion of quantum entanglement~\cite{einstein:35}---the fundamental resource in quantum communication~\cite{ekert:91}, quantum metrology~\cite{dowling:08}, and quantum computing~\cite{kok2007linear}. Quantum entanglement inherently deals with the interference of two-photon states, thus hinting to the fact that the HOM effect might play an important role in the engineering and analysis of entangled states. Although the extension of the HOM effect to multi-photon (more than two photons) will be discussed in Section~\ref{sec:multi}, we will now concentrate on the case of two photons.

Two-photon bidimensional entanglement is conveniently built upon the basis of maximally entangled states, also known as the \emph{Bell states}. Using once again the example of polarization, the Bell states are given by:
\begin{eqnarray}\label{eq:equation5}
| \Psi^{\pm} \rangle_{ab} = \frac{1}{\sqrt{2}} \left( |H\rangle_a |V\rangle_b \pm |V\rangle_a |H\rangle_b \right) \\ \label{eq:equation6}
| \Phi^{\pm} \rangle_{ab} = \frac{1}{\sqrt{2}} \left( |H\rangle_a |H\rangle_b \pm |V\rangle_a |V\rangle_b \right),
\end{eqnarray}
where $|H\rangle$ and $|V\rangle$ represent a one-photon state in the mode of horizontal and vertical linear polarization, respectively. As a first step, let us directly apply a similar calculation, as previously shown, to determine the output state at a 50:50 BS when the Bell states are considered as the input states. After some calculations, we obtain,
\begin{eqnarray} \label{eq:equation7}
| \Psi^{+} \rangle_{ab} &\stackrel{\mathrm{BS\,\,}}{\longrightarrow}& \frac{1}{\sqrt{2}} \left( \hat{c}_H^\dagger \hat{c}_V^\dagger -\hat{d}_H^\dagger \hat{d}_V^\dagger \right) |0\rangle, \\
| \Psi^{-} \rangle_{ab} &\stackrel{\mathrm{BS\,\,}}{\longrightarrow}& \frac{-1}{\sqrt{2}} \left( \hat{c}_H^\dagger \hat{d}_V^\dagger - \hat{c}_V^\dagger \hat{d}_H^\dagger \right) |0\rangle, \nonumber \\
| \Phi^{\pm} \rangle_{ab} &\stackrel{\mathrm{BS\,\,}}{\longrightarrow}& \frac{1}{2 \sqrt{2}} \left( \left(\hat{c}_H^\dagger \right)^2 \pm \left(\hat{c}_V^\dagger \right)^2 - \left( \hat{d}_H^\dagger \right)^2  \mp \left( \hat{d}_V^\dagger \right)^2 \right) |0\rangle. \nonumber
\end{eqnarray}
After examination of the Bell states after passing through the BS, we notice a few interesting features. For the $|\Psi^{+}\rangle$ state, we may notice that the output photons always exit the BS together through the same output ports, just as in the case of indistinguishable photons discussed previously, i.e. $|1\rangle_a |1\rangle_b$. This type of behaviour, know as ``\emph{bunching}'', is also observed for input states $|\Phi^{\pm} \rangle$. However, for the case of $|\Psi^{-}\rangle$, we observe the output photons always exiting the BS in opposite output ports, also known as ``\emph{anti-bunching}''. Interestingly, bunching and anti-bunching are typical \emph{bosonic} and \emph{fermionic} behaviours, respectively. Indeed, those behaviours can be understood from the symmetry of the Bell states~\cite{zeilinger:98}. By definition, a system of bosons is symmetric under the exchange of any pair, whereas a system of fermions is anti-symmetric under the exchange of any pair. Hence, due to their bosonic nature, photon pairs must be in an overall symmetric state, which is clearly the case for $|\Psi^+\rangle$ and $|\Phi^{\pm} \rangle$. 
However, although the $|\Psi^{-}\rangle$ is said to be the ``anti-symmetric state" and exhibits anti-bunching, its overall state, including polarization together with path, \emph{must} be symmetric. Indeed, both the polarization and the path degree of freedom of the photon pair are individually anti-symmetric, leaving the overall state symmetric. Hence, the ability to discriminate the $|\Psi^-\rangle$ Bell state, by taking advantage of its fermionic behaviour, is central to the concept of \emph{Bell state analysis}, as will be discussed in a forthcoming section.

\begin{figure*}[t]
	\centering
	{\includegraphics[width=\textwidth]{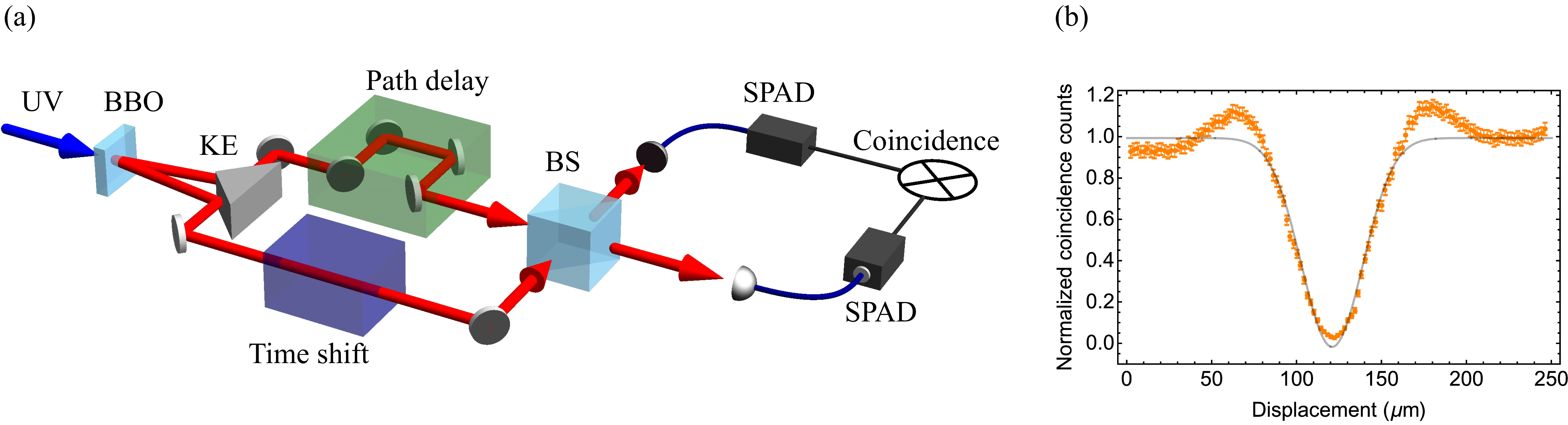}}
	\caption{\textbf{Simplified experimental setup to observe the Hong-Ou-Mandel dip.} (a) We present an experimental setup similar to that presented by Hong, Ou and Mandel. An ultraviolet (UV) laser pumps a nonlinear crystal, e.g. KDP, BBO or ppKTP. Pairs of photons are generated with anti-correlated linear momentum and separated using a knife-edge (KE) mirror. The photons are brought back together at a 50:50 BS, where a variable path delay is scanned to control the arrival time of one of the photons. The photons exiting the output ports of the BS are detected using single-photon avalanche diode (SPAD) detectors and coincidence counts are recorded. (b) Example of experimental results showing the two-photon interference dip, dropping to zero when the two photons enter the BS simultaneously. Solid line indicated expected theoretical coincidence counts, and dots indicate experimental measurements. The peak in counts on either side of the dip is caused by the use of a rectangular bandpass filter in experiment, as compared to a Gaussian filter in theory. Figure legends: UV, ultraviolet beam; BBO, Beta barium borate nonlinear crystal; KE, knife edge; BS, 50:50 beam splitter; SPAD, single photon avalanche diode.}
	\label{fig:timing}
\end{figure*}

\subsection{Two-photon interference with qudits}

Here, we extend the HOM effect to higher dimensions. We use the notation $|n\rangle_i$ to describe an $n$ photon number state of mode $i$, see Eq.~(\ref{eq:photonnumberop}). Recall that a mode can be defined by polarization, frequency and spatial degrees of freedom. To define a $d$-dimensional basis, we pick $d$ modes, $\{\text{`0',`1'},\ldots,\text{`d-1'}\}$, and we populate them with one photon in total. We are therefore working within the 1-photon multiplet of the total photon number operator $\hat n = \sum_i\hat{a}^\dagger_i \hat{a}_i$.

Let us consider two qudit photon states ($|\phi\rangle=\sum_{i=0}^{d-1}\alpha_{i}\ket{1}_\text{`i'}$ and $|\psi\rangle=\sum_{j=0}^{d-1}\beta_{j}\ket{1}_\text{`j'}$) incident on a 50:50 BS, where the notation implies that the omitted modes are all in vacuum. If we input these two states at different input ports, we obtain:
\begin{eqnarray} \label{eq:qudit}
     \ket{\phi}\ket{\psi}&=&\sum_{i=j=0}^{d-1}\alpha_{i}\beta_{j}\left(\frac{\ket{2}_\text{`i'}\ket{0}_\text{`j'}-\ket{0}_\text{`i'}\ket{2}_\text{`j'}}{\sqrt{2}}\right)\nonumber\\
     &+&\sum_{i\neq j=0}^{d-1}
     \left(\frac{\alpha_{i}\beta_{j}-\alpha_j\beta_i}{2}\right)\ket{1}_\text{'i`}\ket{1}_\text{`j'}.
\end{eqnarray}
The coincidence probability is given by the last term in Eq.~(\ref{eq:qudit}), i.e.,
\begin{eqnarray}
     \sum_{i\neq j=0}^{d-1}\frac{1}{4}(\alpha_{i}\beta_{j}-\alpha_{j}\beta_{i})(\alpha_{i}\beta_{j}-\alpha_{j}\beta_{i})^{*}=\frac{1-|\braket{\phi}{\psi}|^2}{2},
\end{eqnarray}
and the two-photon probability is then $\left( 1+|\braket{\phi}{\psi}|^2 \right) /2$. We note that when $\ket{\phi}$ and $\ket{\psi}$ are identical there will be no coincidences and the last term in Eq~(\ref{eq:qudit}) will vanish. While for completely distinguishable states, i.e. $\braket{\phi}{\psi}=0$, the two photons will bunch half of the time. Thus, HOM can be adopted as a powerful tool to directly measure the overlap between two single photon wavepackets.

\section{Precision Measurements}

In the original paper by Hong, Ou and Mandel~\cite{hong:87} (as well as Rarity and Tapster~\cite{rarity:88}), the quantum optical formalism of the two-photon interference is presented and the famous HOM dip is experimentally observed. As will be discussed in the upcoming sections, the two-photon interference is fundamental to a plethora of quantum information tasks. Nevertheless, Hong, Ou and Mandel present the HOM dip and discuss its applications in terms of precise timing measurement. In particular, they draw the link between their scheme and the well known techniques used to measure short pulses using nonlinear materials, e.g. intensity auto-correlation. Subpicosecond time intervals can now be measured for photon pairs from parametric down-conversion going far beyond the previous timing resolution of standard photodetectors, i.e. larger than 100~ps. Since then, several experiments have investigated the use of the HOM effect to perform precise timing measurements.

\subsection{Precise Timing Measurement}


In recent years, there has been a renewed interest in using the HOM effect to measure minute time variations. Motivated by the body of work on structured light that has emerged in the last few decades~\cite{rubinsztein:16}, an interest in the propagation of structured beams in free-space led to a series of experiments measuring time shifts due to the modal structure of the propagating beams~\cite{giovannini:15,bouchard:16,lyons:18,richard:18}. In some of these experiments, a shift in the HOM dip was used to evaluate a temporal shift due to the spatial structure of the beam~\cite{rubinsztein:16}. 

A general setup to perform precise timing measurements using HOM is shown in Fig.~\ref{fig:timing}a). In the first of these experiments~\cite{giovannini:15}, the group velocity of single photons with the profile of a Bessel beam and a focused Gaussian beam is compared to that of a collimated Gaussian beam. In this experiment, entangled photon pairs are generated through spontaneous parametric down-conversion (SPDC), and the signal and idler photons are separated using a knife-edge mirror. It is the signal photon that experiences the time shift, due to a variation in its group velocity. In~\cite{giovannini:15}, the variation in group velocity is induced by structuring the transverse profile of the signal photon. This is achieved by sending the signal photon onto a spatial light modulator (SLM) which can be programmed to act as a diffractive optical element implementing axicons or lenses to modulate the photons into a Bessel beam or a focused Gaussian beam. The signal photon is then allowed to propagate in free space before arriving at a second SLM which reverses the structure of the photon introduced by the first SLM. The signal photon is then made incident onto one of the input ports of a 50:50 BS with the idler photon incident in the other input port. By varying the path length of one of the photons and recording coincidence counts at the output port of the BS, the HOM dip is measured where the minimum of the dip sets a time reference. As a next step, the signal photon's group velocity can be varied using the SLMs, resulting in a different time shift which can be observed by a shifted HOM dip. The shift of the HOM dip is then related to a direct time shift in the arm of the signal photon. In another experiment, variations in the time of flight is also investigated using a similar experimental setup~\cite{lyons:18}. In this case, time shifts arise by varying the optical orbital angular momentum (OAM) value of the signal photon. The addition of OAM in the phase of the beam is compared to a beam with the same intensity profile at the SLM, but no net OAM. Once again, the HOM effect is used as a highly precise timing measurement tool.

In the works mentioned above, a noncommon-path HOM interferometer is employed to measure time shifts with resolutions on the order of a few femto-seconds (micrometers), where the time shifts are obtained by assessing the shift of the HOM dip. Thus, one might expect the limit on time resolutions of this apparatus to be given by the width of the HOM dip, or equivalently the coherence time of the photons. According to this argument, larger time resolutions can be achieved by employing ultra-broadband single photon sources. However, it has recently been shown that by considering the limits dictated by statistical estimation theory, it is possible to push the time resolution of a HOM interferometer to sub-femtosecond resolutions~\cite{lyons:18b}. At the heart of estimation theory, is the Cram\'er-Rao bound which poses the ultimate limit on the precision of the estimation of a parameter, in this case the time shift. In the work of~\cite{lyons:18b}, a standard HOM interferometer, similar to the one shown in Fig.~\ref{fig:timing}, is used to measure a time shift. In particular, the Cram\'er-Rao bound is saturated using the maximum-likelihood estimator, and the Fisher information---which depends on the coherence time of the photons, the visibility of the HOM dip, and losses---is maximized. By doing so, an average accuracy and average precision of 6 attoseconds and 16 attoseconds were achieved, respectively. Moreover, it is also possible to maximize the Fisher information by custom-tailoring the state of the probe photons using frequency entanglement~\cite{chen2019hong}.

\subsection{Quantum Optical Coherence Tomography}

Another particularity of the two-photon interference is its inherent dispersion cancellation, which does not occur in the case of the one-photon interference, or the interference of classical light. In particular, it is the even-order effects of dispersion, such as group-velocity dispersion, that are cancelled~\cite{steinberg:92}. Thus, the two-photon interference effect holds great potential for applications where dispersion can seriously limit optical applications. A great example of such an application is the medical imaging technique known as optical coherence tomography (OCT), where a low-coherence light source is used in an interferometric setup to reconstruct the 2-dimensional or 3-dimensional profile of tissues~\cite{povazay:02}. In theory, the resolution of OCT is limited by the coherence length of the light source; however, in practice, it is the material dispersion that limits the resolution.

\begin{figure}[t]	
	\centering
	{\includegraphics[width=0.45\textwidth]{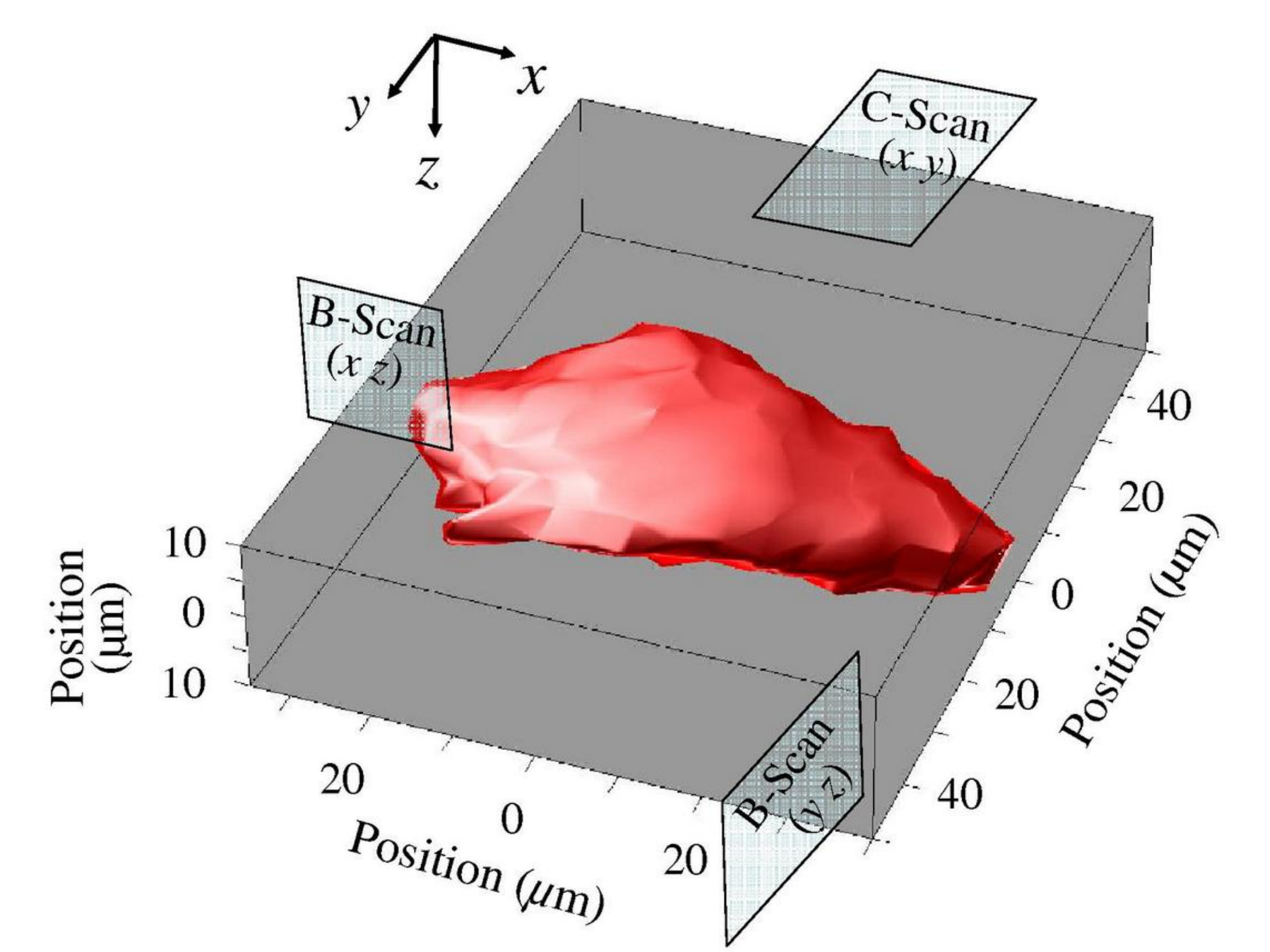}}
	\caption{\textbf{Example of quantum optical coherence tomography}. 3-dimensional image of onion-skin tissue coated with gold nano-particles taken with QOCT. Image taken from \cite{Nasr2009}.\label{fig:QOCT}}
\end{figure}

In order to overcome the resolution limitations, a technique known as quantum optical coherence tomography (QOCT) has been proposed to exploit the automatic dispersion cancellation involved in the two-photon interference effect~\cite{abouraddy:02}. Moreover, for the same bandwidth, QOCT benefits from an extra factor of 2 in resolution compared to classical OCT. The experimental configuration of QOCT would be similar to that presented in Fig.~\ref{fig:timing}, where the sample is introduced in one of the arms of the two-photon interferometer and coincidence counts are recorded while varying the path of the other arm using a variable delay stage. Not long after its proposal, QOCT was experimentally demonstrated, achieving an improvement in resolution by a factor of 5 compared to OCT~\cite{nasr:03}. Further experimental demonstrations of QOCT have also been reported, where micron-sized features are detected even for the case of biological samples~\cite{nasr:04,nasr:09,lopez:12}. Finally, although QOCT offers potential advantages in terms of resolution, new challenges appear, such as low signal and signal artifacts. Thus, new techniques are being investigated in order to exploit the full potential of QOCT~\cite{erkmen:06,banaszek:07,resch:07,mazurek:13}. An example of a 3-dimensional image taken via QOCT for the skin of an onion coated in gold is shown in Fig.~\ref{fig:QOCT}; this result indicates a resolution of 1~$\mu$m.


\subsection{Quantum metrology}
Finally, another field of research where the HOM effect plays an essential role for precision measurement is \textit{quantum metrology}, where quantum resources, such as entanglement and non-classical states of light, are used to achieve higher precision on the measurement of a physical property~\cite{Boto2000, giovannetti:11}. The most well-known class of useful states for quantum metrology is the N00N states~\cite{dowling:08}, given by,
\begin{eqnarray}
|\psi_\mathrm{N00N} \rangle = \frac{1}{\sqrt{2}} \left( |N\rangle_a |0\rangle_b + |0\rangle_a |N\rangle_b \right).
\end{eqnarray}
Here, $\ket{N}$ represents $N$ photons in the Fock basis.

For the trivial case of $N=1$, such state is realized by sending a single photon through a 50:50 BS. For the case of $N=2$~\cite{rarity1990two,edamatsu2002measurement}, we recognize the state from Eq.~(\ref{eq:dist}) where two indistinguishable photons are made incident onto both input ports of a 50:50 BS, resulting in HOM interference. Thus, one can already appreciate the importance of the HOM effect in quantum metrology relying on the entanglement of multi-photon states. The resolution in space and phase is enhanced by a factor of $N$, i.e. scales as $1/N$. Thus, in order to achieve the fundamental quantum limit, also known as the Heisenberg limit, for phase sensitivity, one must consider high photon number N00N states. Several schemes have thus been proposed to achieve high photon number N00N states experimentally~\cite{mccusker2009efficient,kapale2007bootstrapping,cable2007efficient,pezze2008mach} along with experimental demonstrations~\cite{d2001two,mitchell2004super,walther2004broglie,kim2009three,liu2008demonstration,resch2007time,okamoto2008beating,nagata2007beating,afek2010high,zhou2017superresolving}, where most of these schemes are in the spirit of the two-photon, or more generally multi-photon, interference set by the HOM effect.

\section{Quantum state analysis}

In the previous section, we have reviewed several experiments where the two-photon interference effect can be taken advantage of in order to perform precision measurements in terms of timing or phase sensitivity. Another case in which the two-photon interference is of paramount importance is the analysis of quantum states, such as pairs of indistinguishable and entangled photons. In particular, the measurement of the HOM visibility is an invaluable diagnostic tool to characterize single photon sources and states. Furthermore, Bell state measurements are the gold standard for experimentally measuring maximally entangled states. 

\subsection{Mode distinguishability}

At the heart of the HOM effect, there is the concept of distinguishability of the input photons. In other words, two photons are said to be indistinguishable if they are in the same mode of the electromagnetic field, i.e. polarization, time, frequency, position and momentum. In a two-photon interference experiment, maximal interference occurs when the input photons are indistinguishable at both input ports of the BS. The level of indistinguishability is typically determined using the HOM interference visibility, ${\cal V}$, given by,
\begin{eqnarray}
{\cal V}=\frac{{\cal C}_\mathrm{max} - {\cal C}_\mathrm{min}}{{\cal C}_\mathrm{max}},
\end{eqnarray}
where ${\cal C}_\mathrm{max}$ and ${\cal C}_\mathrm{min}$ correspond respectively to the maximal and minimal coincidence count rates at the output of the BS by varying the arrival time of one of the input photon, see Fig.~\ref{fig:timing}b). Indeed, the HOM visibility has been used to characterize the level of mode indistinguishability for a wide range of single photon sources, i.e. spontaneous parametric down-conversion~\cite{kaltenbaek2006experimental,mosley2008heralded}, quantum dots~\cite{sanaka2009indistinguishable,flagg2010interference,patel2010two,wei2014deterministic,senellart:17}, atomic vapours~\cite{felinto2006conditional,chaneliere2007quantum,yuan2007synchronized,yuan2008experimental,chen2008memory}, nitrogen-vacancy centres in diamond~\cite{bernien2012two,sipahigil2012quantum,sipahigil2014indistinguishable}, trapped ions~\cite{maunz2007quantum}, trapped neutral atoms~\cite{beugnon2006quantum,specht2011single}, and molecules~\cite{kiraz2005indistinguishable,lettow2010quantum}. The experimental realization of indistinguishable single photon sources with high purity is a crucial component for engineering large-scale entangled quantum states from independent sources, as will be seen in forthcoming sections.

\subsection{\label{section:bellstate} Bell state measurements}
In general, the discrimination of Bell states with linear optics comes with fundamental limitations. For example, if we use exclusively linear optics and no other ancillary modes, we are bound to a 50\% discrimination success rate \cite{calsamiglia2001maximum}. Higher success rates can be achieved with nonlinear optics \cite{kim2001quantum}, entanglement in auxiliary modes \cite{kwiat1998embedded, barreiro2008beating}, or feed-forward techniques \cite{knill2001scheme}. The simplest implementation uses only one interference at a single beam splitter and two detectors. We start with two maximally entangled modes in the Bell basis, $\ket{\Psi^\pm}_{ab}$ and $\ket{\Phi^\pm}_{ab}$ from Eq.~(\ref{eq:equation5}) and ~(\ref{eq:equation6}), respectively. Recall that the BS transforms the Bell states according to Eq.~\ref{eq:equation7}, explicitly given as,
\begin{eqnarray}
| \Psi^{+} \rangle_{ab} &\stackrel{\mathrm{BS\,\,}}{\longrightarrow}& \frac{1}{\sqrt{2}} \left( \ket{1}_{c,H} \ket{1}_{c,V} - \ket{1}_{d,H} \ket{1}_{d,V} \right), \\
| \Psi^{-} \rangle_{ab} &\stackrel{\mathrm{BS\,\,}}{\longrightarrow}& \frac{-1}{\sqrt{2}} \left(\ket{1}_{c,H} \ket{1}_{d,V}-\ket{1}_{c,V} \ket{1}_{d,H} \right), \nonumber \\
| \Phi^{\pm} \rangle_{ab} &\stackrel{\mathrm{BS\,\,}}{\longrightarrow}& \frac{1}{2} \left( \ket{2}_{c,H}\pm\ket{2}_{c,V}-\ket{2}_{d,H}\mp\ket{2}_{d,V}\right), \nonumber
\end{eqnarray}
where $\ket{N}_{p,s}$ is $N$ photons in the path $p$ with polarization $s$. Again, we see the bunching of the HOM effect acting on the last two states. In the first case, both photons end up in the same detector, but they have opposite polarization; in the second case, we have one photon per detector, and in the last case they also reach the same detector, but they have the same polarizations. We cannot distinguish the last two states, $\ket{\Phi^\pm}$, from each other with a simple pair of photon detectors; hence, the success rate of 50\%. As anticipated, this figure can be improved by supplying some additional entangled modes. As outlined in \cite{grice2011arbitrarily}, an example can be to arrange four beam splitters so that we interfere the original state with another entangled ancillary pair of modes first, and then each of the two pairs of output modes goes through a simple Bell state analyzer. In this way, we can distinguish one of the two states that we could not distinguish before, raising the success rate to $75\%$. It is possible to combine $N$ interferences and raise the success rate to $1-2^{-N}$.

\section{Quantum communication}

A major and growing part of quantum technologies is the field of quantum communications and quantum cryptography~\cite{gisin:02}. Carriers of information made of single quanta of energy, such as photons, possess a surprisingly large potential for secure communications. The Heisenberg uncertainty principle and the no-cloning theorem~\cite{wootters:82} are the basis of security when encoding information on single photons, since any type of eavesdropping results in disturbances of the photonic quantum states. Another key physical principle that is at the heart of quantum communications is entanglement~\cite{einstein:35}. In particular, quantum cryptography takes an elegant form when explained in terms of entanglement-based schemes~\cite{ekert:91} and can further simplify security analyses~\cite{shor:00}. Beyond quantum cryptography, many quantum communication schemes---such as entanglement swapping~\cite{zukowski:93}, quantum teleportation~\cite{bennett:93}, and dense-coding~\cite{bennett:92a}---are also based on the concept of entanglement.

In the following section, we review the role of the two-photon interference effect in established schemes such as quantum teleportation and entanglement swapping. Subsequently, we review recently introduced quantum cryptographic schemes that are based on the two-photon interference effect. In particular, the \emph{measurement-device-independent}~\cite{lo:12} and the passive \emph{round-robin differential phase-shift}~\cite{guan:15} quantum key distribution (QKD) protocols are considered, which are milestones for the emerging field of \emph{practical} QKD. 

\subsection{\label{section:entanglementswapping} Teleportation and entanglement swapping}
\begin{figure}[t]	
	\centering
	{\includegraphics[width=0.45\textwidth]{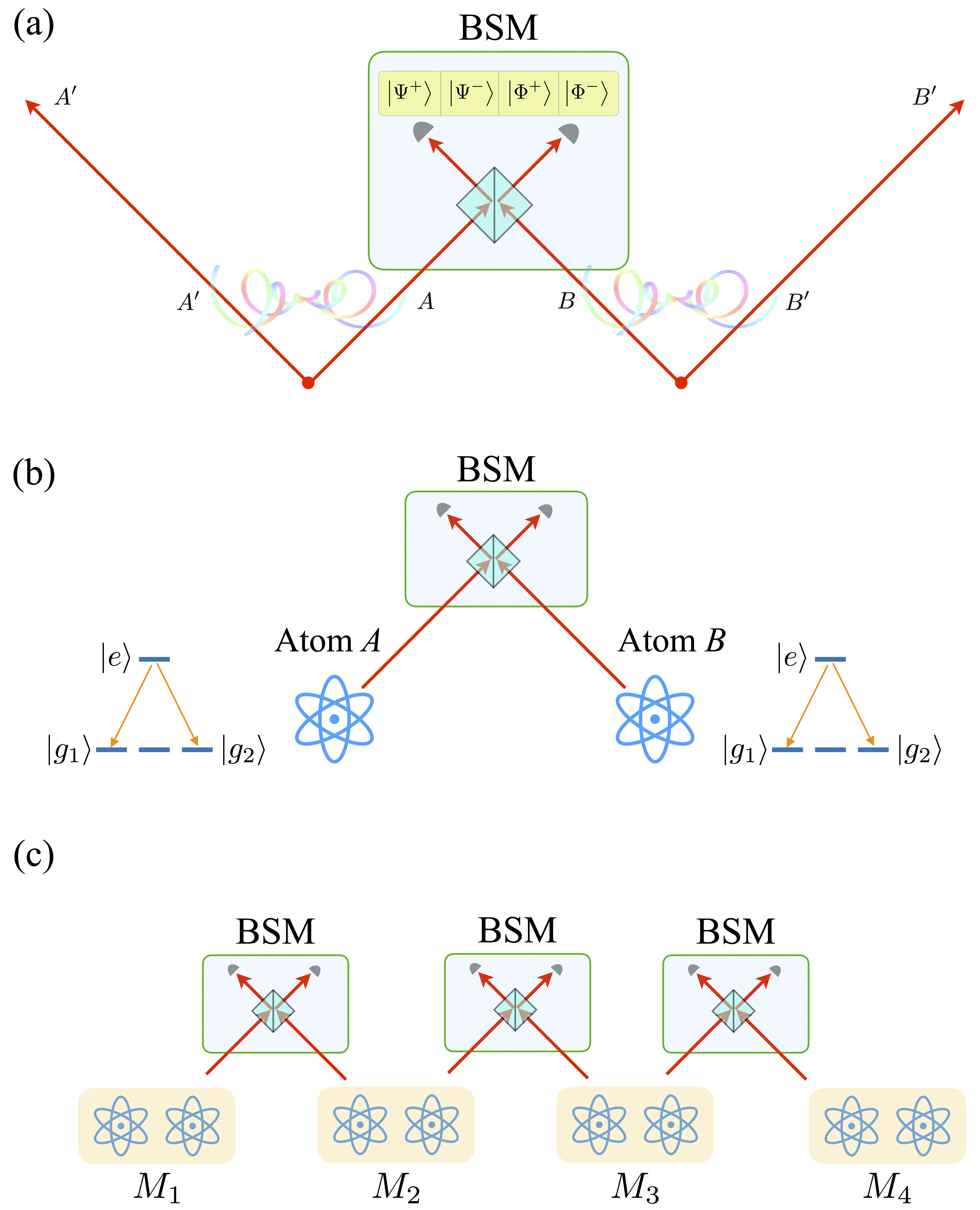}}
	\caption{\textbf{Schematic of performing entanglement swapping.} a) Given two entangled systems $AA'$ and $BB'$, $A$ and $B$ are made to interfere at a 50:50 BS in the form of a Bell state measurement (BSM). Thus, $A'$ and $B'$ are consequently entangled based on the outcome of the BSM. b) A BSM is performed on two photons created from two separate excited atoms. Similar to a), the two atoms become entangled based on the outcome of the BSM. c) A quantum repeater scheme with four memories in which each memory contains two separate atoms. Following b), a BSM is made on two photons from neighbouring memories to entangle the neighbouring atoms. By cascading such a process, the entanglement can be distributed over long distances. \label{fig:entanglementSwapping}}
\end{figure}
The idea of entanglement swapping is to start with a pair of entangled systems, $AA'$ and $BB'$, and to measure systems $AB$ in the Bell basis, with the consequence of leaving systems $A'B'$ entangled with each other. This is entanglement swapping in a nutshell. The way we can achieve it is indeed to perform a Bell state measurement, whose outcome determines in which Bell state the other two systems are left.

For example \cite{hofmann2012heralded}, consider a single trapped excited atom with a degenerate ground state corresponding to two distinct photon polarizations. In such a way, when the atom emits a photon, the atom-photon state is $\frac{1}{\sqrt{2}}(|g_1,H\rangle+|g_2,V\rangle)$, where $g_1$ and $g_2$ are the two orthogonal ground states. When there are two such events from two different atoms (from different cells), the emitted photons can be interfered at a beam splitter to perform a linear optical Bell state measurement, as discussed in the previous section. If successful, this procedure has swapped the entanglement from being between each atom and its photon to being between the two atoms alone. We have thus created an entangled state of two atoms, which will display non-local correlations as one can verify for example by violating Bell's inequalities~\cite{Bell:64}.
This idea can be applied to other domains as well, for example to entangle a pair of transmon qubits by interfering two microwave photons at a microwave beam splitter junction \cite{narla2016robust}. 

Another domain of utilization is for quantum repeaters \cite{sangouard2011quantum}: the so-called ``two-way architecture'' works by having quantum memories evenly distributed along a communication channel, each memory hosting two atoms that emit photons as described above, one forwards and one backwards along the channel. Mid-way between two memories $M_1M_2$ sits a Bell state analyzer that measures the two photons coming from $M_1$ and $M_2$ in the Bell basis. If the measurement is successful, we have entangled a pair of atoms, one in $M_1$ and one in $M_2$. As soon as this procedure is successful between the second atom in $M_2$ and an atom in $M_3$, the memory $M_2$ can itself measure its two atoms in the Bell basis, with the consequence of entangling a pair of atoms, one in $M_1$ and one in  $M_3$. One can continue with this protocol until the first and the last memories are entangled, and we have thus effectively obtained an entangled pair of atoms among potentially very distant locations \cite{briegel1998quantum}. These can then be used to perform any kind of protocol, from quantum key distribution to distributed quantum computations \cite{sangouard2011quantum}.

\subsection{Measurement device independent QKD}

In theory, quantum key distribution promises the exchange of secret information, where the security relies on fundamental physical principles. Nevertheless, practical implementations do not necessarily follow the assumptions made in the security proofs. Thus, security loopholes have, early on, been recognized to pose a threat to the security of realistic QKD implementations~\cite{fung:07,zhao:08}. For instance, several side channel attacks have been demonstrated experimentally, showing the vulnerability of the commercial QKD systems under study~\cite{lydersen:10,gerhardt:11}. In order to overcome these security issues, several solutions have been proposed. One such solution is known as the full device independent QKD~\cite{mayers:98,acin:07}, where the security can be proven without any knowledge of the devices used in the implementation. However, this scheme yields very low secret key rates, and is highly impractical due to several stringent requirements such as near unity detection efficiencies. In order to overcome the impracticality of this scheme, the measurement-device-independent (MDI) QKD protocol has been proposed~\cite{lo:12}, where all detector side channels attacks are now removed. In contrast to the full device independent scheme, MDI QKD protocol still necessitates the assumptions of Alice's and Bob's states generation to be perfect, which is still reasonable for current implementations.

The two-photon interference and Bell state analysis are central concepts in MDI-QKD. The protocol goes as follows: Alice and Bob randomly prepare their individual photons in a particular state, then distribute their photons to a third untrusted partner, named Charlie. Upon receiving Alice's and Bob's photons, Charlie performs a Bell state measurement by making each photon incident on a beam splitter. The two photons are made to arrive simultaneously on the beam splitter, in order to observe two-photon interference. Finally, Charlie publicly announces the result of his Bell state measurement, allowing Alice and Bob to establish a shared raw key. Usual classical post-processing, such as error correction and privacy amplification, is then applied in order to obtain a secure secret key between Alice and Bob. The MDI QKD protocol has been experimentally demonstrated using time-bin phase-encoding~\cite{liu:13}, and polarization encoding~\cite{tang:14}. Moreover, the MDI scheme has also been demonstrated over long distances through optical fibres~\cite{tang:14b,yin:16}, showing the feasibility of the protocol and its potential for unconditional security. In the aforementioned implementations of MDI QKD, to achieve high secret key rates, attenuated laser pulses were employed along with the decoy state protocol~\cite{lo:05}. In order to achieve an optimal HOM interference, the attenuated lasers are phase randomized, achieving a maximum HOM visibility of 1/2. A heralded single photon source achieving a unity HOM visibility would lead to a higher secret key rate per transmitted photon.  

\subsection{Passive round-robin differential phase shift QKD}

Another milestone in QKD is the round-robin differential-phase shift (RRDPS) protocol~\cite{sasaki:14}. An important task in any practical QKD implementation is the assessment of the information leakage to an adversary eavesdropper. By determining this quantity, one may then perform the appropriate amount of classical post-processing to the raw key shared between Alice and Bob. However, determining the amount of information leakage requires active monitoring of the quantum bit error rate, or other quantities such as interference visibilities. This stringent requirement may render several QKD implementations impractical or lower its key rate efficiency. The RRDPS protocol removes the requirement for signal disturbance monitoring by bounding the amount of information leaked to the eavesdropper, where the bound depends entirely on Alice's generation stage. Hence, the RRDPS scheme is another major step towards practical implementation of unconditionally secure quantum communication systems.

Several realizations of the RRDPS protocol have recently been demonstrated experimentally using time-bin phase encoding~\cite{takesue:15,wang:15,li:16,Yin:18} and using transverse spatial modes~\cite{bouchard:18c}. Although a single bit of information per transmitted pulse is encoded, the RRDPS QKD can be considered as a high-dimensional protocol. Dimensions ranging between 5 and 128 have been achieved in the time-bin configuration using interferometers with variable delays. In particular, the security of the RRDPS protocol is largely enhanced when considering large dimensions; therefore, it is worth exploring different implementations achieving larger dimensions. In order to overcome the difficulties in achieving stable variable-delay interferometers, a passive version of the RRDPS QKD was proposed~\cite{guan:15}. In this scheme, the two-photon interference is exploited to achieve extremely large dimensions such as $10^5$. Instead of measuring the interference of a train of pulses at Bob's stage, Bob generates a reference pulse train, and sends Alice's and his state to a beam splitter. By using such a configuration, Bob takes advantage of the phase stability of the two-photon interference effect. Finally, it has recently been shown that the passive RRDPS protocol could also be extended to other high-dimensional QKD protocols~\cite{islam2019scalable}.

The passive RRDPS QKD is another demonstration of the advantage of using the two-photon interference effect in quantum communications protocols. As a result, such a configuration typically yields larger security, better stability, and faster communications.

\section{Quantum state engineering}
One of the necessary ingredients for quantum technologies is the ability to manipulate quantum systems to bring them to the desired state. For example, in metrological applications, certain states are more sensitive than others to the effect that we want to measure; or in quantum computation, the system should be initialized in a subspace that offers protection from errors. 
Furthermore, in light of the Choi-Jamio\l kowski isomorphism which allows us to associate a bipartite state to a quantum channel~\cite{choi:75, jam:72}, generating the desired quantum states also has consequences on our ability to implement full quantum channels through the technique of gate teleportation.

In this section, we describe some of the achievements of quantum state engineering: the measurement and generation of entangled states, quantum cloning, and the technique of state merging---all through the lens of the HOM effect.

\subsection{Entanglement engineering}
Starting from a highly entangled pair of particles, such as the photon pairs generated in SPDC, it is possible to engineer the state to one's needs using the HOM effect. 
Any input state of the beam splitter with one photon in each input port can be written as $\ket{\Psi} = \sum_{i,j=1}^d c_{ij}\ket{1}_{a,i}\ket{1}_{b,j}$, with $\ket{1}_{p,k}$ representing one photon in path $p$ with mode $k$. It is shown in~\cite{zhang2016engineering} that this can be rewritten as a superposition of symmetric and antisymmetric Bell states $\ket{\Psi}=\sum_{i\neq j} \tfrac{c_{ij}}{\sqrt{2}} \left(  \ket{\Psi^+_{ij}} + \ket{\Psi^-_{ij}} \right) + \sum_i c_{ii}\ket{\Psi^+_{ii}}$, where $\ket{\Psi^\pm_{ij}}=\ket{1}_{a,i}\ket{1}_{b,j}\pm\ket{1}_{a,j}\ket{1}_{b,i}$. Upon exiting the beam splitter, the symmetric component, $\ket{\Psi^+_{ij}}$, of the input state will reseult in two photons being detected in one of the output ports, while the antisymmetric component, $\ket{\Psi^-_{ij}}$, will result in one photon being detected in each output port. Therefore, when conditioning on coincidence detection between the two output ports, the HOM effect acts as a filter for the antisymmetric component for any arbitrary high-dimensional input state. 

A second example is a way of entangling photons with different frequencies \cite{zhao2014entangling}. The idea here is to create two pairs of photons and make a joint Bell state measurement, implementing what is known as ``entanglement swapping'' (see Sec.~\ref{section:entanglementswapping}). The frequency difference $\Delta \omega$ generates a phase of $\exp(i\Delta\omega\Delta t)$ between the components, which will render the state mixed if unaccounted for because the detection time difference $\Delta t$ is random. However, one can solve this issue either by postselecting on only the measurements that fall within a very narrow time window (but this reduces the total number of successful measurements), or by recording the time difference and feed-forward a phase correction so that every single photon pair will have the same phase.

\subsection{Universal optimal quantum cloning and NOT gate}
Another application of the HOM effect is \emph{quantum cloning}. Considering the well-known fact that quantum mechanics does not allow the possibility to clone an unknown state~\cite{wootters:82, peres:03}, the ability to perform quantum cloning may seem counterintuitive. The key is that, in quantum mechanics, we are allowed to perform a task as long as we cannot use it to violate the linearity of the theory. The no-cloning theorem states that we cannot \emph{always} clone an unknown state, which is indeed correct; however, this does not prevent us from creating a clone some of the time with random success.
Optimal $1\rightarrow 2$ cloning (i.e. two clones of one original state) of a qudit can be achieved with a fidelity of $F_{cloning} = \tfrac{1}{2} + \tfrac{1}{d+1}$~\cite{navez:03,nagali:09,nagali:10, bouchard:17}.

For example, an arbitrary polarization qubit $\ket{1}_{s}=\alpha\ket{1}_{H}+\beta\ket{1}_{V}$, where $\abs{\alpha}^2+\abs{\beta}^2=1$, can be optimally cloned using the HOM effect by interfering it on a beam splitter with an ancilla that is in a maximally mixed polarization state $\hat{\rho} = \left( \ket{1}_{s}\!\bra{1} + \ket{1}_{s_\perp}\!\bra{1}\right)/2$, where $s_\perp$ is the orthogonal polarization state. The total dimension of the system is $d+1=3$, since the mixed state has dimension 2, while the arbitrary polarization qubit is an independent vector, adding an additional dimension.
When the two photons exit through the same port, 2/3 of the time they will have the same polarization state; while 1/3 of the time, they will have orthogonal polarization states, $\ket{1}_s\ket{1}_{s_\perp}$. Overall, the cloning fidelity is then $F_{cloning} = \frac{2}{3}\times1 + \frac{1}{3}\times\frac{1}{2} = \frac{5}{6}$.


Suppose we created the maximally mixed polarization state by using one of the photons from a maximally entangled singlet state. Then 2/3 of the time, the other photon in the singlet state will have the orthogonal polarization of the cloned states, and 1/3 of the time the same polarization. This effectively implements an optimal universal NOT gate~\cite{demartini:02,ricci:04}---itself a non-unitary operation.


\subsection{Quantum state joining}
Quantum state joining is the process of transferring the state of two particle onto a third one (replacing its state). We thus require a particle to have enough degrees of freedom to accommodate the information of both incoming states. For example, when working with photons, this can be achieved by exploiting the spatial degree of freedom in addition to polarization, but one could also use the frequency/time degree of freedom.

The Knill-Laflamme-Milburn (KLM) linear optical implementation of the CNOT quantum gate, which will be discussed in detail in Sec.~\ref{section:KLM}, plays the key role for achieving quantum state joining. Experimentally, quantum state joining has been achieved for the polarization state of two photons, in which the information was transferred onto a third single photon by exploiting the fact that a single particle can occupy different spatial modes and polarizations~\cite{vitelli2013joining}. This process is probabilistic and requires an ancilla state, with a success probability of 1/32, or 1/8 if a suitable feed-forward is used. Explicitly, the polarization state of two particles
\begin{eqnarray}
\ket{\phi}_{\text{I}}\otimes\ket{\psi}_{\text{II}} &=& (\alpha\ket{1}_{\text{I},H}+\beta\ket{1}_{\text{I},V})\otimes(\gamma\ket{1}_{\text{II},H}+\delta\ket{1}_{\text{II},V})\quad \\
    &=& \alpha\gamma\ket{1_{H},1_{H}} + \beta\gamma\ket{1_{H},1_{V}}\cr
    & &  + \alpha\delta\ket{1_{V},1_{H}}
     + \beta\delta\ket{1_{V},1_{V}}, \nonumber
\end{eqnarray}
where $\ket{1_s,1_{s'}}=\ket{1}_{\text{I},s}\otimes\ket{1}_{\text{II},s'}$, respectively, can be joined onto the state of a third single photon,
\begin{equation}
    \ket{\varphi}_{\text{III}}= \alpha\gamma\ket{1}_{\text{`0'}} + \beta\gamma\ket{1}_{\text{`1'}}+ \alpha\delta\ket{1}_{\text{`2'}} + \beta\delta\ket{1}_{\text{`3'}}.
\end{equation}
Here, the subscripts \{`0',`1',`2',`3'\} represents a 4-dimensional logical basis, e.g. path-polarization or OAM.
This scheme relies on the interaction of two photons, which would require nonlinear optics. 
The inverse process to joining states, known as quantum state splitting, is such that the information on a ququart state (for the above experiment) is split onto two photonic qubits~\cite{passaro:13}.

\section{Quantum computation}
After having covered a few examples of engineering quantum states, we can now move on to engineering the dynamics of a system. In particular, we want to focus on those dynamics that can be considered a ``computation'' on quantum data. Of course, a complete treatise of quantum computing would require much more space than that of this section, so our aim here is to summarize the main idea and then show how the HOM effect can be harnessed to perform some meaningful quantum state processing.

\subsection{\label{section:KLM}Linear quantum computing}
Universal quantum computation can be achieved with linear optics, post-selection, single photon sources and detectors, and feed-forward, as was proposed by the KLM protocol \cite{knill2001scheme}. 
Here, qubits are defined in the dual-rail encoding, i.e. the computational states are defined as a single photon that occupies one of a pair of optical modes. In our notation, a single photon qubit can be written simply as $\ket{\phi} = \alpha \ket{0,1} + \beta \ket{1,0}$, where the first and second labels in each ket corresponds to the number of photons in each rail. The logical qubits in the dual-rail encoding could, for example, be $\ket{\text{`0'}}:=\ket{1,0}$ and $\ket{\text{`1'}}:=\ket{0,1}$. Single qubit gates are then optical components that couple pairs of modes, such as a beam splitter. For two-qubit gates where interaction is required, the HOM effect is relied heavily upon. Two such gates are the control-Z (CZ) and control-NOT (CNOT) gates, which are main components for universal quantum computation.

\begin{figure}[t]	
	\centering
	{\includegraphics[width=0.45\textwidth]{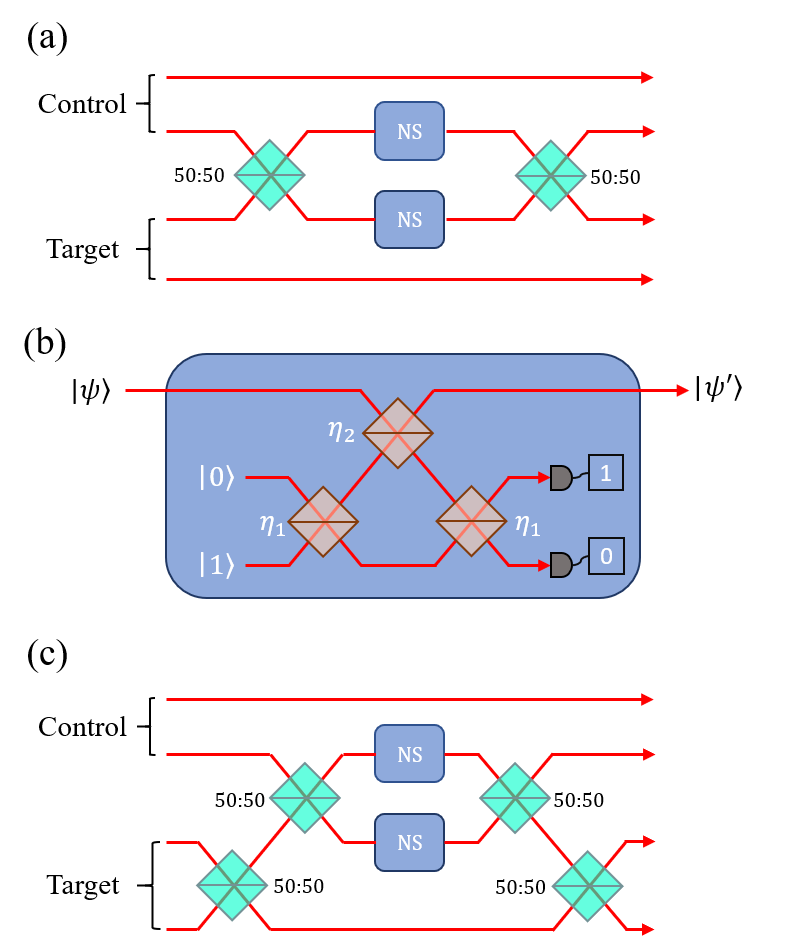}}
	\caption{\textbf{Schematic of photonic quantum gates.} a) Control-Z gate. b) Nonlinear sign gate with $\eta_1= 1/(4-2\sqrt{2})$ and $\eta_2=3-2\sqrt{2}$. c) Control-NOT gate. \label{fig:gates}}
\end{figure}

The two-qubit control-Z (CZ) gate performs the following operation, 
\begin{equation}
	\ket{q_1,q_2} \rightarrow (-1)^{q_1q_2}\ket{q_1,q_2},
\end{equation}
where $q_1$ and $q_2$ are the control and target logical qubits, respectively, taking values of `0' or `1'. It was found that a CZ gate can be constructed using two 50:50 BS and two so-called nonlinear sign (NS) gates~\cite{knill2001scheme}, see Fig.~\ref{fig:gates}a). A NS gate applies a $\pi$ phase shift to only the two-photon Fock state, i.e.,
\begin{equation} \label{eq:NS}
	\ket{\psi} = \alpha \ket{0} + \beta \ket{1} + \gamma \ket{2} \xrightarrow{NS} \ket{\psi'} = \alpha \ket{0} + \beta \ket{1} - \gamma \ket{2}.
\end{equation}
As shown in Fig.\ref{fig:gates}b), the NS gate uses two ancillary modes (vacuum and a single photon), three BS with transmission amplitudes of $\eta_1=1/(4-2\sqrt{2})$ and $\eta_2 = 3-2\sqrt{2}$. The NS gate is successful, i.e., Eq.~(\ref{eq:NS}) occurs, when the photon number resolving detectors detect 1 and 0 photons, respectively. This post-selection process renders the NS gate to be probabilistic, with a success probability of $p_{NS}=1/4$ for an arbitrary input state without feedforward. The upper bound for the NS gate was found to be 1/2~\cite{knill:03}. Variations of the NS gate have been reported: one proposal shows the possibility to implement it using only two BS but with slightly lower success probability of $(3-\sqrt{2})/7$~\cite{ralph:01}; other schemes with again two ancillary photons with success probabilities 1/5~\cite{zou:02} and 1/4~\cite{scheel:04}. It is also possible to implement the NS gate in the polarization basis, wherein polarization rotations are used instead of the variable BS; this also has a success probability of $(3-\sqrt{2})/7$.
Overall, we see that the CZ gate heavily relies on the HOM effect, the output depending on whether the control and target photons bunched or not. Since two NS gates are required, the total success probability of this CZ gate is $p_\text{CZ} = p_{NS}^2$. The most efficient CZ gate has a success probability of 2/27, utilizing two single photon ancillary modes and only four BS~\cite{knill:02}. 

The CNOT gate performs the operation  $\ket{q_1,q_2} \rightarrow \ket{q_1,q_2\oplus q_1}$, where $\oplus$ denotes addition modulo 2. In this way, the target qubit is flipped if and only if the control qubit is `1'. The implementation of the CNOT gate is similar to that of the CZ gate in that it requires two NS gates; however, two additional 50:50 BS are required acting on the target qubit~\cite{okamoto:11}, as shown in Fig.~\ref{fig:gates}. Again, since the CNOT gate is based around the NS gate, it is a probabilistic process with success probability $p_\text{CNOT}=p_{NS}^2=1/16$.

It is possible to achieve a useful quantum computation without resorting to a universal quantum computer. In this case, it is sufficient to have a device that performs a single algorithm.
There are several examples of specific quantum optical implementations that rely on the HOM effect, to list a few: solving systems of linear equations \cite{harrow2009quantum, cai2013experimental}, computation on encrypted data \cite{fisher2014quantum}, computing discrete and fractional Fourier transform \cite{weimann2016implementation}, and computations on a single spatial mode (using temporal modes) \cite{humphreys2013linear}.

Once the experimentation phase of a new technology becomes mature enough, it is followed by a phase of better production methods which may include miniaturization into embedded devices. This is a trend that quantum optical technologies are experiencing at the moment. Waveguides in photonics chips allow for higher portability, higher cost efficiency, and higher robustness \cite{o2009photonic, wang:2019}.

\subsection{The SWAP test}

The Hong-Ou-Mandel effect is colloquially referred to as a test for the distinguishability between two photons: if we witness a coincidence, the two photons had different states. Similarly, the SWAP test is a quantum computing primitive operation that tests for the inequality between two quantum states. Perhaps surprisingly, the analogy between them goes all the way down to formal equivalence: the HOM effect is an optical implementation of the (destructive) SWAP test \cite{garcia2013swap}. In this section, we will explain this formidable equivalence.

\begin{figure}[]	
	\centering
	{\includegraphics[width=0.35\textwidth]{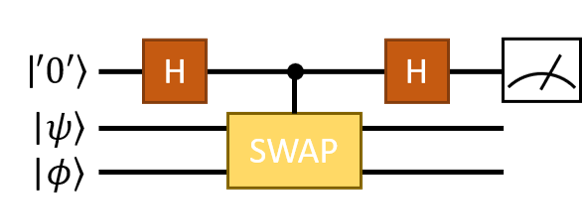}}
	\caption{\textbf{SWAP test.} The quantum circuit implementing the SWAP test on the states $|\psi\rangle$ and $|\phi\rangle$. A measurement of `1' implies that the states are different, while a measurement of `0' is inconclusive. \label{fig:swap}}
\end{figure}

Let's analyse what occurs in the SWAP test, see Fig.~\ref{fig:swap} for the circuit. At the  input, we have two states $|\phi\rangle$ and $|\psi\rangle$, and an ancillary logical qubit initialized in $\ket{\text{`0'}}$. The Hadamard (H) gate converts the state $\ket{\text{`0'}}$ into a superposition $\frac{\ket{\text{`0'}}+\ket{\text{`1'}}}{\sqrt{2}}$ (and the state $\ket{\text{`1'}}$ into $\frac{\ket{\text{`0'}}-\ket{\text{`1'}}}{\sqrt{2}}$). The controlled-SWAP (CSWAP) gate swaps the states $|\phi\rangle$ and $|\psi\rangle$ if the ancillary qubit is in state $\ket{\text{`1'}}$, and does nothing otherwise. The evolution of the input throughout the SWAP test is,
\begin{eqnarray}
\begin{split}
    \ket{\text{`0'}}\ket{\phi}\ket{\psi}&\stackrel{\text{H}}{\longrightarrow}\frac{\ket{\text{`0'}}+\ket{\text{`1'}}}{\sqrt{2}}\ket{\phi}\ket{\psi}\\
    &\stackrel{\text{CSWAP}}{\longrightarrow}\frac{\ket{\text{`0'}}\ket{\phi}\ket{\psi}+\ket{\text{`1'}}\ket{\psi}\ket{\phi}}{\sqrt{2}}\\
    &\stackrel{\text{H}}{\longrightarrow}\frac{\ket{\text{`0'}}\bigl(\ket{\phi}\ket{\psi}+\ket{\psi}\ket{\phi}\bigr)+\ket{\text{`1'}}\bigl(\ket{\phi}\ket{\psi}-\ket{\psi}\ket{\phi}\bigr)}{2}.
\end{split}
    \label{eq:evolution}
\end{eqnarray}
In terms of the measurement outcome, if $\ket{\text{`1'}}$ is measured, we can confirm that the two input states are different; however, if the outcome is the state $\ket{\text{`0'}}$, they may or may not be equal. Equivalently in terms of the states $\ket{\phi}$ and $\ket{\psi}$, if they are identically equal, the second term in the last line of Eq.~\eqref{eq:evolution} vanishes and the outcome of the measurement must be $\ket{\text{`0'}}$ with probability 1. However, if they are not equal, either of the two possible outcomes could be measured. If we demonstrate their inequality, we say the states ``fail'' the test; otherwise, we say that the states ``pass'' the test. As mentioned above, passing the test does not mean that the states are equal: we can only say that the more copies of the initial pair pass the test, the more certain we are about their equality.

From Eq.~\eqref{eq:evolution}, we can find the probability for passing the test is $\left( 1+|\braket{\phi}{\psi}|^2 \right)/2$, and consequently the probability of failing the test is $\left(1-|\braket{\phi}{\psi}|^2 \right)/2$. This means that the more similar the two states are, the harder it is to demonstrate their inequality.
For two maximally distinguishable (orthogonal) states, the overlap is 0 and $P=1/2$, while for maximally indistinguishable states the overlap is 1 and the probability to pass the test is also 1. This justifies the notion that we are certain that the states are not equal if they fail the test.

The eigenstates of the SWAP operator are the Bell states: three Bell states correspond to the eigenvalue +1 (the symmetric subspace), and one corresponds to the eigenvalue -1 (the singlet state, i.e. the antisymmetric subspace). This suggests that one could also implement a \emph{destructive} SWAP test by measuring directly the two systems in the Bell basis. In order to make a full equivalence with the HOM effect, we need to get rid of the ancillary qubit. This can be done by exploiting the observation that the SWAP test projects the two input states onto the symmetric/antisymmetric subspaces of the space of states, which is what a Bell state measurement does. For the two-dimensional case, we can implement a Bell state measurement with a single beamsplitter followed by a detector in each output port, which is equivalent to a \emph{destructive} SWAP test in the sense that the states get measured and are not available afterwards. 
In higher dimensions, one needs the equivalent circuit for projecting onto the symmetric/antisymmetric subspaces. This is sufficient to conclude with a full analogy between the HOM effect and the SWAP test.

\section{\label{sec:multi} Generalization to multipartite and multimode systems}
So far, we have only discussed two-photon interferences appearing in a 2-input/2-output device, i.e. a beam splitter, just as it was envisioned in the original experiment by Hong, Ou and Mandel. However, the property of multiphoton interference is not limited to this rather simple (and yet powerful) case, but it can be generalized to more particles in more complicated beam splitter networks with multiple-input and multiple-output ports, i.e. so-called multiports. Although such studies have been started soon after Hong, Ou and Mandel's  seminal paper \cite{zeilinger1993einstein,reck1994experimental,weihs1996two}, the recent development of more efficient photon sources and the shift to integrated quantum optics experiments has led to a revival and extension of these investigations. 

\subsection{Three-photon interferences in a Bell-tritter}
\begin{figure}[htb]
\centering
\includegraphics[width=\columnwidth]{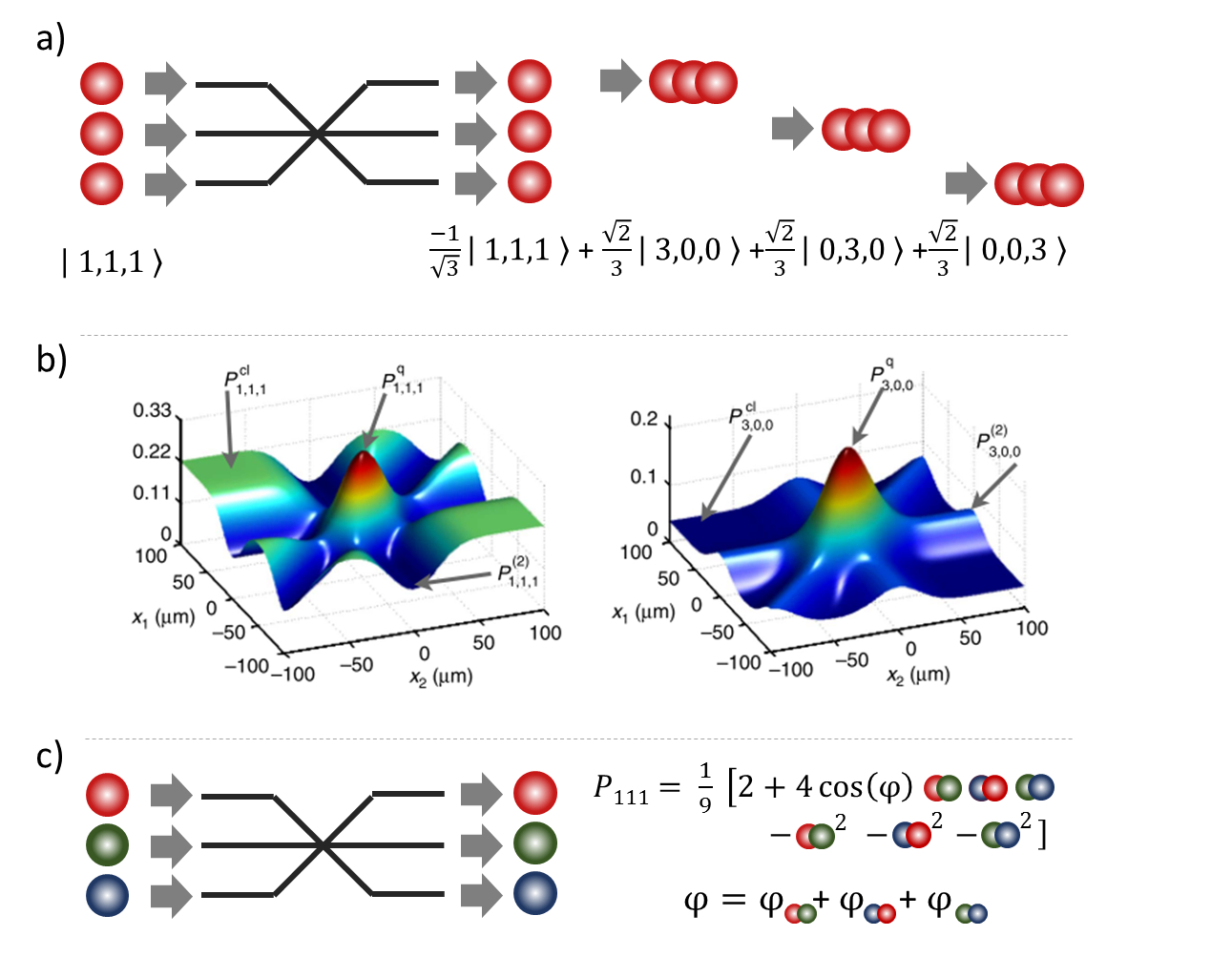}
\caption{\textbf{Three photon interference in a tritter.} a) Three photon analog to the original HOM experiment. If three indistinguishable photons are sent to a Bell tritter (see unitary (\ref{eq:tritter}), only 4 out of 10 possible outcome are allowed. b) Theoretical output probabilities for detecting all photons in different output ports $P_{1,1,1}$ (left) and the same output port $P_{3,0,0}$ (right) as a function of the delay of the input photons (adapted from \cite{spagnolo2013three}). c) If the photons are partially distinguishable (depicted by the different colors), the output probability $P_{1,1,1}$ also depends on the triad phase $\varphi$, which is the sum of pairwise phase differences of the photons (adapted from \cite{menssen2017distinguishability}). }
\label{fig:tritter}
\end{figure}

It is suggestive to generalize the two photon beam splitter arrangement at first to a three photon and three-mode analog to a beam splitter, a so-called tritter \cite{zukowski1997realizable,campos2000three}. Similar to the two photon HOM-interference, three photons can interfere in a tritter if they are indistinguishable. In general, the output probability distribution of the three photons depends on the input distribution as well as the exact unitary transformation of the tritter. Although any unitary transformation between the input and output modes might be possible and of some interest \cite{reck1994experimental}, it is instructive to look into the simple case of a so-called Bell-tritter \cite{zukowski1997realizable,campos2000three}, which redistributes the three incoming photons to the three output ports in an unbiased way. The ideal unitary transformation $U^{\text{tritter}}$ of a Bell-tritter is mapping the input field operators $a_i^\dagger$ to the output field operators $b_i^\dagger = \Sigma_j U_{ij}a_j^\dagger$ and can be described \cite{spagnolo2013three} by, 
\begin{equation}
U^\text{tritter} = 
\frac{1}{\sqrt{3}}
\begin{pmatrix}
  1 & 1 & 1 \\
  1 & e^{i2\pi / 3} & e^{i4\pi / 3} \\
  1 & e^{i4\pi / 3}  & e^{i8\pi / 3}   
 \end{pmatrix} .
 \label{eq:tritter}
\end{equation}
Bell-ports have been discussed theoretically since many years in connection to multipartite and high-dimensional quantum information \cite{zukowski1997realizable,campos2000three,reck1994experimental}. They have also been experimentally realized with fused optical fibers \cite{weihs1996two} and more recently with passive \cite{spagnolo2013three} and active integrated waveguide circuits \cite{schaeff2015experimental}. Although tritters have been used to demonstrate quantum features in the two-photon regime \cite{weihs1996two,schaeff2015experimental,meany2012non}, only recently have three photon HOM coalescences been observed \cite{spagnolo2013three,menssen2017distinguishability}. Analog to the two-photon HOM effect, if three indistinguishable photons are sent into the three input ports, not all possible output-distributions will be realized. As can be seen in Figure \ref{fig:tritter}a for a three-dimensional Bell-port, bosonic coalescence leads to only 4 out of 10 possible output distributions. By making the photons pairwise distinguishable, e.g. by delaying independently two photons, one can map a probability surface with non-trivial features (see Fig.\ref{fig:tritter}b). This surface can be seen as the three-dimensional extension to the 2-dimensional HOM interference dip \cite{spagnolo2013three}. Importantly, the complex features of three-partite interference experiments were also discussed in terms of quantum metrology and phase sensing to enhance precision \cite{spagnolo2012quantum}.

While these results can be seen as a direct extension to the standard two-photon HOM effect, another recent experiment demonstrated that there are additional features unique to multi-partite interferences \cite{menssen2017distinguishability}. In contrast to the two-photon case, where the interference is solely dictated by the distinguishability of the two photons, the interference becomes more complex when more particles are involved. For three photons for example, an additional collective phase is required to describe the photons' scattering behaviour \cite{tichy2014interference,de2014coincidence,shchesnovich2015partial}. This so-called triad phase is the sum of the three relative phases between pairwise inputs (see Fig.\ref{fig:tritter}c) and only effects the outcome if the photons are partially distinguishable. It leads to a change in the detection probabilities when all three photons are detected in separate output ports and does not affect the bi-photon transmission probabilities. Thus, it has to be considered a genuine three-photon interference effect with no analog in the standard HOM experiment. Furthermore, similar concepts and results, such as a zero-transmission law for various output distributions, can also be generalized to $n$ bosons sent in the $n$ ports of a Bell-multiport \cite{tichy2010zero,tichy2011four,tichy2012many,ra2013nonmonotonic}.  

\subsection{Mulitphotons in general multiports}
After this initial generalization step, one can extend the discussion by enlarging the number of modes of the multiport device and by increasing the number of involved particles. The increasing complexity opens a myriad of different effects, which have been or will be studied and that are all relying on the interference of two or more particles. When two particles for example are sent into a multiport, e.g. multimode waveguide \cite{lahini2012quantum,poem2012two}, multimode fiber \cite{defienne2016two} or integrated waveguide structures \cite{politi2009integrated}, the transmission can be described as a complex bi-partite quantum walk that gives rise to interesting quantum correlation patterns \cite{peruzzo2010quantum,poulios2014quantum}, can be used to perform quantum simulations \cite{owens2011two,crespi2013anderson} or can probe the statistics in quantum walks \cite{sansoni2012two}. 

As discussed in the last paragraph, if more than two photons are interfered, a probability surface or coincidence landscape can be investigated \cite{tillmann2015generalized}, which is useful to probe indistinguishability or to predict the complex device's quantum behaviour \cite{metcalf2013multiphoton}. Interferences of many bosons in complex network structures have also found various applications in quantum information science. Most recently, a multiphoton interference in a multiport has enabled the first quantum teleportation of a high-dimensional quantum state~\cite{luo2019quantum}. The procedure for high-dimensional quantum teleporation is very similar to the one for qubits; however, a high-dimensionally entangled pair has to be shared between the two parties first. In addition, depending on the dimensionality of the teleported state, $d$-1 ancillary photons are required to perform the high-dimensional Bell-state measurement.  In the experiment, a three-dimensional quantum state was teleported using one additional ancillary photon. After appropriate postselection, the three-dimensional quantum state was found to be teleported to the photon, which was initially a part of the entangled pair. The complex interference of mutliple photon in multiports cannot only be used in teleportation schemes, it also enables the generation of multi-partite entanglement~\cite{tichy2013limits}. 

Moreover, the enormous complexity of the underlying physics is nicely illustrated in terms of the so-called boson sampling problem \cite{aaronson2011computational,brod:19}. It has been shown that calculating the output probability distribution for a sufficiently large number of photons interfering in a large multiport arrangement is exponentially hard to solve classically, and as such it is considered to be a promising candidate to show quantum supremacy. In addition, it is also suggested that this challenges a fundamental principle of computer science, namely the extended Church-Turing thesis. Because of the complexity of this topic and as well as the enormous attention and progress that has been done over the last few years \cite{broome2013photonic,spring2012boson,tillmann2013experimental,crespi2013integrated,spagnolo2014experimental,bentivegna2015experimental,lund2014boson,wang2017high}, boson sampling would justify a review on its own and we refrain from discussing it in more detail. Instead, we conclude with the note that, in all boson sampling tasks, multi-photon as well as multimodal interferences, i.e. generalizations of the original HOM idea, play a crucial role.

\section{Implementations in non-photonic systems}\label{section8}
While most of the experiments studying and utilizing the HOM interference use single photons and follow the original proposal, it has also been investigated extensively with other quantum systems. In this section, we will discuss the cases of plasmons, phonons, atoms, and electrons.

\subsection{Plasmons}
Using plasmons as quantum carriers has been an increasingly popular field in quantum science that pushed forward the advantages of light-matter interactions at the nano-scale and investigates its quantum physical features. So-called surface plasmon polaritons (SPP) are the quanta of the surface plasma wave, analog to photons in the electromagnetic field. Their theoretical prediction dates back to the 1950s; however, the experimental investigations in the quantum realm has only been started recently (for a more detailed review about quantum plasmonics, we refer the interested reader to \cite{tame2013quantum,marquier2017revisiting}). Because SPPs are described as quasiparticles with a bosonic nature, they were also expected to bunch in a HOM-type experiment. While first experiments involving HOM interference only tested the persistence of the indistinguishability when photons are coupled to plasmons and back to photons \cite{fujii2012preservation,wang2012hong}, actual plasmonic bunching was experimentally observed in 2013 \cite{heeres2013quantum}. 
\begin{figure}[htb]
\centering
\includegraphics[width=\columnwidth]{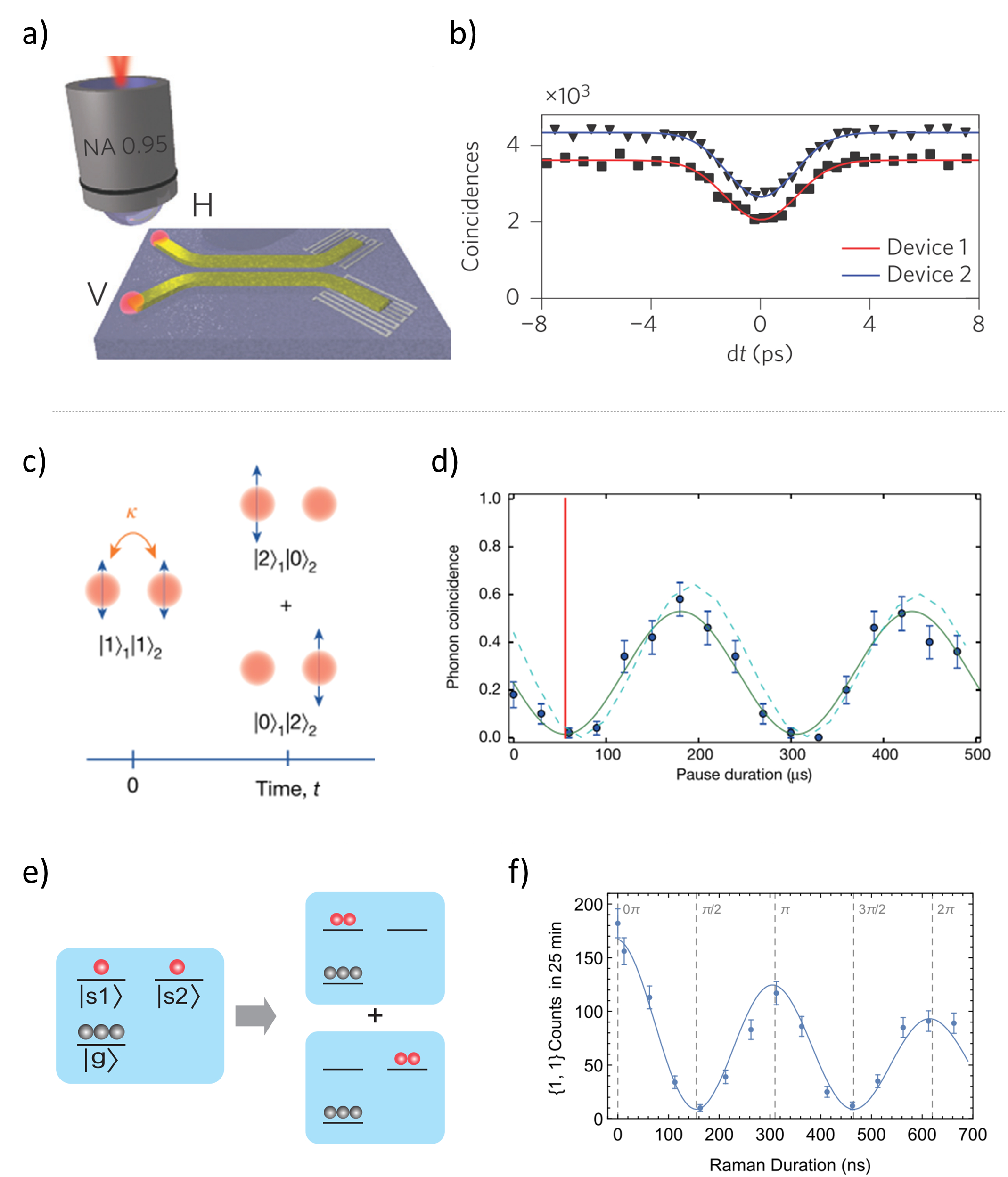}
\caption{\textbf{Non-photonic HOM interference.} HOM interference with plasmons (a,b from \cite{heeres2013quantum}), phonons (c,d from \cite{toyoda2015hong}) and collective excitations (e,f from \cite{li2016hong}). a) Two plasmons are excited by two single photons, brought to interference by a waveguide beam splitter and detected on chip by a superconducting detector. b) The coincidence counts between the two detectors as a function of the delay of one input photon show the expected HOM dip. c) Two phonons are excited on each site of a two-ion crystal and are allowed to hop to the neighboring ion. After a specific time $t$ corresponding to a 50-50 beam splitter, the phonons are only found at one of the two ions. d) Phonon coincidence deduced from the read-out of the internal state as a function of waiting time until the read out is performed. The red vertical line depicts the time of 50\% hopping probability. e) Rb atoms were collectively excited into two different Zeeman levels and coupled through a stimulated Raman pulse. If pulse length corresponds to a 50:50 coupling, the excitations are bunched together. f) Coincidence measurements of the two different excitations show a clear dip for multiple of $\pi/2$ coupling ratios.}
\label{fig:plasmon}
\end{figure}
The experiment was performed in very close analogy to integrated photon experiments using plasmonic waveguides to guide the plasmons to a dielectric bi-directional coupler acting as the 50:50 beam splitter (see Fig.\ref{fig:plasmon}a). At first, photon pairs were generated from a SPDC source, which were then transferred to SPPs and brought to interference. After mixing on the beam splitter, the plasmons were directly detected using on-chip superconducting single plasmon detectors. Analog to photon experiments, the arrival time of one plasmon at the beam splitter was scanned by means of a photonic delay line while the coincidence counts were registered. Although the observed dip had only a visibility of 0.43$\pm$0.02 (see Fig.\ref{fig:plasmon}b) and thus below the classical limit of 0.5, the result is a nice indication of a two plasmons quantum interference. Moreover, soon after this initial demonstration, different experiments using different plasmonic configurations, and off-chip photon detection schemes achieved visibilities far beyond the classical limit up to 0.95 \cite{fakonas2014two,di2014observation,cai2014high,fujii2014direct}, thus fully certifying the quantumness of the two plasmon interferences. Moreover, in another recent study, it was shown that the usually deleterious losses associated with plasmonic systems might be turned into a advantageous feature when the nanostructures are carefully tuned. At the plasmonics beam splitter, anti-coalescence or non-linear absorption can be observed \cite{vest2017anti}. Since two plasmon HOM interference is now well-established, it can be considered one of the building blocks of future quantum plasmonic circuits, and first studies to apply it in more complicated schemes, e.g. to generate entanglement \cite{dieleman2017experimental}, have been performed. 

\subsection{Phonons}
In addition to photons and plasmons, there are other quasiparticles resulting from a quantum mechanical description of excited modes, e.g. phonons that are the quantized excitation of mechanical motions. Phonons also follow Bose-Einstein statistics, and as such they are expected to bunch in a HOM-like arrangement. Recently, such a two-phonon interference effect has been observed involving two trapped ions \cite{toyoda2015hong}. In the experiment, the excitations of the radial modes of the trapped Ca$^+$ ions  served as the phonons. After cooling the ions to the ground state, a two-pulse sequence brought the ions to an excited state followed by a re-initialization back to the ground state leaving phononic excitations at both ions. A subsequent time period with no laser irradiation allowed the phonons to hop to the neighbouring ion. Finally, a read-out pulse filtering states of phonons at two different ions was applied. For waiting times when the hopping probability of the phonons was 50\%, bunching was observed (see Fig.~\ref{fig:plasmon}c and d). Although the statistical significance of the fidelity was too small to exclude classical explanations with certainty (0.52$\pm$0.03), the experiment opens the path to phonon quantum information experiments, such as boson sampling or entanglement generation in larger ion crystals, relying on a phononic HOM interference. A nice example recently demonstrated a quantum walk with phonons implemented in a trapped ion crystal~\cite{tamura2020quantum}.

\subsection{Collective atomic excitations}
In the previous paragraph, excitations of atoms were used to generate phonons in an ion crystal in a controlled manner. However, collective excitations have also been tested with respect to interference effects in a recent experiment using a Rb ensemble \cite{li2016hong}. By using the Rydberg blockade, two distinct collective excitations in different Zeeman levels have been generated. Using a stimulating Raman pulse that acts as a beam splitter, the excitations can be interfered and bunched together in either of the two Zeeman states (see Fig.~\ref{fig:plasmon}e). A final Raman readout pulse leads to an emission of the two photons. Similar to photonic experiments, bunching was confirmed through a dip with a visibility up to 0.89$\pm$0.06 in the correlation measurements of the two photons (see Fig.~\ref{fig:plasmon}f) by varying the distinguishability and splitting ratio.

\subsection{Atoms}
\begin{figure}[htb]
\centering
\includegraphics[width=\columnwidth]{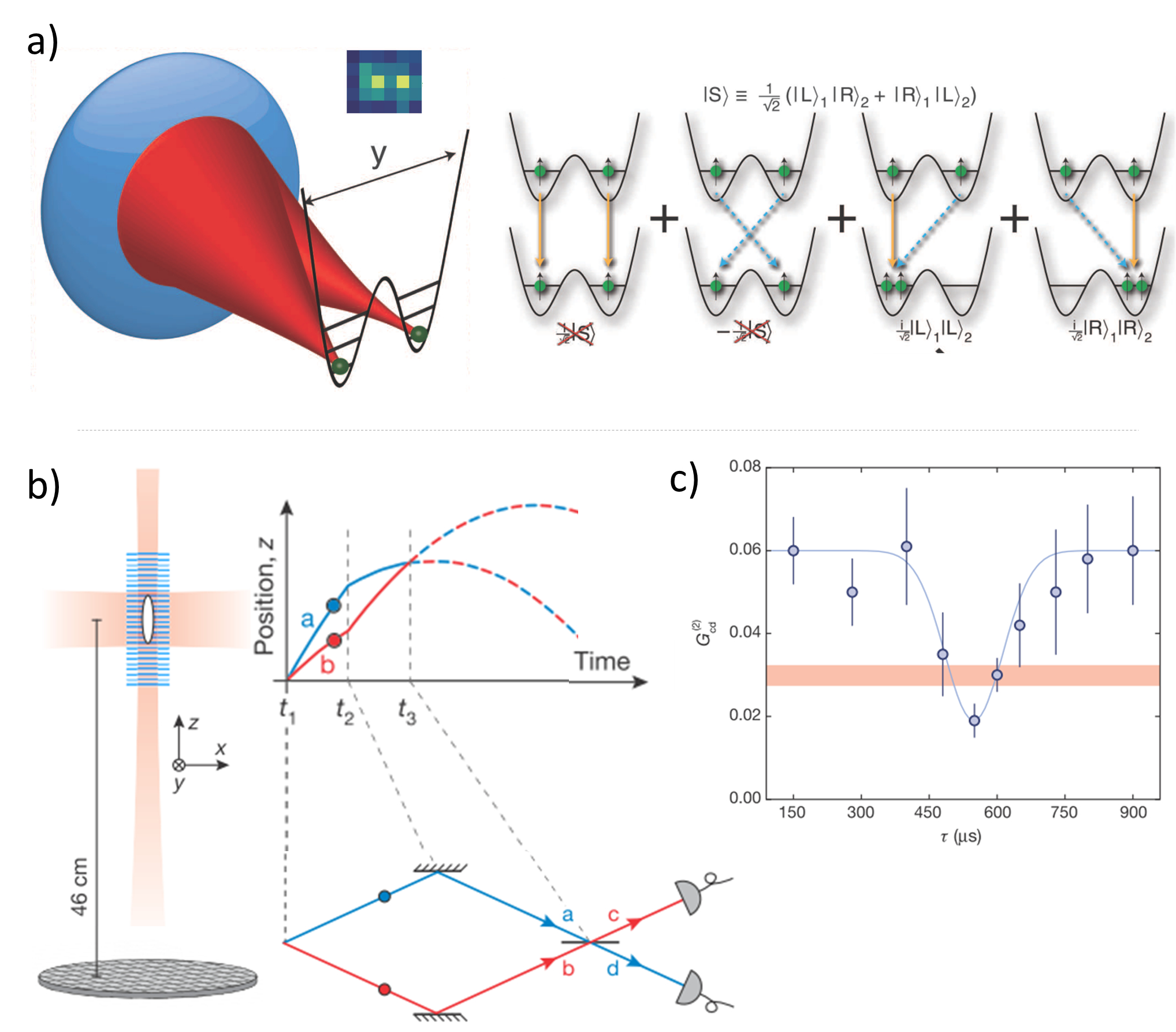}
\caption{\textbf{Atomic HOM interference.} a) Two single atoms were trapped in an optical tweezer and made indistinguishable by ground state cooling. Inset shows a camera picture of the atoms. Coupling between the trapping potentials allows for a tunneling probability of 50\%. Since the first two cases cancel out each other, the atoms bunch and are always found in one trap (taken from \cite{kaufman2014two}). b) HOM experiment with a twin-pair source of $^4$He atoms. The twin-pairs are produced by imprinting optical gratings leading to a four-wave mixing process, beam reflection and splitting. With the help of multichannel detectors, spatio-temporal imaging and momentum distribution measurements were performed. c) Measured cross-correlation between atoms found in different output ports depending on the time between mirror and beam splitter, i.e. their distinguishability (b and c taken from \cite{lopes2015atomic}).}\label{fig:atom}
\end{figure}

So far, we have only discussed quasi-particles, i.e. quantized excitation of photonic, plasmonic or phononic modes. However, following quantum mechanics, even massive particles should be able to exhibit two-particle interferences. In fact, two recent experiments, both using very different approaches, have shown exactly this: a two-atom HOM interference. In one experiment, two laser-cooled Rb atoms trapped in an optical tweezer were perfectly controlled in all internal and external degrees of freedom, thereby making them indistinguishable \cite{kaufman2014two}. By careful tuning the position and depth of the tweezers, the bosonic atoms were allowed to tunnel between the two traps. After a specific amount of time, a 50\% probability of tunneling of the atoms to the neighbouring trap can be realized (see Fig.~\ref{fig:atom}a). The resulting probability of finding two indistinguishable atoms in one trap exceeded the probability for distinguishable atoms by 6 standard deviations, which can be considered a clear indication for bunching of massive systems. Shortly after this initial step, these results were extended to a complex quantum walk of atoms in a two-dimensional optical lattice \cite{preiss2015strongly}.

After this HOM experiment with trapped single atoms, another experiment was performed using metastable $^4$He atoms released from a Bose-Einstein condensate (BEC) consisting of around $5\times10^4$ atoms \cite{lopes2015atomic}. At first, a moving optical lattice was superimposed on the BEC, which induced a scattering of the atoms along the vertical direction (see Fig.~\ref{fig:atom}b) through a process similar to optical spontaneous four-wave mixing. The scattering led to a beam of twin-atoms at two different velocities that moved apart under the influence of gravity. A few hundred microseconds later, another optical lattice imposed Bragg diffraction on both beams to swap their velocities and bring them back together. Finally, when the two trajectories met again, a last grating imposed a 50:50 beam splitter operation on the twin atoms and enabled the interference. Using a micro-channel plate to detect the falling atoms, the cross-correlation between the two detected velocities, i.e. the two output channels of the beam splitter, was measured while changing the time of the last grating, i.e. the splitting ratio. Again, the results show a clear dip in the cross-correlation measurements with a visibility of 0.65$\pm $0.07, which is above the classical limit of 0.5 and as such a proof for atomic bunching \cite{lewis2014proposal}. 
While both experiments show impressive results for atomic HOM experiments, future research in this direction could enable multi-particle interferences, entanglement operations and quantum information tasks \cite{dussarrat2017two,kaufman2018hong} known from the mature field of quantum optics.

\subsection{Electrons}
\begin{figure}[htb]
\centering
\includegraphics[width=\columnwidth]{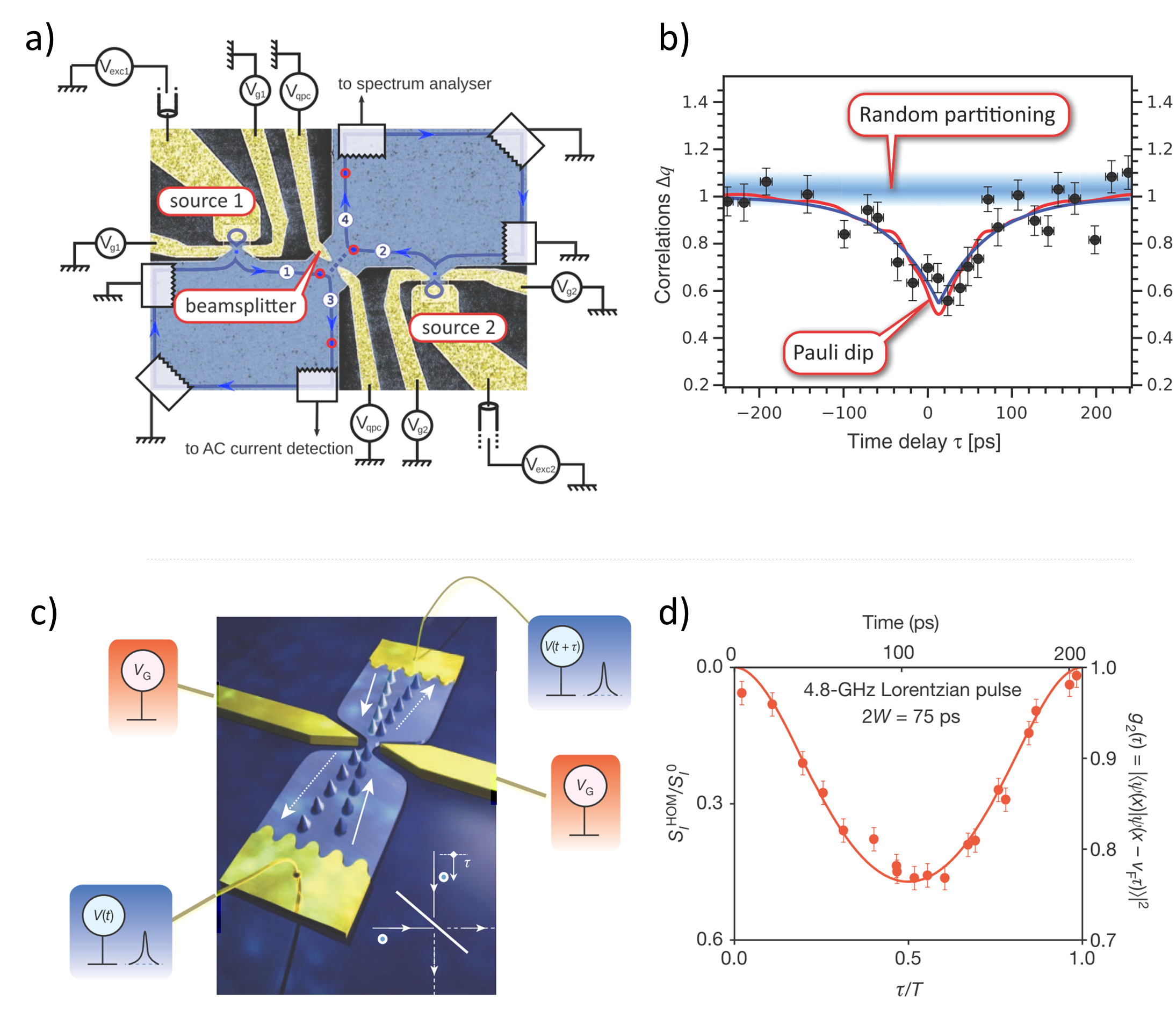}
\caption{\textbf{Electronic HOM experiments.} a) Two quantum dots (source 1 and 2) emit single electrons into chiral edges states in a two-dimensional electron gas. A quantum point contact acts as a 50:50 beam splitter. b) The joint measurement of the noise properties (labeled as correlation $\Delta q$) are found to be lower than classically allowed due to anti-bunching of the electrons; thus, a Pauli-dip appears if the delay between the two electrons is zero at the beam splitter (a and b taken from \cite{feve2007demand}). c) Two levitons are emitted into a electronic gas by applying a Lorentzian pulse to two opposing contacts. d) Again, noise properties corresponding to correlation measurements reveal anti-bunching depending on the delay between the two emissions (c and d taken from \cite{dubois2013minimal}).}
\label{fig:electron}
\end{figure}

Finally, it has been shown that not only bosons but also fermions in the form of electrons can interfere in an HOM-like arrangement. The major difference between indistinguishable bosons and fermions is that the latter obey Fermi statistics, which leads to anti-bunching at a beam splitter. Fermionic two-particle interferences are a manifestation of the Pauli exclusion-principle. The first such interference was studied in 1998 with a constant stream of colliding electrons produced in a two-dimensional electron gas \cite{liu1998quantum}. A small gate finger in the centre of the scattering area enabled the tuning of the splitting ratio to around 50\% transmission. Due to anti-bunching, a reduction in the noise properties of the output ports was observed; thus, an indirect verification of two-electron anti-bunching was achieved. However, because these initial findings used a continuous stream of electrons, the interference might not exclusively be interpreted as a result from the overlap between two single-electron wave packets. Recently, this deficiency was resolved by an improved experiment, using two separate sources of single electrons emitted from a triggered quantum dot \cite{bocquillon2012electron,bocquillon2013coherence}. By applying a voltage pulse to the dot with an energy far above the Fermi level, single electrons are generated followed by the emission of a single hole \cite{feve2007demand}. The electrons are emitted into chiral edge states and propagate along the edges of a two-dimensional electron gas in the quantum Hall regime. The electronic beam splitter is realized by a narrow constriction, also called quantum point contact. As single coincidence detections are not possible, the correlation noise properties at the outputs were studied and found to be below the random partition noise, i.e. classically expected noise (see Fig.~\ref{fig:electron}a). If one electron is delayed, the noise properties change such that a Pauli-dip (analog to the HOM-dip for coincidences measurements) of around 50\% of the classical noise level was observed. Following these experiments, the HOM interference was further applied to study an electron gas in the integer quantum Hall regime at a filling factor of 2, where a charge transport along two co-propagating edge channels with opposite spin occurs \cite{freulon2015hong}. 

Interestingly, electronic systems also permit the generation and manipulation of quasi-particles, so-called levitons. A leviton is the quantized minimal excitation of the electronic Fermi sea in a conductor for which no hole is generated. Compared to the electron described above, which can be seen as energy-resolved sources, levitons can be described as a time-resolved excitation. They are triggered by a well-adjusted Lorentzian voltage pulse during which only a positive energy boost is given to the Fermi sea. Although such electronic quasi-particles have been predicted a long time ago \cite{levitov1996electron,keeling2006minimal}, they have been observed and even brought to two-fermionic quantum interference only recently \cite{dubois2013minimal}. After the emission from two ohmic contacts into a two-dimensional electron gas, levitons propagate through a quantum point contact that acts as a 50:50 beam splitter. From the noise measurements, a $g^{(2)}$ value up to 1 was deduced if both levitons were emitted simultaneously, which is a signature for anti-bunching. Because levitons are propagating with less disturbance, the visibility of the interference is significantly higher than the one of the electronic HOM interference described in the last paragraph (see Fig.~\ref{fig:electron}b)). This feature also led to an intensive study of levitons in the quantum regime \cite{glattli2017levitons}, for example by investigating the temperature dependence on the indistinguishability \cite{glattli2016hanbury} or by performing quantum tomography on an electron \cite{jullien2014quantum}.

\section{Conclusion}
Two-photon interference is fundamentally interesting because it has no classical counterpart, with many applications ranging from precision measurement and state determination to quantum computations and quantum communication. In metrology, two photon interference allows for femtosecond and even attosecond time-resolution, state determination (quantum state analysis), and generating non-classical light for resolution enhancement with linear optics. In quantum state analysis, it gives the possibility to distinguish between modes, since only indistinguishable states will bunch. In particular, two-photon interference enables the measurement of maximally entangled states through Bell states measurements, or equivalently, the projection onto the Bell basis. Indeed, Bell state measurements are central to the concepts of teleportation and entanglement swapping. In quantum communication, it can be used to circumvent the problem of untrusted measurement devices, and build novel protocols. In quantum computation, the interaction between two photons is often required to create the building blocks of certain quantum gates, for example CNOT and CZ gates. Two-photon interference acts as the linear optical solution to this challenge, and thus provides a scalable technique to build photonic quantum computers. Two particle interferences is not only the basis for most of optical quantum information science and a powerful tool for metrology and other fields, but it has also led to many fundamental demonstrations in non-photonic systems such as plasmons, phonons, atoms and electrons. Moreover, it still inspires physicists to study its connection to novel systems such as quantized spin waves, i.e. magnons~\cite{ahmed2017guided}, Bogoliubov quasiparticles~\cite{ferraro2015nonlocal}, massless Dirac fermions in topological insulators and graphene \cite{khan2014two} or to transfer the idea to analog situations, e.g. with bright solitons~\cite{sun2014mean}.

\section{Acknowledgements} 
E.K. acknowledges the fruitful conversation with Gerd Leuchs, Miles J. Padgett, and Sir Peter Knight. This work was supported by Canada Research Chairs (CRC), Canada First Excellence Research Fund (CFREF), and Ontario's Early Researcher Award (ERA).


\begin{thebibliography}{100}
\expandafter\ifx\csname url\endcsname\relax
  \def\url#1{\texttt{#1}}\fi
\expandafter\ifx\csname urlprefix\endcsname\relax\def\urlprefix{URL }\fi
\providecommand{\bibinfo}[2]{#2}
\providecommand{\eprint}[2][]{\url{#2}}

\bibitem{dilip:20}
\bibinfo{author}{Paneru, D.}, \bibinfo{author}{Cohen, E.},
  \bibinfo{author}{Fickler, R.}, \bibinfo{author}{Boyd, R.~W.} \&
  \bibinfo{author}{Karimi, E.}
\newblock \bibinfo{title}{Entanglement: Quantum or classical?}
\newblock \emph{\bibinfo{journal}{Reports on Progress in Physics}}
 \textbf{\bibinfo{volume}{83}}, \bibinfo{pages}{064001}
  (\bibinfo{year}{2020}).

\bibitem{prasad:87}
\bibinfo{author}{Prasad, S.}, \bibinfo{author}{Scully, M.~O.} \&
  \bibinfo{author}{Martienssen, W.}
\newblock \bibinfo{title}{A quantum description of the beam splitter}.
\newblock \emph{\bibinfo{journal}{Optics Communications}}
  \textbf{\bibinfo{volume}{62}}, \bibinfo{pages}{139--145}
  (\bibinfo{year}{1987}).

\bibitem{ou:87}
\bibinfo{author}{Ou, Z.}, \bibinfo{author}{Hong, C.} \&
  \bibinfo{author}{Mandel, L.}
\newblock \bibinfo{title}{Relation between input and output states for a beam
  splitter}.
\newblock \emph{\bibinfo{journal}{Optics Communications}}
  \textbf{\bibinfo{volume}{63}}, \bibinfo{pages}{118--122}
  (\bibinfo{year}{1987}).

\bibitem{hong:87}
\bibinfo{author}{Hong, C.~K.}, \bibinfo{author}{Ou, Z.~Y.} \&
  \bibinfo{author}{Mandel, L.}
\newblock \bibinfo{title}{Measurement of subpicosecond time intervals between
  two photons by interference}.
\newblock \emph{\bibinfo{journal}{Physical Review Letters}}
  \textbf{\bibinfo{volume}{59}}, \bibinfo{pages}{2044--2046}
  (\bibinfo{year}{1987}).

\bibitem{fearn:87}
\bibinfo{author}{Fearn, H.} \& \bibinfo{author}{Loudon, R.}
\newblock \bibinfo{title}{Quantum theory of the lossless beam splitter}.
\newblock \emph{\bibinfo{journal}{Optics Communications}}
  \textbf{\bibinfo{volume}{64}}, \bibinfo{pages}{485--490}
  (\bibinfo{year}{1987}).

\bibitem{rarity:88}
\bibinfo{author}{Rarity, J.} \& \bibinfo{author}{Tapster, P.}
\newblock \bibinfo{title}{Nonclassical effects in parametric}.
\newblock \emph{\bibinfo{journal}{Photons and Quantum Fluctuations}}
  \textbf{\bibinfo{volume}{5}}, \bibinfo{pages}{122} (\bibinfo{year}{1988}).

\bibitem{rarity:89}
\bibinfo{author}{Rarity, J.~G.} \& \bibinfo{author}{Tapster, P.}
\newblock \bibinfo{title}{Fourth-order interference in parametric
  downconversion}.
\newblock \emph{\bibinfo{journal}{JOSA B}} \textbf{\bibinfo{volume}{6}},
  \bibinfo{pages}{1221--1226} (\bibinfo{year}{1989}).

\bibitem{shih:88}
\bibinfo{author}{Shih, Y.} \& \bibinfo{author}{Alley, C.}
\newblock \bibinfo{title}{New type of einstein-podoolsky-rosen-bohm experiment
  using pairs of light quanta produced by optical parametric down conversion}.
\newblock \emph{\bibinfo{journal}{Physical Review Letters}}
  \textbf{\bibinfo{volume}{61}}, \bibinfo{pages}{2921} (\bibinfo{year}{1988}).

\bibitem{hamilton:00}
\bibinfo{author}{Hamilton, M.~W.}
\newblock \bibinfo{title}{Phase shifts in multilayer dielectric beam
  splitters}.
\newblock \emph{\bibinfo{journal}{American Journal of Physics}}
  \textbf{\bibinfo{volume}{68}}, \bibinfo{pages}{186--191}
  (\bibinfo{year}{2000}).

\bibitem{einstein:35}
\bibinfo{author}{Einstein, A.}, \bibinfo{author}{Podolsky, B.} \&
  \bibinfo{author}{Rosen, N.}
\newblock \bibinfo{title}{Can quantum-mechanical description of physical
  reality be considered complete?}
\newblock \emph{\bibinfo{journal}{Physical Review}}
  \textbf{\bibinfo{volume}{47}}, \bibinfo{pages}{777} (\bibinfo{year}{1935}).

\bibitem{ekert:91}
\bibinfo{author}{Ekert, A.~K.}
\newblock \bibinfo{title}{Quantum cryptography based on bell's theorem}.
\newblock \emph{\bibinfo{journal}{Phys. Rev. Lett.}}
  \textbf{\bibinfo{volume}{67}}, \bibinfo{pages}{661} (\bibinfo{year}{1991}).

\bibitem{dowling:08}
\bibinfo{author}{Dowling, J.~P.}
\newblock \bibinfo{title}{Quantum optical metrology--the lowdown on high-N00N
  states}.
\newblock \emph{\bibinfo{journal}{Contemporary Physics}}
  \textbf{\bibinfo{volume}{49}}, \bibinfo{pages}{125--143}
  (\bibinfo{year}{2008}).

\bibitem{kok2007linear}
\bibinfo{author}{Kok, P.} \emph{et~al.}
\newblock \bibinfo{title}{Linear optical quantum computing with photonic
  qubits}.
\newblock \emph{\bibinfo{journal}{Reviews of Modern Physics}}
  \textbf{\bibinfo{volume}{79}}, \bibinfo{pages}{135} (\bibinfo{year}{2007}).

\bibitem{zeilinger:98}
\bibinfo{author}{Zeilinger, A.}
\newblock \bibinfo{title}{Quantum entanglement: a fundamental concept finding
  its applications}.
\newblock \emph{\bibinfo{journal}{Physica Scripta}}
  \textbf{\bibinfo{volume}{T76}}, \bibinfo{pages}{203--209}
  (\bibinfo{year}{1998}).

\bibitem{rubinsztein:16}
\bibinfo{author}{Rubinsztein-Dunlop, H.} \emph{et~al.}
\newblock \bibinfo{title}{Roadmap on structured light}.
\newblock \emph{\bibinfo{journal}{Journal of Optics}}
  \textbf{\bibinfo{volume}{19}}, \bibinfo{pages}{013001}
  (\bibinfo{year}{2016}).

\bibitem{giovannini:15}
\bibinfo{author}{Giovannini, D.} \emph{et~al.}
\newblock \bibinfo{title}{Spatially structured photons that travel in free
  space slower than the speed of light}.
\newblock \emph{\bibinfo{journal}{Science}} \textbf{\bibinfo{volume}{347}},
  \bibinfo{pages}{857--860} (\bibinfo{year}{2015}).

\bibitem{bouchard:16}
\bibinfo{author}{Bouchard, F.}, \bibinfo{author}{Harris, J.},
  \bibinfo{author}{Mand, H.}, \bibinfo{author}{Boyd, R.~W.} \&
  \bibinfo{author}{Karimi, E.}
\newblock \bibinfo{title}{Observation of subluminal twisted light in vacuum}.
\newblock \emph{\bibinfo{journal}{Optica}} \textbf{\bibinfo{volume}{3}},
  \bibinfo{pages}{351--354} (\bibinfo{year}{2016}).

\bibitem{lyons:18}
\bibinfo{author}{Lyons, A.} \emph{et~al.}
\newblock \bibinfo{title}{How fast is a twisted photon?}
\newblock \emph{\bibinfo{journal}{Optica}} \textbf{\bibinfo{volume}{5}},
  \bibinfo{pages}{682--686} (\bibinfo{year}{2018}).

\bibitem{richard:18}
\bibinfo{author}{Richard, G.} \emph{et~al.}
\newblock \bibinfo{title}{Twisting waves increase the visibility of nonlinear
  behaviour}.
\newblock \emph{\bibinfo{journal}{New Journal of Physics}} \bibinfo{pages}{in
  press} (\bibinfo{year}{2020}).

\bibitem{lyons:18b}
\bibinfo{author}{Lyons, A.} \emph{et~al.}
\newblock \bibinfo{title}{Attosecond-resolution hong-ou-mandel interferometry}.
\newblock \emph{\bibinfo{journal}{Science Advances}}
  \textbf{\bibinfo{volume}{4}}, \bibinfo{pages}{eaap9416}
  (\bibinfo{year}{2018}).

\bibitem{chen2019hong}
\bibinfo{author}{Chen, Y.}, \bibinfo{author}{Fink, M.},
  \bibinfo{author}{Steinlechner, F.}, \bibinfo{author}{Torres, J.~P.} \&
  \bibinfo{author}{Ursin, R.}
\newblock \bibinfo{title}{Hong-ou-mandel interferometry on a biphoton beat
  note}.
\newblock \emph{\bibinfo{journal}{npj Quantum Information}}
  \textbf{\bibinfo{volume}{5}}, \bibinfo{pages}{1--6} (\bibinfo{year}{2019}).

\bibitem{steinberg:92}
\bibinfo{author}{Steinberg, A.~M.}, \bibinfo{author}{Kwiat, P.~G.} \&
  \bibinfo{author}{Chiao, R.~Y.}
\newblock \bibinfo{title}{Dispersion cancellation and high-resolution time
  measurements in a fourth-order optical interferometer}.
\newblock \emph{\bibinfo{journal}{Physical Review A}}
  \textbf{\bibinfo{volume}{45}}, \bibinfo{pages}{6659} (\bibinfo{year}{1992}).

\bibitem{povazay:02}
\bibinfo{author}{Povazay, B.} \emph{et~al.}
\newblock \bibinfo{title}{Submicrometer axial resolution optical coherence
  tomography}.
\newblock \emph{\bibinfo{journal}{Optics Letters}}
  \textbf{\bibinfo{volume}{27}}, \bibinfo{pages}{1800--1802}
  (\bibinfo{year}{2002}).

\bibitem{abouraddy:02}
\bibinfo{author}{Abouraddy, A.~F.}, \bibinfo{author}{Nasr, M.~B.},
  \bibinfo{author}{Saleh, B.~E.}, \bibinfo{author}{Sergienko, A.~V.} \&
  \bibinfo{author}{Teich, M.~C.}
\newblock \bibinfo{title}{Quantum-optical coherence tomography with dispersion
  cancellation}.
\newblock \emph{\bibinfo{journal}{Physical Review A}}
  \textbf{\bibinfo{volume}{65}}, \bibinfo{pages}{053817}
  (\bibinfo{year}{2002}).

\bibitem{nasr:03}
\bibinfo{author}{Nasr, M.~B.}, \bibinfo{author}{Saleh, B.~E.},
  \bibinfo{author}{Sergienko, A.~V.} \& \bibinfo{author}{Teich, M.~C.}
\newblock \bibinfo{title}{Demonstration of dispersion-canceled quantum-optical
  coherence tomography}.
\newblock \emph{\bibinfo{journal}{Physical Review Letters}}
  \textbf{\bibinfo{volume}{91}}, \bibinfo{pages}{083601}
  (\bibinfo{year}{2003}).

\bibitem{nasr:04}
\bibinfo{author}{Nasr, M.~B.}, \bibinfo{author}{Saleh, B.~E.},
  \bibinfo{author}{Sergienko, A.~V.} \& \bibinfo{author}{Teich, M.~C.}
\newblock \bibinfo{title}{Dispersion-cancelled and dispersion-sensitive quantum
  optical coherence tomography}.
\newblock \emph{\bibinfo{journal}{Optics Express}}
  \textbf{\bibinfo{volume}{12}}, \bibinfo{pages}{1353--1362}
  (\bibinfo{year}{2004}).

\bibitem{nasr:09}
\bibinfo{author}{Nasr, M.~B.} \emph{et~al.}
\newblock \bibinfo{title}{Quantum optical coherence tomography of a biological
  sample}.
\newblock \emph{\bibinfo{journal}{Optics Communications}}
  \textbf{\bibinfo{volume}{282}}, \bibinfo{pages}{1154--1159}
  (\bibinfo{year}{2009}).

\bibitem{lopez:12}
\bibinfo{author}{Lopez-Mago, D.} \& \bibinfo{author}{Novotny, L.}
\newblock \bibinfo{title}{Quantum-optical coherence tomography with collinear
  entangled photons}.
\newblock \emph{\bibinfo{journal}{Optics Letters}}
  \textbf{\bibinfo{volume}{37}}, \bibinfo{pages}{4077--4079}
  (\bibinfo{year}{2012}).

\bibitem{erkmen:06}
\bibinfo{author}{Erkmen, B.~I.} \& \bibinfo{author}{Shapiro, J.~H.}
\newblock \bibinfo{title}{Phase-conjugate optical coherence tomography}.
\newblock \emph{\bibinfo{journal}{Physical Review A}}
  \textbf{\bibinfo{volume}{74}}, \bibinfo{pages}{041601}
  (\bibinfo{year}{2006}).

\bibitem{banaszek:07}
\bibinfo{author}{Banaszek, K.}, \bibinfo{author}{Radunsky, A.~S.} \&
  \bibinfo{author}{Walmsley, I.~A.}
\newblock \bibinfo{title}{Blind dispersion compensation for optical coherence
  tomography}.
\newblock \emph{\bibinfo{journal}{Optics Communications}}
  \textbf{\bibinfo{volume}{269}}, \bibinfo{pages}{152--155}
  (\bibinfo{year}{2007}).

\bibitem{resch:07}
\bibinfo{author}{Resch, K.}, \bibinfo{author}{Puvanathasan, P.},
  \bibinfo{author}{Lundeen, J.}, \bibinfo{author}{Mitchell, M.} \&
  \bibinfo{author}{Bizheva, K.}
\newblock \bibinfo{title}{Classical dispersion-cancellation interferometry}.
\newblock \emph{\bibinfo{journal}{Optics Express}}
  \textbf{\bibinfo{volume}{15}}, \bibinfo{pages}{8797--8804}
  (\bibinfo{year}{2007}).

\bibitem{mazurek:13}
\bibinfo{author}{Mazurek, M.}, \bibinfo{author}{Schreiter, K.},
  \bibinfo{author}{Prevedel, R.}, \bibinfo{author}{Kaltenbaek, R.} \&
  \bibinfo{author}{Resch, K.}
\newblock \bibinfo{title}{Dispersion-cancelled biological imaging with
  quantum-inspired interferometry}.
\newblock \emph{\bibinfo{journal}{Scientific Reports}}
  \textbf{\bibinfo{volume}{3}}, \bibinfo{pages}{1582} (\bibinfo{year}{2013}).

\bibitem{Nasr2009}
\bibinfo{author}{Nasr, M.~B.} \emph{et~al.}
\newblock \bibinfo{title}{Quantum optical coherence tomography of a biological
  sample}.
\newblock \emph{\bibinfo{journal}{Optics Communications}}
  \textbf{\bibinfo{volume}{282}}, \bibinfo{pages}{1154 -- 1159}
  (\bibinfo{year}{2009}).

\bibitem{Boto2000}
\bibinfo{author}{Boto, A.~N.} \emph{et~al.}
\newblock \bibinfo{title}{Quantum interferometric optical lithography:
  Exploiting entanglement to beat the diffraction limit}.
\newblock \emph{\bibinfo{journal}{Physical Review Letters}}
  \textbf{\bibinfo{volume}{85}}, \bibinfo{pages}{2733--2736}
  (\bibinfo{year}{2000}).

\bibitem{giovannetti:11}
\bibinfo{author}{Giovannetti, V.}, \bibinfo{author}{Lloyd, S.} \&
  \bibinfo{author}{Maccone, L.}
\newblock \bibinfo{title}{Advances in quantum metrology}.
\newblock \emph{\bibinfo{journal}{Nature Photonics}}
  \textbf{\bibinfo{volume}{5}}, \bibinfo{pages}{222} (\bibinfo{year}{2011}).

\bibitem{rarity1990two}
\bibinfo{author}{Rarity, J.} \emph{et~al.}
\newblock \bibinfo{title}{Two-photon interference in a Mach-Zehnder
  interferometer}.
\newblock \emph{\bibinfo{journal}{Physical Review Letters}}
  \textbf{\bibinfo{volume}{65}}, \bibinfo{pages}{1348} (\bibinfo{year}{1990}).

\bibitem{edamatsu2002measurement}
\bibinfo{author}{Edamatsu, K.}, \bibinfo{author}{Shimizu, R.} \&
  \bibinfo{author}{Itoh, T.}
\newblock \bibinfo{title}{Measurement of the photonic de Broglie wavelength of
  entangled photon pairs generated by spontaneous parametric down-conversion}.
\newblock \emph{\bibinfo{journal}{Physical Review Letters}}
  \textbf{\bibinfo{volume}{89}}, \bibinfo{pages}{213601}
  (\bibinfo{year}{2002}).

\bibitem{mccusker2009efficient}
\bibinfo{author}{McCusker, K.~T.} \& \bibinfo{author}{Kwiat, P.~G.}
\newblock \bibinfo{title}{Efficient optical quantum state engineering}.
\newblock \emph{\bibinfo{journal}{Physical Review Letters}}
  \textbf{\bibinfo{volume}{103}}, \bibinfo{pages}{163602}
  (\bibinfo{year}{2009}).

\bibitem{kapale2007bootstrapping}
\bibinfo{author}{Kapale, K.~T.} \& \bibinfo{author}{Dowling, J.~P.}
\newblock \bibinfo{title}{Bootstrapping approach for generating maximally
  path-entangled photon states}.
\newblock \emph{\bibinfo{journal}{Physical Review Letters}}
  \textbf{\bibinfo{volume}{99}}, \bibinfo{pages}{053602}
  (\bibinfo{year}{2007}).

\bibitem{cable2007efficient}
\bibinfo{author}{Cable, H.} \& \bibinfo{author}{Dowling, J.~P.}
\newblock \bibinfo{title}{Efficient generation of large number-path
  entanglement using only linear optics and feed-forward}.
\newblock \emph{\bibinfo{journal}{Physical Review Letters}}
  \textbf{\bibinfo{volume}{99}}, \bibinfo{pages}{163604}
  (\bibinfo{year}{2007}).

\bibitem{pezze2008mach}
\bibinfo{author}{Pezz{\'e}, L.} \& \bibinfo{author}{Smerzi, A.}
\newblock \bibinfo{title}{Mach-zehnder interferometry at the Heisenberg limit
  with coherent and squeezed-vacuum light}.
\newblock \emph{\bibinfo{journal}{Physical Review Letters}}
  \textbf{\bibinfo{volume}{100}}, \bibinfo{pages}{073601}
  (\bibinfo{year}{2008}).

\bibitem{d2001two}
\bibinfo{author}{D'Angelo, M.}, \bibinfo{author}{Chekhova, M.~V.} \&
  \bibinfo{author}{Shih, Y.}
\newblock \bibinfo{title}{Two-photon diffraction and quantum lithography}.
\newblock \emph{\bibinfo{journal}{Physical Review Letters}}
  \textbf{\bibinfo{volume}{87}}, \bibinfo{pages}{013602}
  (\bibinfo{year}{2001}).

\bibitem{mitchell2004super}
\bibinfo{author}{Mitchell, M.~W.}, \bibinfo{author}{Lundeen, J.~S.} \&
  \bibinfo{author}{Steinberg, A.~M.}
\newblock \bibinfo{title}{Super-resolving phase measurements with a multiphoton
  entangled state}.
\newblock \emph{\bibinfo{journal}{Nature}} \textbf{\bibinfo{volume}{429}},
  \bibinfo{pages}{161} (\bibinfo{year}{2004}).

\bibitem{walther2004broglie}
\bibinfo{author}{Walther, P.} \emph{et~al.}
\newblock \bibinfo{title}{De broglie wavelength of a non-local four-photon
  state}.
\newblock \emph{\bibinfo{journal}{Nature}} \textbf{\bibinfo{volume}{429}},
  \bibinfo{pages}{158} (\bibinfo{year}{2004}).

\bibitem{kim2009three}
\bibinfo{author}{Kim, H.}, \bibinfo{author}{Park, H.~S.} \&
  \bibinfo{author}{Choi, S.-K.}
\newblock \bibinfo{title}{Three-photon n00n states generated by photon
  subtraction from double photon pairs}.
\newblock \emph{\bibinfo{journal}{Optics Express}}
  \textbf{\bibinfo{volume}{17}}, \bibinfo{pages}{19720--19726}
  (\bibinfo{year}{2009}).

\bibitem{liu2008demonstration}
\bibinfo{author}{Liu, B.} \emph{et~al.}
\newblock \bibinfo{title}{Demonstration of the three-photon de broglie
  wavelength by projection measurement}.
\newblock \emph{\bibinfo{journal}{Physical Review A}}
  \textbf{\bibinfo{volume}{77}}, \bibinfo{pages}{023815}
  (\bibinfo{year}{2008}).

\bibitem{resch2007time}
\bibinfo{author}{Resch, K.~J.} \emph{et~al.}
\newblock \bibinfo{title}{Time-reversal and super-resolving phase
  measurements}.
\newblock \emph{\bibinfo{journal}{Physical Review Letters}}
  \textbf{\bibinfo{volume}{98}}, \bibinfo{pages}{223601}
  (\bibinfo{year}{2007}).

\bibitem{okamoto2008beating}
\bibinfo{author}{Okamoto, R.} \emph{et~al.}
\newblock \bibinfo{title}{Beating the standard quantum limit: phase
  super-sensitivity of n-photon interferometers}.
\newblock \emph{\bibinfo{journal}{New Journal of Physics}}
  \textbf{\bibinfo{volume}{10}}, \bibinfo{pages}{073033}
  (\bibinfo{year}{2008}).

\bibitem{nagata2007beating}
\bibinfo{author}{Nagata, T.}, \bibinfo{author}{Okamoto, R.},
  \bibinfo{author}{O'Brien, J.~L.}, \bibinfo{author}{Sasaki, K.} \&
  \bibinfo{author}{Takeuchi, S.}
\newblock \bibinfo{title}{Beating the standard quantum limit with
  four-entangled photons}.
\newblock \emph{\bibinfo{journal}{Science}} \textbf{\bibinfo{volume}{316}},
  \bibinfo{pages}{726--729} (\bibinfo{year}{2007}).

\bibitem{afek2010high}
\bibinfo{author}{Afek, I.}, \bibinfo{author}{Ambar, O.} \&
  \bibinfo{author}{Silberberg, Y.}
\newblock \bibinfo{title}{High-N00N states by mixing quantum and classical
  light}.
\newblock \emph{\bibinfo{journal}{Science}} \textbf{\bibinfo{volume}{328}},
  \bibinfo{pages}{879--881} (\bibinfo{year}{2010}).

\bibitem{zhou2017superresolving}
\bibinfo{author}{Zhou, Z.-Y.} \emph{et~al.}
\newblock \bibinfo{title}{Superresolving phase measurement with
  short-wavelength noon states by quantum frequency up-conversion}.
\newblock \emph{\bibinfo{journal}{Physical Review Applied}}
  \textbf{\bibinfo{volume}{7}}, \bibinfo{pages}{064025} (\bibinfo{year}{2017}).

\bibitem{kaltenbaek2006experimental}
\bibinfo{author}{Kaltenbaek, R.}, \bibinfo{author}{Blauensteiner, B.},
  \bibinfo{author}{{\.Z}ukowski, M.}, \bibinfo{author}{Aspelmeyer, M.} \&
  \bibinfo{author}{Zeilinger, A.}
\newblock \bibinfo{title}{Experimental interference of independent photons}.
\newblock \emph{\bibinfo{journal}{Physical Review Letters}}
  \textbf{\bibinfo{volume}{96}}, \bibinfo{pages}{240502}
  (\bibinfo{year}{2006}).

\bibitem{mosley2008heralded}
\bibinfo{author}{Mosley, P.~J.} \emph{et~al.}
\newblock \bibinfo{title}{Heralded generation of ultrafast single photons in
  pure quantum states}.
\newblock \emph{\bibinfo{journal}{Physical Review Letters}}
  \textbf{\bibinfo{volume}{100}}, \bibinfo{pages}{133601}
  (\bibinfo{year}{2008}).

\bibitem{sanaka2009indistinguishable}
\bibinfo{author}{Sanaka, K.}, \bibinfo{author}{Pawlis, A.},
  \bibinfo{author}{Ladd, T.~D.}, \bibinfo{author}{Lischka, K.} \&
  \bibinfo{author}{Yamamoto, Y.}
\newblock \bibinfo{title}{Indistinguishable photons from independent
  semiconductor nanostructures}.
\newblock \emph{\bibinfo{journal}{Physical Review Letters}}
  \textbf{\bibinfo{volume}{103}}, \bibinfo{pages}{053601}
  (\bibinfo{year}{2009}).

\bibitem{flagg2010interference}
\bibinfo{author}{Flagg, E.~B.} \emph{et~al.}
\newblock \bibinfo{title}{Interference of single photons from two separate
  semiconductor quantum dots}.
\newblock \emph{\bibinfo{journal}{Physical Review Letters}}
  \textbf{\bibinfo{volume}{104}}, \bibinfo{pages}{137401}
  (\bibinfo{year}{2010}).

\bibitem{patel2010two}
\bibinfo{author}{Patel, R.~B.} \emph{et~al.}
\newblock \bibinfo{title}{Two-photon interference of the emission from
  electrically tunable remote quantum dots}.
\newblock \emph{\bibinfo{journal}{Nature Photonics}}
  \textbf{\bibinfo{volume}{4}}, \bibinfo{pages}{632} (\bibinfo{year}{2010}).

\bibitem{wei2014deterministic}
\bibinfo{author}{Wei, Y.-J.} \emph{et~al.}
\newblock \bibinfo{title}{Deterministic and robust generation of single photons
  from a single quantum dot with 99.5\% indistinguishability using adiabatic
  rapid passage}.
\newblock \emph{\bibinfo{journal}{Nano Letters}} \textbf{\bibinfo{volume}{14}},
  \bibinfo{pages}{6515--6519} (\bibinfo{year}{2014}).

\bibitem{senellart:17}
\bibinfo{author}{Senellart, P.}, \bibinfo{author}{Solomon, G.} \&
  \bibinfo{author}{White, A.}
\newblock \bibinfo{title}{High-performance semiconductor quantum-dot
  single-photon sources}.
\newblock \emph{\bibinfo{journal}{Nature Nanotechnology}}
  \textbf{\bibinfo{volume}{12}}, \bibinfo{pages}{1026--1039}
  (\bibinfo{year}{2017}).

\bibitem{felinto2006conditional}
\bibinfo{author}{Felinto, D.} \emph{et~al.}
\newblock \bibinfo{title}{Conditional control of the quantum states of remote
  atomic memories for quantum networking}.
\newblock \emph{\bibinfo{journal}{Nature Physics}}
  \textbf{\bibinfo{volume}{2}}, \bibinfo{pages}{844} (\bibinfo{year}{2006}).

\bibitem{chaneliere2007quantum}
\bibinfo{author}{Chaneliere, T.} \emph{et~al.}
\newblock \bibinfo{title}{Quantum interference of electromagnetic fields from
  remote quantum memories}.
\newblock \emph{\bibinfo{journal}{Physical Review Letters}}
  \textbf{\bibinfo{volume}{98}}, \bibinfo{pages}{113602}
  (\bibinfo{year}{2007}).

\bibitem{yuan2007synchronized}
\bibinfo{author}{Yuan, Z.-S.} \emph{et~al.}
\newblock \bibinfo{title}{Synchronized independent narrow-band single photons
  and efficient generation of photonic entanglement}.
\newblock \emph{\bibinfo{journal}{Physical Review Letters}}
  \textbf{\bibinfo{volume}{98}}, \bibinfo{pages}{180503}
  (\bibinfo{year}{2007}).

\bibitem{yuan2008experimental}
\bibinfo{author}{Yuan, Z.-S.} \emph{et~al.}
\newblock \bibinfo{title}{Experimental demonstration of a bdcz quantum repeater
  node}.
\newblock \emph{\bibinfo{journal}{Nature}} \textbf{\bibinfo{volume}{454}},
  \bibinfo{pages}{1098} (\bibinfo{year}{2008}).

\bibitem{chen2008memory}
\bibinfo{author}{Chen, Y.-A.} \emph{et~al.}
\newblock \bibinfo{title}{Memory-built-in quantum teleportation with photonic
  and atomic qubits}.
\newblock \emph{\bibinfo{journal}{Nature Physics}}
  \textbf{\bibinfo{volume}{4}}, \bibinfo{pages}{103} (\bibinfo{year}{2008}).

\bibitem{bernien2012two}
\bibinfo{author}{Bernien, H.} \emph{et~al.}
\newblock \bibinfo{title}{Two-photon quantum interference from separate
  nitrogen vacancy centers in diamond}.
\newblock \emph{\bibinfo{journal}{Physical Review Letters}}
  \textbf{\bibinfo{volume}{108}}, \bibinfo{pages}{043604}
  (\bibinfo{year}{2012}).

\bibitem{sipahigil2012quantum}
\bibinfo{author}{Sipahigil, A.} \emph{et~al.}
\newblock \bibinfo{title}{Quantum interference of single photons from remote
  nitrogen-vacancy centers in diamond}.
\newblock \emph{\bibinfo{journal}{Physical Review Letters}}
  \textbf{\bibinfo{volume}{108}}, \bibinfo{pages}{143601}
  (\bibinfo{year}{2012}).

\bibitem{sipahigil2014indistinguishable}
\bibinfo{author}{Sipahigil, A.} \emph{et~al.}
\newblock \bibinfo{title}{Indistinguishable photons from separated
  silicon-vacancy centers in diamond}.
\newblock \emph{\bibinfo{journal}{Physical Review Letters}}
  \textbf{\bibinfo{volume}{113}}, \bibinfo{pages}{113602}
  (\bibinfo{year}{2014}).

\bibitem{maunz2007quantum}
\bibinfo{author}{Maunz, P.} \emph{et~al.}
\newblock \bibinfo{title}{Quantum interference of photon pairs from two remote
  trapped atomic ions}.
\newblock \emph{\bibinfo{journal}{Nature Physics}}
  \textbf{\bibinfo{volume}{3}}, \bibinfo{pages}{538} (\bibinfo{year}{2007}).

\bibitem{beugnon2006quantum}
\bibinfo{author}{Beugnon, J.} \emph{et~al.}
\newblock \bibinfo{title}{Quantum interference between two single photons
  emitted by independently trapped atoms}.
\newblock \emph{\bibinfo{journal}{Nature}} \textbf{\bibinfo{volume}{440}},
  \bibinfo{pages}{779} (\bibinfo{year}{2006}).

\bibitem{specht2011single}
\bibinfo{author}{Specht, H.~P.} \emph{et~al.}
\newblock \bibinfo{title}{A single-atom quantum memory}.
\newblock \emph{\bibinfo{journal}{Nature}} \textbf{\bibinfo{volume}{473}},
  \bibinfo{pages}{190} (\bibinfo{year}{2011}).

\bibitem{kiraz2005indistinguishable}
\bibinfo{author}{Kiraz, A.} \emph{et~al.}
\newblock \bibinfo{title}{Indistinguishable photons from a single molecule}.
\newblock \emph{\bibinfo{journal}{Physical Review Letters}}
  \textbf{\bibinfo{volume}{94}}, \bibinfo{pages}{223602}
  (\bibinfo{year}{2005}).

\bibitem{lettow2010quantum}
\bibinfo{author}{Lettow, R.} \emph{et~al.}
\newblock \bibinfo{title}{Quantum interference of tunably indistinguishable
  photons from remote organic molecules}.
\newblock \emph{\bibinfo{journal}{Physical Review Letters}}
  \textbf{\bibinfo{volume}{104}}, \bibinfo{pages}{123605}
  (\bibinfo{year}{2010}).

\bibitem{calsamiglia2001maximum}
\bibinfo{author}{Calsamiglia, J.} \& \bibinfo{author}{L{\"u}tkenhaus, N.}
\newblock \bibinfo{title}{Maximum efficiency of a linear-optical bell-state
  analyzer}.
\newblock \emph{\bibinfo{journal}{Applied Physics B}}
  \textbf{\bibinfo{volume}{72}}, \bibinfo{pages}{67--71}
  (\bibinfo{year}{2001}).

\bibitem{kim2001quantum}
\bibinfo{author}{Kim, Y.-H.}, \bibinfo{author}{Kulik, S.~P.} \&
  \bibinfo{author}{Shih, Y.}
\newblock \bibinfo{title}{Quantum teleportation of a polarization state with a
  complete bell state measurement}.
\newblock \emph{\bibinfo{journal}{Physical Review Letters}}
  \textbf{\bibinfo{volume}{86}}, \bibinfo{pages}{1370} (\bibinfo{year}{2001}).

\bibitem{kwiat1998embedded}
\bibinfo{author}{Kwiat, P.~G.} \& \bibinfo{author}{Weinfurter, H.}
\newblock \bibinfo{title}{Embedded bell-state analysis}.
\newblock \emph{\bibinfo{journal}{Physical Review A}}
  \textbf{\bibinfo{volume}{58}}, \bibinfo{pages}{R2623} (\bibinfo{year}{1998}).

\bibitem{barreiro2008beating}
\bibinfo{author}{Barreiro, J.~T.}, \bibinfo{author}{Wei, T.-C.} \&
  \bibinfo{author}{Kwiat, P.~G.}
\newblock \bibinfo{title}{Beating the channel capacity limit for linear
  photonic superdense coding}.
\newblock \emph{\bibinfo{journal}{Nature physics}}
  \textbf{\bibinfo{volume}{4}}, \bibinfo{pages}{282} (\bibinfo{year}{2008}).

\bibitem{knill2001scheme}
\bibinfo{author}{Knill, E.}, \bibinfo{author}{Laflamme, R.} \&
  \bibinfo{author}{Milburn, G.~J.}
\newblock \bibinfo{title}{A scheme for efficient quantum computation with
  linear optics}.
\newblock \emph{\bibinfo{journal}{Nature}} \textbf{\bibinfo{volume}{409}},
  \bibinfo{pages}{46} (\bibinfo{year}{2001}).

\bibitem{grice2011arbitrarily}
\bibinfo{author}{Grice, W.~P.}
\newblock \bibinfo{title}{Arbitrarily complete bell-state measurement using
  only linear optical elements}.
\newblock \emph{\bibinfo{journal}{Physical Review A}}
  \textbf{\bibinfo{volume}{84}}, \bibinfo{pages}{042331}
  (\bibinfo{year}{2011}).

\bibitem{gisin:02}
\bibinfo{author}{Gisin, N.}, \bibinfo{author}{Ribordy, G.},
  \bibinfo{author}{Tittel, W.} \& \bibinfo{author}{Zbinden, H.}
\newblock \bibinfo{title}{Quantum cryptography}.
\newblock \emph{\bibinfo{journal}{Reviews of Modern Physics}}
  \textbf{\bibinfo{volume}{74}}, \bibinfo{pages}{145} (\bibinfo{year}{2002}).

\bibitem{wootters:82}
\bibinfo{author}{Wootters, W.~K.} \& \bibinfo{author}{Zurek, W.~H.}
\newblock \bibinfo{title}{A single quantum cannot be cloned}.
\newblock \emph{\bibinfo{journal}{Nature}} \textbf{\bibinfo{volume}{299}},
  \bibinfo{pages}{802--803} (\bibinfo{year}{1982}).

\bibitem{shor:00}
\bibinfo{author}{Shor, P.~W.} \& \bibinfo{author}{Preskill, J.}
\newblock \bibinfo{title}{Simple proof of security of the bb84 quantum key
  distribution protocol}.
\newblock \emph{\bibinfo{journal}{Physical Review Letters}}
  \textbf{\bibinfo{volume}{85}}, \bibinfo{pages}{441} (\bibinfo{year}{2000}).

\bibitem{zukowski:93}
\bibinfo{author}{Zukowski, M.}, \bibinfo{author}{Zeilinger, A.},
  \bibinfo{author}{Horne, M.~A.} \& \bibinfo{author}{Ekert, A.~K.}
\newblock \bibinfo{title}{``event-ready-detectors''bell experiment via
  entanglement swapping}.
\newblock \emph{\bibinfo{journal}{Physical Review Letters}}
  \textbf{\bibinfo{volume}{71}}, \bibinfo{pages}{4287--4290}
  (\bibinfo{year}{1993}).

\bibitem{bennett:93}
\bibinfo{author}{Bennett, C.~H.} \emph{et~al.}
\newblock \bibinfo{title}{Teleporting an unknown quantum state via dual
  classical and einstein-podolsky-rosen channels}.
\newblock \emph{\bibinfo{journal}{Physical Review Letters}}
  \textbf{\bibinfo{volume}{70}}, \bibinfo{pages}{1895} (\bibinfo{year}{1993}).

\bibitem{bennett:92a}
\bibinfo{author}{Bennett, C.~H.} \& \bibinfo{author}{Wiesner, S.~J.}
\newblock \bibinfo{title}{Communication via one-and two-particle operators on
  einstein-podolsky-rosen states}.
\newblock \emph{\bibinfo{journal}{Physical Review Letters}}
  \textbf{\bibinfo{volume}{69}}, \bibinfo{pages}{2881} (\bibinfo{year}{1992}).

\bibitem{lo:12}
\bibinfo{author}{Lo, H.-K.}, \bibinfo{author}{Curty, M.} \&
  \bibinfo{author}{Qi, B.}
\newblock \bibinfo{title}{Measurement-device-independent quantum key
  distribution}.
\newblock \emph{\bibinfo{journal}{Physical Review Letters}}
  \textbf{\bibinfo{volume}{108}}, \bibinfo{pages}{130503}
  (\bibinfo{year}{2012}).

\bibitem{guan:15}
\bibinfo{author}{Guan, J.-Y.} \emph{et~al.}
\newblock \bibinfo{title}{Experimental passive round-robin differential
  phase-shift quantum key distribution}.
\newblock \emph{\bibinfo{journal}{Physical Review Letters}}
  \textbf{\bibinfo{volume}{114}}, \bibinfo{pages}{180502}
  (\bibinfo{year}{2015}).

\bibitem{hofmann2012heralded}
\bibinfo{author}{Hofmann, J.} \emph{et~al.}
\newblock \bibinfo{title}{Heralded entanglement between widely separated
  atoms}.
\newblock \emph{\bibinfo{journal}{Science}} \textbf{\bibinfo{volume}{337}},
  \bibinfo{pages}{72--75} (\bibinfo{year}{2012}).

\bibitem{Bell:64}
\bibinfo{author}{Bell, J.~S.}
\newblock \bibinfo{title}{On the einstein podolsky rosen paradox}.
\newblock \emph{\bibinfo{journal}{Physics Physique Fizika}}
  \textbf{\bibinfo{volume}{1}}, \bibinfo{pages}{195--200}
  (\bibinfo{year}{1964}).

\bibitem{narla2016robust}
\bibinfo{author}{Narla, A.} \emph{et~al.}
\newblock \bibinfo{title}{Robust concurrent remote entanglement between two
  superconducting qubits}.
\newblock \emph{\bibinfo{journal}{Physical Review X}}
  \textbf{\bibinfo{volume}{6}}, \bibinfo{pages}{031036} (\bibinfo{year}{2016}).

\bibitem{sangouard2011quantum}
\bibinfo{author}{Sangouard, N.}, \bibinfo{author}{Simon, C.},
  \bibinfo{author}{De~Riedmatten, H.} \& \bibinfo{author}{Gisin, N.}
\newblock \bibinfo{title}{Quantum repeaters based on atomic ensembles and
  linear optics}.
\newblock \emph{\bibinfo{journal}{Reviews of Modern Physics}}
  \textbf{\bibinfo{volume}{83}}, \bibinfo{pages}{33} (\bibinfo{year}{2011}).

\bibitem{briegel1998quantum}
\bibinfo{author}{Briegel, H.-J.}, \bibinfo{author}{D{\"u}r, W.},
  \bibinfo{author}{Cirac, J.~I.} \& \bibinfo{author}{Zoller, P.}
\newblock \bibinfo{title}{Quantum repeaters: the role of imperfect local
  operations in quantum communication}.
\newblock \emph{\bibinfo{journal}{Physical Review Letters}}
  \textbf{\bibinfo{volume}{81}}, \bibinfo{pages}{5932} (\bibinfo{year}{1998}).

\bibitem{fung:07}
\bibinfo{author}{Fung, C.-H.~F.}, \bibinfo{author}{Qi, B.},
  \bibinfo{author}{Tamaki, K.} \& \bibinfo{author}{Lo, H.-K.}
\newblock \bibinfo{title}{Phase-remapping attack in practical
  quantum-key-distribution systems}.
\newblock \emph{\bibinfo{journal}{Physical Review A}}
  \textbf{\bibinfo{volume}{75}}, \bibinfo{pages}{032314}
  (\bibinfo{year}{2007}).

\bibitem{zhao:08}
\bibinfo{author}{Zhao, Y.}, \bibinfo{author}{Fung, C.-H.~F.},
  \bibinfo{author}{Qi, B.}, \bibinfo{author}{Chen, C.} \& \bibinfo{author}{Lo,
  H.-K.}
\newblock \bibinfo{title}{Quantum hacking: Experimental demonstration of
  time-shift attack against practical quantum-key-distribution systems}.
\newblock \emph{\bibinfo{journal}{Physical Review A}}
  \textbf{\bibinfo{volume}{78}}, \bibinfo{pages}{042333}
  (\bibinfo{year}{2008}).

\bibitem{lydersen:10}
\bibinfo{author}{Lydersen, L.} \emph{et~al.}
\newblock \bibinfo{title}{Hacking commercial quantum cryptography systems by
  tailored bright illumination}.
\newblock \emph{\bibinfo{journal}{Nature Photonics}}
  \textbf{\bibinfo{volume}{4}}, \bibinfo{pages}{686} (\bibinfo{year}{2010}).

\bibitem{gerhardt:11}
\bibinfo{author}{Gerhardt, I.} \emph{et~al.}
\newblock \bibinfo{title}{Full-field implementation of a perfect eavesdropper
  on a quantum cryptography system}.
\newblock \emph{\bibinfo{journal}{Nature Communications}}
  \textbf{\bibinfo{volume}{2}}, \bibinfo{pages}{349} (\bibinfo{year}{2011}).

\bibitem{mayers:98}
\bibinfo{author}{Mayers, D.} \& \bibinfo{author}{Yao, A.}
\newblock \bibinfo{title}{Proceedings of the 39th annual symposium on
  foundations of computer science (focs98)}  (\bibinfo{year}{1998}).

\bibitem{acin:07}
\bibinfo{author}{Ac{\'\i}n, A.} \emph{et~al.}
\newblock \bibinfo{title}{Device-independent security of quantum cryptography
  against collective attacks}.
\newblock \emph{\bibinfo{journal}{Physical Review Letters}}
  \textbf{\bibinfo{volume}{98}}, \bibinfo{pages}{230501}
  (\bibinfo{year}{2007}).

\bibitem{liu:13}
\bibinfo{author}{Liu, Y.} \emph{et~al.}
\newblock \bibinfo{title}{Experimental measurement-device-independent quantum
  key distribution}.
\newblock \emph{\bibinfo{journal}{Physical Review Letters}}
  \textbf{\bibinfo{volume}{111}}, \bibinfo{pages}{130502}
  (\bibinfo{year}{2013}).

\bibitem{tang:14}
\bibinfo{author}{Tang, Z.} \emph{et~al.}
\newblock \bibinfo{title}{Experimental demonstration of polarization encoding
  measurement-device-independent quantum key distribution}.
\newblock \emph{\bibinfo{journal}{Physical Review Letters}}
  \textbf{\bibinfo{volume}{112}}, \bibinfo{pages}{190503}
  (\bibinfo{year}{2014}).

\bibitem{tang:14b}
\bibinfo{author}{Tang, Y.-L.} \emph{et~al.}
\newblock \bibinfo{title}{Measurement-device-independent quantum key
  distribution over 200 km}.
\newblock \emph{\bibinfo{journal}{Physical Review Letters}}
  \textbf{\bibinfo{volume}{113}}, \bibinfo{pages}{190501}
  (\bibinfo{year}{2014}).

\bibitem{yin:16}
\bibinfo{author}{Yin, H.-L.} \emph{et~al.}
\newblock \bibinfo{title}{Measurement-device-independent quantum key
  distribution over a 404 km optical fiber}.
\newblock \emph{\bibinfo{journal}{Physical Review Letters}}
  \textbf{\bibinfo{volume}{117}}, \bibinfo{pages}{190501}
  (\bibinfo{year}{2016}).

\bibitem{lo:05}
\bibinfo{author}{Lo, H.-K.}, \bibinfo{author}{Ma, X.} \& \bibinfo{author}{Chen,
  K.}
\newblock \bibinfo{title}{Decoy state quantum key distribution}.
\newblock \emph{\bibinfo{journal}{Physical Review Letters}}
  \textbf{\bibinfo{volume}{94}}, \bibinfo{pages}{230504}
  (\bibinfo{year}{2005}).

\bibitem{sasaki:14}
\bibinfo{author}{Sasaki, T.}, \bibinfo{author}{Yamamoto, Y.} \&
  \bibinfo{author}{Koashi, M.}
\newblock \bibinfo{title}{Practical quantum key distribution protocol without
  monitoring signal disturbance}.
\newblock \emph{\bibinfo{journal}{Nature}} \textbf{\bibinfo{volume}{509}},
  \bibinfo{pages}{475} (\bibinfo{year}{2014}).

\bibitem{takesue:15}
\bibinfo{author}{Takesue, H.}, \bibinfo{author}{Sasaki, T.},
  \bibinfo{author}{Tamaki, K.} \& \bibinfo{author}{Koashi, M.}
\newblock \bibinfo{title}{Experimental quantum key distribution without
  monitoring signal disturbance}.
\newblock \emph{\bibinfo{journal}{Nature Photonics}}
  \textbf{\bibinfo{volume}{9}}, \bibinfo{pages}{827} (\bibinfo{year}{2015}).

\bibitem{wang:15}
\bibinfo{author}{Wang, S.} \emph{et~al.}
\newblock \bibinfo{title}{Experimental demonstration of a quantum key
  distribution without signal disturbance monitoring}.
\newblock \emph{\bibinfo{journal}{Nature Physics}} \textbf{\bibinfo{volume}{9}},
  \bibinfo{pages}{832} (\bibinfo{year}{2015}).

\bibitem{li:16}
\bibinfo{author}{Li, Y.-H.} \emph{et~al.}
\newblock \bibinfo{title}{Experimental round-robin differential phase-shift
  quantum key distribution}.
\newblock \emph{\bibinfo{journal}{Physical Review A}} \textbf{\bibinfo{volume}{93}},
  \bibinfo{pages}{030302} (\bibinfo{year}{2016}).

\bibitem{Yin:18}
\bibinfo{author}{Yin, Z.-Q.} \emph{et~al.}
\newblock \bibinfo{title}{Improved security bound for the
  round-robin-differential-phase-shift quantum key distribution}.
\newblock \emph{\bibinfo{journal}{Nature Communications}} \textbf{\bibinfo{volume}{9}},
  \bibinfo{pages}{457} (\bibinfo{year}{2018}).

\bibitem{bouchard:18c}
\bibinfo{author}{Bouchard, F.}, \bibinfo{author}{Sit, A.}, \bibinfo{author}{Heshami, K.},
  \bibinfo{author}{Fickler, R.} \& \bibinfo{author}{Karimi, E.}
\newblock \bibinfo{title}{Round-robin differential-phase-shift quantum key
  distribution with twisted photons}.
\newblock \emph{\bibinfo{journal}{Physical Review A}}
  \textbf{\bibinfo{volume}{98}} (\bibinfo{year}{2018}).

\bibitem{islam2019scalable}
\bibinfo{author}{Islam, N.~T.} \emph{et~al.}
\newblock \bibinfo{title}{Scalable high-rate, high-dimensional time-bin
  encoding quantum key distribution}.
\newblock \emph{\bibinfo{journal}{Quantum Science and Technology}}
  \textbf{\bibinfo{volume}{4}}, \bibinfo{pages}{035008} (\bibinfo{year}{2019}).

\bibitem{choi:75}
\bibinfo{author}{Choi, M.-D.}
\newblock \bibinfo{title}{Completely positive linear maps on complex matrices}.
\newblock \emph{\bibinfo{journal}{Linear Algebra and its Applications}}
  \textbf{\bibinfo{volume}{10}}, \bibinfo{pages}{285 -- 290}
  (\bibinfo{year}{1975}).

\bibitem{jam:72}
\bibinfo{author}{Jamiolkowski, A.}
\newblock \bibinfo{title}{Linear transformations which preserve trace and
  positive semidefiniteness of operators}.
\newblock \emph{\bibinfo{journal}{Reports on Mathematical Physics}}
  \textbf{\bibinfo{volume}{3}}, \bibinfo{pages}{275 -- 278}
  (\bibinfo{year}{1972}).

\bibitem{zhang2016engineering}
\bibinfo{author}{Zhang, Y.} \emph{et~al.}
\newblock \bibinfo{title}{Engineering two-photon high-dimensional states
  through quantum interference}.
\newblock \emph{\bibinfo{journal}{Science Advances}}
  \textbf{\bibinfo{volume}{2}}, \bibinfo{pages}{e1501165}
  (\bibinfo{year}{2016}).

\bibitem{zhao2014entangling}
\bibinfo{author}{Zhao, T.-M.} \emph{et~al.}
\newblock \bibinfo{title}{Entangling different-color photons via time-resolved
  measurement and active feed forward}.
\newblock \emph{\bibinfo{journal}{Physical Review Letters}}
  \textbf{\bibinfo{volume}{112}}, \bibinfo{pages}{103602}
  (\bibinfo{year}{2014}).

\bibitem{peres:03}
\bibinfo{author}{Peres, A.}
\newblock \bibinfo{title}{How the no-cloning theorem got its name}.
\newblock \emph{\bibinfo{journal}{Fortschritte der Physik}}
  \textbf{\bibinfo{volume}{51}}, \bibinfo{pages}{458--461}
  (\bibinfo{year}{2003}).

\bibitem{navez:03}
\bibinfo{author}{Navez, P.} \& \bibinfo{author}{Cerf, N.~J.}
\newblock \bibinfo{title}{Cloning a real d-dimensional quantum state on the
  edge of the no-signaling condition}.
\newblock \emph{\bibinfo{journal}{Physical Review A}}
  \textbf{\bibinfo{volume}{68}}, \bibinfo{pages}{032313}
  (\bibinfo{year}{2003}).

\bibitem{nagali:09}
\bibinfo{author}{Nagali, E.} \emph{et~al.}
\newblock \bibinfo{title}{Optimal quantum cloning of orbital angular momentum
  photon qubits through Hong--Ou--Mandel coalescence}.
\newblock \emph{\bibinfo{journal}{Nature Photonics}}
  \textbf{\bibinfo{volume}{3}}, \bibinfo{pages}{720} (\bibinfo{year}{2009}).

\bibitem{nagali:10}
\bibinfo{author}{Nagali, E.} \emph{et~al.}
\newblock \bibinfo{title}{Experimental optimal cloning of four-dimensional
  quantum states of photons}.
\newblock \emph{\bibinfo{journal}{Physical Review Letters}}
  \textbf{\bibinfo{volume}{105}}, \bibinfo{pages}{073602}
  (\bibinfo{year}{2010}).

\bibitem{bouchard:17}
\bibinfo{author}{Bouchard, F.}, \bibinfo{author}{Fickler, R.},
  \bibinfo{author}{Boyd, R.~W.} \& \bibinfo{author}{Karimi, E.}
\newblock \bibinfo{title}{High-dimensional quantum cloning and applications to
  quantum hacking}.
\newblock \emph{\bibinfo{journal}{Science Advances}} \textbf{\bibinfo{volume}{3}},
  \bibinfo{pages}{e1601915} (\bibinfo{year}{2017}).

\bibitem{demartini:02}
\bibinfo{author}{De~Martini, F.}, \bibinfo{author}{Bu{\v{z}}ek, V.},
  \bibinfo{author}{Sciarrino, F.} \& \bibinfo{author}{Sias, C.}
\newblock \bibinfo{title}{Experimental realization of the quantum universal not
  gate}.
\newblock \emph{\bibinfo{journal}{Nature}} \textbf{\bibinfo{volume}{419}},
  \bibinfo{pages}{815--818} (\bibinfo{year}{2002}).

\bibitem{ricci:04}
\bibinfo{author}{Ricci, M.}, \bibinfo{author}{Sciarrino, F.},
  \bibinfo{author}{Sias, C.} \& \bibinfo{author}{De~Martini, F.}
\newblock \bibinfo{title}{Teleportation scheme implementing the universal
  optical quantum cloning machine and the universal not gate}.
\newblock \emph{\bibinfo{journal}{Physical Review Letters}}
  \textbf{\bibinfo{volume}{92}}, \bibinfo{pages}{047901}
  (\bibinfo{year}{2004}).

\bibitem{vitelli2013joining}
\bibinfo{author}{Vitelli, C.} \emph{et~al.}
\newblock \bibinfo{title}{Joining the quantum state of two photons into one}.
\newblock \emph{\bibinfo{journal}{Nature Photonics}}
  \textbf{\bibinfo{volume}{7}}, \bibinfo{pages}{521} (\bibinfo{year}{2013}).

\bibitem{passaro:13}
\bibinfo{author}{Passaro, E.} \emph{et~al.}
\newblock \bibinfo{title}{Joining and splitting the quantum states of photons}.
\newblock \emph{\bibinfo{journal}{Physical Review A}} \textbf{\bibinfo{volume}{88}},
  \bibinfo{pages}{062321} (\bibinfo{year}{2013}).

\bibitem{knill:03}
\bibinfo{author}{Knill, E.}
\newblock \bibinfo{title}{Bounds on the probability of success of postselected
  nonlinear sign shifts implemented with linear optics}.
\newblock \emph{\bibinfo{journal}{Physical Review A}}
  \textbf{\bibinfo{volume}{68}}, \bibinfo{pages}{064303}
  (\bibinfo{year}{2003}).

\bibitem{ralph:01}
\bibinfo{author}{Raplh, T.}, \bibinfo{author}{White, A.},
  \bibinfo{author}{Munro, W.} \& \bibinfo{author}{Milburn, G.}
\newblock \bibinfo{title}{Simple scheme for efficient linear optics quantum
  gates}.
\newblock \emph{\bibinfo{journal}{Physical Review A}}
  \textbf{\bibinfo{volume}{65}}, \bibinfo{pages}{012314}
  (\bibinfo{year}{2001}).

\bibitem{zou:02}
\bibinfo{author}{Zou, P.~K., X.B.} \& \bibinfo{author}{Mathis, W.}
\newblock \bibinfo{title}{Teleportation implementation of nondeterministic
  quantum logic operations by using linear optical elements}.
\newblock \emph{\bibinfo{journal}{Physical Review A}}
  \textbf{\bibinfo{volume}{65}}, \bibinfo{pages}{064305}
  (\bibinfo{year}{2002}).

\bibitem{scheel:04}
\bibinfo{author}{Scheel, S.}, \bibinfo{author}{Pachos, J.},
  \bibinfo{author}{Hinds, E.} \& \bibinfo{author}{Knight, P.}
\newblock \bibinfo{title}{Quantum gates and decoherence}.
\newblock \emph{\bibinfo{journal}{arXiv preprint arXiv:quant-ph/0403152}}
  (\bibinfo{year}{2004}).

\bibitem{knill:02}
\bibinfo{author}{Knill, E.}
\newblock \bibinfo{title}{Quantum gates using linear optics and postselection}.
\newblock \emph{\bibinfo{journal}{Physical Review A}}
  \textbf{\bibinfo{volume}{66}}, \bibinfo{pages}{052306}
  (\bibinfo{year}{2002}).

\bibitem{okamoto:11}
\bibinfo{author}{Okamoto, R.}, \bibinfo{author}{O'Brien, J.},
  \bibinfo{author}{Hofmann, H.} \& \bibinfo{author}{Takeuchi, S.}
\newblock \bibinfo{title}{Realization of a Knill-Laflamme-Milburn
  controlled-not photonic quantum circuit combining effective optical
  nonlinearities}.
\newblock \emph{\bibinfo{journal}{PNAS}} \textbf{\bibinfo{volume}{108}},
  \bibinfo{pages}{10067--10071} (\bibinfo{year}{2011}).

\bibitem{harrow2009quantum}
\bibinfo{author}{Harrow, A.~W.}, \bibinfo{author}{Hassidim, A.} \&
  \bibinfo{author}{Lloyd, S.}
\newblock \bibinfo{title}{Quantum algorithm for linear systems of equations}.
\newblock \emph{\bibinfo{journal}{Physical Review Letters}}
  \textbf{\bibinfo{volume}{103}}, \bibinfo{pages}{150502}
  (\bibinfo{year}{2009}).

\bibitem{cai2013experimental}
\bibinfo{author}{Cai, X.-D.} \emph{et~al.}
\newblock \bibinfo{title}{Experimental quantum computing to solve systems of
  linear equations}.
\newblock \emph{\bibinfo{journal}{Physical Review letters}}
  \textbf{\bibinfo{volume}{110}}, \bibinfo{pages}{230501}
  (\bibinfo{year}{2013}).

\bibitem{fisher2014quantum}
\bibinfo{author}{Fisher, K.} \emph{et~al.}
\newblock \bibinfo{title}{Quantum computing on encrypted data}.
\newblock \emph{\bibinfo{journal}{Nature Communications}}
  \textbf{\bibinfo{volume}{5}}, \bibinfo{pages}{3074} (\bibinfo{year}{2014}).

\bibitem{weimann2016implementation}
\bibinfo{author}{Weimann, S.} \emph{et~al.}
\newblock \bibinfo{title}{Implementation of quantum and classical discrete
  fractional fourier transforms}.
\newblock \emph{\bibinfo{journal}{Nature Communications}}
  \textbf{\bibinfo{volume}{7}}, \bibinfo{pages}{11027} (\bibinfo{year}{2016}).

\bibitem{humphreys2013linear}
\bibinfo{author}{Humphreys, P.~C.} \emph{et~al.}
\newblock \bibinfo{title}{Linear optical quantum computing in a single spatial
  mode}.
\newblock \emph{\bibinfo{journal}{Physical Review Letters}}
  \textbf{\bibinfo{volume}{111}}, \bibinfo{pages}{150501}
  (\bibinfo{year}{2013}).

\bibitem{o2009photonic}
\bibinfo{author}{O'brien, J.~L.}, \bibinfo{author}{Furusawa, A.} \&
  \bibinfo{author}{Vu{\v{c}}kovi{\'c}, J.}
\newblock \bibinfo{title}{Photonic quantum technologies}.
\newblock \emph{\bibinfo{journal}{Nature Photonics}}
  \textbf{\bibinfo{volume}{3}}, \bibinfo{pages}{687} (\bibinfo{year}{2009}).

\bibitem{wang:2019}
\bibinfo{author}{Wang, J.}, \bibinfo{author}{Sciarrino, F.},
  \bibinfo{author}{Laing, A.} \& \bibinfo{author}{Thompson, M.}
\newblock \bibinfo{title}{Integrated photonic quantum technologies}.
\newblock \emph{\bibinfo{journal}{Nature Photonics}}
  \textbf{\bibinfo{volume}{14}}, \bibinfo{pages}{273--284}
  (\bibinfo{year}{2020}).

\bibitem{garcia2013swap}
\bibinfo{author}{Garcia-Escartin, J.~C.} \& \bibinfo{author}{Chamorro-Posada,
  P.}
\newblock \bibinfo{title}{Swap test and hong-ou-mandel effect are equivalent}.
\newblock \emph{\bibinfo{journal}{Physical Review A}}
  \textbf{\bibinfo{volume}{87}}, \bibinfo{pages}{052330}
  (\bibinfo{year}{2013}).

\bibitem{zeilinger1993einstein}
\bibinfo{author}{Zeilinger, A.}, \bibinfo{author}{Zukowski, M.},
  \bibinfo{author}{Horne, M.}, \bibinfo{author}{Bernstein, H.} \&
  \bibinfo{author}{Greenberger, D.}
\newblock \bibinfo{title}{Einstein-Podolsky-Rosen correlations in higher
  dimensions}.
\newblock \emph{\bibinfo{journal}{Fundamental Aspects of Quantum Theory,
  Singapore: World Scientific}}  (\bibinfo{year}{1993}).

\bibitem{reck1994experimental}
\bibinfo{author}{Reck, M.}, \bibinfo{author}{Zeilinger, A.},
  \bibinfo{author}{Bernstein, H.~J.} \& \bibinfo{author}{Bertani, P.}
\newblock \bibinfo{title}{Experimental realization of any discrete unitary
  operator}.
\newblock \emph{\bibinfo{journal}{Physical Review Letters}}
  \textbf{\bibinfo{volume}{73}}, \bibinfo{pages}{58} (\bibinfo{year}{1994}).

\bibitem{weihs1996two}
\bibinfo{author}{Weihs, G.}, \bibinfo{author}{Reck, M.},
  \bibinfo{author}{Weinfurter, H.} \& \bibinfo{author}{Zeilinger, A.}
\newblock \bibinfo{title}{Two-photon interference in optical fiber multiports}.
\newblock \emph{\bibinfo{journal}{Physical Review A}}
  \textbf{\bibinfo{volume}{54}}, \bibinfo{pages}{893} (\bibinfo{year}{1996}).

\bibitem{zukowski1997realizable}
\bibinfo{author}{{\.Z}ukowski, M.}, \bibinfo{author}{Zeilinger, A.} \&
  \bibinfo{author}{Horne, M.~A.}
\newblock \bibinfo{title}{Realizable higher-dimensional two-particle
  entanglements via multiport beam splitters}.
\newblock \emph{\bibinfo{journal}{Physical Review A}}
  \textbf{\bibinfo{volume}{55}}, \bibinfo{pages}{2564} (\bibinfo{year}{1997}).

\bibitem{campos2000three}
\bibinfo{author}{Campos, R.~A.}
\newblock \bibinfo{title}{Three-photon Hong-Ou-Mandel interference at a
  multiport mixer}.
\newblock \emph{\bibinfo{journal}{Physical Review A}}
  \textbf{\bibinfo{volume}{62}}, \bibinfo{pages}{013809}
  (\bibinfo{year}{2000}).

\bibitem{spagnolo2013three}
\bibinfo{author}{Spagnolo, N.} \emph{et~al.}
\newblock \bibinfo{title}{Three-photon bosonic coalescence in an integrated
  tritter}.
\newblock \emph{\bibinfo{journal}{Nature Communications}}
  \textbf{\bibinfo{volume}{4}}, \bibinfo{pages}{1606} (\bibinfo{year}{2013}).

\bibitem{schaeff2015experimental}
\bibinfo{author}{Schaeff, C.}, \bibinfo{author}{Polster, R.},
  \bibinfo{author}{Huber, M.}, \bibinfo{author}{Ramelow, S.} \&
  \bibinfo{author}{Zeilinger, A.}
\newblock \bibinfo{title}{Experimental access to higher-dimensional entangled
  quantum systems using integrated optics}.
\newblock \emph{\bibinfo{journal}{Optica}} \textbf{\bibinfo{volume}{2}},
  \bibinfo{pages}{523--529} (\bibinfo{year}{2015}).

\bibitem{meany2012non}
\bibinfo{author}{Meany, T.} \emph{et~al.}
\newblock \bibinfo{title}{Non-classical interference in integrated 3d
  multiports}.
\newblock \emph{\bibinfo{journal}{Optics Express}}
  \textbf{\bibinfo{volume}{20}}, \bibinfo{pages}{26895--26905}
  (\bibinfo{year}{2012}).

\bibitem{menssen2017distinguishability}
\bibinfo{author}{Menssen, A.~J.} \emph{et~al.}
\newblock \bibinfo{title}{Distinguishability and many-particle interference}.
\newblock \emph{\bibinfo{journal}{Physical Review Letters}}
  \textbf{\bibinfo{volume}{118}}, \bibinfo{pages}{153603}
  (\bibinfo{year}{2017}).

\bibitem{spagnolo2012quantum}
\bibinfo{author}{Spagnolo, N.} \emph{et~al.}
\newblock \bibinfo{title}{Quantum interferometry with three-dimensional
  geometry}.
\newblock \emph{\bibinfo{journal}{Scientific Reports}}
  \textbf{\bibinfo{volume}{2}}, \bibinfo{pages}{862} (\bibinfo{year}{2012}).

\bibitem{tichy2014interference}
\bibinfo{author}{Tichy, M.~C.}
\newblock \bibinfo{title}{Interference of identical particles from entanglement
  to boson-sampling}.
\newblock \emph{\bibinfo{journal}{Journal of Physics B: Atomic, Molecular and
  Optical Physics}} \textbf{\bibinfo{volume}{47}}, \bibinfo{pages}{103001}
  (\bibinfo{year}{2014}).

\bibitem{de2014coincidence}
\bibinfo{author}{de~Guise, H.}, \bibinfo{author}{Tan, S.-H.},
  \bibinfo{author}{Poulin, I.~P.} \& \bibinfo{author}{Sanders, B.~C.}
\newblock \bibinfo{title}{Coincidence landscapes for three-channel linear
  optical networks}.
\newblock \emph{\bibinfo{journal}{Physical Review A}}
  \textbf{\bibinfo{volume}{89}}, \bibinfo{pages}{063819}
  (\bibinfo{year}{2014}).

\bibitem{shchesnovich2015partial}
\bibinfo{author}{Shchesnovich, V.}
\newblock \bibinfo{title}{Partial indistinguishability theory for multiphoton
  experiments in multiport devices}.
\newblock \emph{\bibinfo{journal}{Physical Review A}}
  \textbf{\bibinfo{volume}{91}}, \bibinfo{pages}{013844}
  (\bibinfo{year}{2015}).

\bibitem{tichy2010zero}
\bibinfo{author}{Tichy, M.~C.}, \bibinfo{author}{Tiersch, M.},
  \bibinfo{author}{de~Melo, F.}, \bibinfo{author}{Mintert, F.} \&
  \bibinfo{author}{Buchleitner, A.}
\newblock \bibinfo{title}{Zero-transmission law for multiport beam splitters}.
\newblock \emph{\bibinfo{journal}{Physical Review Letters}}
  \textbf{\bibinfo{volume}{104}}, \bibinfo{pages}{220405}
  (\bibinfo{year}{2010}).

\bibitem{tichy2011four}
\bibinfo{author}{Tichy, M.~C.} \emph{et~al.}
\newblock \bibinfo{title}{Four-photon indistinguishability transition}.
\newblock \emph{\bibinfo{journal}{Physical Review A}}
  \textbf{\bibinfo{volume}{83}}, \bibinfo{pages}{062111}
  (\bibinfo{year}{2011}).

\bibitem{tichy2012many}
\bibinfo{author}{Tichy, M.~C.}, \bibinfo{author}{Tiersch, M.},
  \bibinfo{author}{Mintert, F.} \& \bibinfo{author}{Buchleitner, A.}
\newblock \bibinfo{title}{Many-particle interference beyond many-boson and
  many-fermion statistics}.
\newblock \emph{\bibinfo{journal}{New Journal of Physics}}
  \textbf{\bibinfo{volume}{14}}, \bibinfo{pages}{093015}
  (\bibinfo{year}{2012}).

\bibitem{ra2013nonmonotonic}
\bibinfo{author}{Ra, Y.-S.} \emph{et~al.}
\newblock \bibinfo{title}{Nonmonotonic quantum-to-classical transition in
  multiparticle interference}.
\newblock \emph{\bibinfo{journal}{Proceedings of the National Academy of
  Sciences}} \textbf{\bibinfo{volume}{110}}, \bibinfo{pages}{1227--1231}
  (\bibinfo{year}{2013}).

\bibitem{lahini2012quantum}
\bibinfo{author}{Lahini, Y.} \emph{et~al.}
\newblock \bibinfo{title}{Quantum walk of two interacting bosons}.
\newblock \emph{\bibinfo{journal}{Physical Review A}}
  \textbf{\bibinfo{volume}{86}}, \bibinfo{pages}{011603}
  (\bibinfo{year}{2012}).

\bibitem{poem2012two}
\bibinfo{author}{Poem, E.}, \bibinfo{author}{Gilead, Y.} \&
  \bibinfo{author}{Silberberg, Y.}
\newblock \bibinfo{title}{Two-photon path-entangled states in multimode
  waveguides}.
\newblock \emph{\bibinfo{journal}{Physical Review Letters}}
  \textbf{\bibinfo{volume}{108}}, \bibinfo{pages}{153602}
  (\bibinfo{year}{2012}).

\bibitem{defienne2016two}
\bibinfo{author}{Defienne, H.}, \bibinfo{author}{Barbieri, M.},
  \bibinfo{author}{Walmsley, I.~A.}, \bibinfo{author}{Smith, B.~J.} \&
  \bibinfo{author}{Gigan, S.}
\newblock \bibinfo{title}{Two-photon quantum walk in a multimode fiber}.
\newblock \emph{\bibinfo{journal}{Science advances}}
  \textbf{\bibinfo{volume}{2}}, \bibinfo{pages}{e1501054}
  (\bibinfo{year}{2016}).

\bibitem{politi2009integrated}
\bibinfo{author}{Politi, A.}, \bibinfo{author}{Matthews, J.~C.},
  \bibinfo{author}{Thompson, M.~G.} \& \bibinfo{author}{O'Brien, J.~L.}
\newblock \bibinfo{title}{Integrated quantum photonics}.
\newblock \emph{\bibinfo{journal}{IEEE Journal of Selected Topics in Quantum
  Electronics}} \textbf{\bibinfo{volume}{15}}, \bibinfo{pages}{1673--1684}
  (\bibinfo{year}{2009}).

\bibitem{peruzzo2010quantum}
\bibinfo{author}{Peruzzo, A.} \emph{et~al.}
\newblock \bibinfo{title}{Quantum walks of correlated photons}.
\newblock \emph{\bibinfo{journal}{Science}} \textbf{\bibinfo{volume}{329}},
  \bibinfo{pages}{1500--1503} (\bibinfo{year}{2010}).

\bibitem{poulios2014quantum}
\bibinfo{author}{Poulios, K.} \emph{et~al.}
\newblock \bibinfo{title}{Quantum walks of correlated photon pairs in
  two-dimensional waveguide arrays}.
\newblock \emph{\bibinfo{journal}{Physical Review Letters}}
  \textbf{\bibinfo{volume}{112}}, \bibinfo{pages}{143604}
  (\bibinfo{year}{2014}).

\bibitem{owens2011two}
\bibinfo{author}{Owens, J.~O.} \emph{et~al.}
\newblock \bibinfo{title}{Two-photon quantum walks in an elliptical
  direct-write waveguide array}.
\newblock \emph{\bibinfo{journal}{New Journal of Physics}}
  \textbf{\bibinfo{volume}{13}}, \bibinfo{pages}{075003}
  (\bibinfo{year}{2011}).

\bibitem{crespi2013anderson}
\bibinfo{author}{Crespi, A.} \emph{et~al.}
\newblock \bibinfo{title}{Anderson localization of entangled photons in an
  integrated quantum walk}.
\newblock \emph{\bibinfo{journal}{Nature Photonics}}
  \textbf{\bibinfo{volume}{7}}, \bibinfo{pages}{322} (\bibinfo{year}{2013}).

\bibitem{sansoni2012two}
\bibinfo{author}{Sansoni, L.} \emph{et~al.}
\newblock \bibinfo{title}{Two-particle bosonic-fermionic quantum walk via
  integrated photonics}.
\newblock \emph{\bibinfo{journal}{Physical Review Letters}}
  \textbf{\bibinfo{volume}{108}}, \bibinfo{pages}{010502}
  (\bibinfo{year}{2012}).

\bibitem{tillmann2015generalized}
\bibinfo{author}{Tillmann, M.} \emph{et~al.}
\newblock \bibinfo{title}{Generalized multiphoton quantum interference}.
\newblock \emph{\bibinfo{journal}{Physical Review X}}
  \textbf{\bibinfo{volume}{5}}, \bibinfo{pages}{041015} (\bibinfo{year}{2015}).

\bibitem{metcalf2013multiphoton}
\bibinfo{author}{Metcalf, B.~J.} \emph{et~al.}
\newblock \bibinfo{title}{Multiphoton quantum interference in a multiport
  integrated photonic device}.
\newblock \emph{\bibinfo{journal}{Nature Communications}}
  \textbf{\bibinfo{volume}{4}}, \bibinfo{pages}{1356} (\bibinfo{year}{2013}).

\bibitem{luo2019quantum}
\bibinfo{author}{Luo, Y.-H.} \emph{et~al.}
\newblock \bibinfo{title}{Quantum teleportation in high dimensions}.
\newblock \emph{\bibinfo{journal}{Physical Review Letters}}
  \textbf{\bibinfo{volume}{123}}, \bibinfo{pages}{070505}
  (\bibinfo{year}{2019}).

\bibitem{tichy2013limits}
\bibinfo{author}{Tichy, M.~C.}, \bibinfo{author}{Mintert, F.} \&
  \bibinfo{author}{Buchleitner, A.}
\newblock \bibinfo{title}{Limits to multipartite entanglement generation with
  bosons and fermions}.
\newblock \emph{\bibinfo{journal}{Physical Review A}}
  \textbf{\bibinfo{volume}{87}}, \bibinfo{pages}{022319}
  (\bibinfo{year}{2013}).

\bibitem{aaronson2011computational}
\bibinfo{author}{Aaronson, S.} \& \bibinfo{author}{Arkhipov, A.}
\newblock \bibinfo{title}{The computational complexity of linear optics}.
\newblock In \emph{\bibinfo{booktitle}{Proceedings of the forty-third annual
  ACM symposium on Theory of computing}}, \bibinfo{pages}{333--342}
  (\bibinfo{organization}{ACM}, \bibinfo{year}{2011}).

\bibitem{brod:19}
\bibinfo{author}{Brod, D.} \emph{et~al.}
\newblock \bibinfo{title}{Photonic implementation of boson sampling: a review}.
\newblock \emph{\bibinfo{journal}{Advanced Photonics}}
  \textbf{\bibinfo{volume}{1}}, \bibinfo{pages}{034001} (\bibinfo{year}{2019}).

\bibitem{broome2013photonic}
\bibinfo{author}{Broome, M.~A.} \emph{et~al.}
\newblock \bibinfo{title}{Photonic boson sampling in a tunable circuit}.
\newblock \emph{\bibinfo{journal}{Science}} \textbf{\bibinfo{volume}{339}},
  \bibinfo{pages}{794--798} (\bibinfo{year}{2013}).

\bibitem{spring2012boson}
\bibinfo{author}{Spring, J.~B.} \emph{et~al.}
\newblock \bibinfo{title}{Boson sampling on a photonic chip}.
\newblock \emph{\bibinfo{journal}{Science}} \bibinfo{pages}{1231692}
  (\bibinfo{year}{2012}).

\bibitem{tillmann2013experimental}
\bibinfo{author}{Tillmann, M.} \emph{et~al.}
\newblock \bibinfo{title}{Experimental boson sampling}.
\newblock \emph{\bibinfo{journal}{Nature Photonics}}
  \textbf{\bibinfo{volume}{7}}, \bibinfo{pages}{540} (\bibinfo{year}{2013}).

\bibitem{crespi2013integrated}
\bibinfo{author}{Crespi, A.} \emph{et~al.}
\newblock \bibinfo{title}{Integrated multimode interferometers with arbitrary
  designs for photonic boson sampling}.
\newblock \emph{\bibinfo{journal}{Nature Photonics}}
  \textbf{\bibinfo{volume}{7}}, \bibinfo{pages}{545} (\bibinfo{year}{2013}).

\bibitem{spagnolo2014experimental}
\bibinfo{author}{Spagnolo, N.} \emph{et~al.}
\newblock \bibinfo{title}{Experimental validation of photonic boson sampling}.
\newblock \emph{\bibinfo{journal}{Nature Photonics}}
  \textbf{\bibinfo{volume}{8}}, \bibinfo{pages}{615} (\bibinfo{year}{2014}).

\bibitem{bentivegna2015experimental}
\bibinfo{author}{Bentivegna, M.} \emph{et~al.}
\newblock \bibinfo{title}{Experimental scattershot boson sampling}.
\newblock \emph{\bibinfo{journal}{Science Advances}}
  \textbf{\bibinfo{volume}{1}}, \bibinfo{pages}{e1400255}
  (\bibinfo{year}{2015}).

\bibitem{lund2014boson}
\bibinfo{author}{Lund, A.} \emph{et~al.}
\newblock \bibinfo{title}{Boson sampling from a gaussian state}.
\newblock \emph{\bibinfo{journal}{Physical Review Letters}}
  \textbf{\bibinfo{volume}{113}}, \bibinfo{pages}{100502}
  (\bibinfo{year}{2014}).

\bibitem{wang2017high}
\bibinfo{author}{Wang, H.} \emph{et~al.}
\newblock \bibinfo{title}{High-efficiency multiphoton boson sampling}.
\newblock \emph{\bibinfo{journal}{Nature Photonics}}
  \textbf{\bibinfo{volume}{11}}, \bibinfo{pages}{361} (\bibinfo{year}{2017}).

\bibitem{tame2013quantum}
\bibinfo{author}{Tame, M.~S.} \emph{et~al.}
\newblock \bibinfo{title}{Quantum plasmonics}.
\newblock \emph{\bibinfo{journal}{Nature Physics}}
  \textbf{\bibinfo{volume}{9}}, \bibinfo{pages}{329} (\bibinfo{year}{2013}).

\bibitem{marquier2017revisiting}
\bibinfo{author}{Marquier, F.}, \bibinfo{author}{Sauvan, C.} \&
  \bibinfo{author}{Greffet, J.-J.}
\newblock \bibinfo{title}{Revisiting quantum optics with surface plasmons and
  plasmonic resonators}.
\newblock \emph{\bibinfo{journal}{ACS photonics}} \textbf{\bibinfo{volume}{4}},
  \bibinfo{pages}{2091--2101} (\bibinfo{year}{2017}).

\bibitem{fujii2012preservation}
\bibinfo{author}{Fujii, G.} \emph{et~al.}
\newblock \bibinfo{title}{Preservation of photon indistinguishability after
  transmission through surface-plasmon-polariton waveguide}.
\newblock \emph{\bibinfo{journal}{Optics Letters}}
  \textbf{\bibinfo{volume}{37}}, \bibinfo{pages}{1535--1537}
  (\bibinfo{year}{2012}).

\bibitem{wang2012hong}
\bibinfo{author}{Wang, S.} \emph{et~al.}
\newblock \bibinfo{title}{Hong-Ou-Mandel interference mediated by the magnetic
  plasmon waves in a three-dimensional optical metamaterial}.
\newblock \emph{\bibinfo{journal}{Optics Express}}
  \textbf{\bibinfo{volume}{20}}, \bibinfo{pages}{5213--5218}
  (\bibinfo{year}{2012}).

\bibitem{heeres2013quantum}
\bibinfo{author}{Heeres, R.~W.}, \bibinfo{author}{Kouwenhoven, L.~P.} \&
  \bibinfo{author}{Zwiller, V.}
\newblock \bibinfo{title}{Quantum interference in plasmonic circuits}.
\newblock \emph{\bibinfo{journal}{Nature Nanotechnology}}
  \textbf{\bibinfo{volume}{8}}, \bibinfo{pages}{719} (\bibinfo{year}{2013}).

\bibitem{toyoda2015hong}
\bibinfo{author}{Toyoda, K.}, \bibinfo{author}{Hiji, R.},
  \bibinfo{author}{Noguchi, A.} \& \bibinfo{author}{Urabe, S.}
\newblock \bibinfo{title}{Hong--Ou--Mandel interference of two phonons in
  trapped ions}.
\newblock \emph{\bibinfo{journal}{Nature}} \textbf{\bibinfo{volume}{527}},
  \bibinfo{pages}{74} (\bibinfo{year}{2015}).

\bibitem{li2016hong}
\bibinfo{author}{Li, J.} \emph{et~al.}
\newblock \bibinfo{title}{Hong-ou-mandel interference between two deterministic
  collective excitations in an atomic ensemble}.
\newblock \emph{\bibinfo{journal}{Physical Review Letters}}
  \textbf{\bibinfo{volume}{117}}, \bibinfo{pages}{180501}
  (\bibinfo{year}{2016}).

\bibitem{fakonas2014two}
\bibinfo{author}{Fakonas, J.~S.}, \bibinfo{author}{Lee, H.},
  \bibinfo{author}{Kelaita, Y.~A.} \& \bibinfo{author}{Atwater, H.~A.}
\newblock \bibinfo{title}{Two-plasmon quantum interference}.
\newblock \emph{\bibinfo{journal}{Nature Photonics}}
  \textbf{\bibinfo{volume}{8}}, \bibinfo{pages}{317} (\bibinfo{year}{2014}).

\bibitem{di2014observation}
\bibinfo{author}{Di~Martino, G.} \emph{et~al.}
\newblock \bibinfo{title}{Observation of quantum interference in the plasmonic
  hong-ou-mandel effect}.
\newblock \emph{\bibinfo{journal}{Physical Review Applied}}
  \textbf{\bibinfo{volume}{1}}, \bibinfo{pages}{034004} (\bibinfo{year}{2014}).

\bibitem{cai2014high}
\bibinfo{author}{Cai, Y.-J.} \emph{et~al.}
\newblock \bibinfo{title}{High-visibility on-chip quantum interference of
  single surface plasmons}.
\newblock \emph{\bibinfo{journal}{Physical Review Applied}}
  \textbf{\bibinfo{volume}{2}}, \bibinfo{pages}{014004} (\bibinfo{year}{2014}).

\bibitem{fujii2014direct}
\bibinfo{author}{Fujii, G.}, \bibinfo{author}{Fukuda, D.} \&
  \bibinfo{author}{Inoue, S.}
\newblock \bibinfo{title}{Direct observation of bosonic quantum interference of
  surface plasmon polaritons using photon-number-resolving detectors}.
\newblock \emph{\bibinfo{journal}{Physical Review B}}
  \textbf{\bibinfo{volume}{90}}, \bibinfo{pages}{085430}
  (\bibinfo{year}{2014}).

\bibitem{vest2017anti}
\bibinfo{author}{Vest, B.} \emph{et~al.}
\newblock \bibinfo{title}{Anti-coalescence of bosons on a lossy beam splitter}.
\newblock \emph{\bibinfo{journal}{Science}} \textbf{\bibinfo{volume}{356}},
  \bibinfo{pages}{1373--1376} (\bibinfo{year}{2017}).

\bibitem{dieleman2017experimental}
\bibinfo{author}{Dieleman, F.}, \bibinfo{author}{Tame, M.},
  \bibinfo{author}{Sonnefraud, Y.}, \bibinfo{author}{Kim, M.} \&
  \bibinfo{author}{Maier, S.~A.}
\newblock \bibinfo{title}{Experimental verification of entanglement generated
  in a plasmonic system}.
\newblock \emph{\bibinfo{journal}{Nano letters}} \textbf{\bibinfo{volume}{17}},
  \bibinfo{pages}{7455--7461} (\bibinfo{year}{2017}).

\bibitem{tamura2020quantum}
\bibinfo{author}{Tamura, M.}, \bibinfo{author}{Mukaiyama, T.} \&
  \bibinfo{author}{Toyoda, K.}
\newblock \bibinfo{title}{Quantum walks of a phonon in trapped ions}.
\newblock \emph{\bibinfo{journal}{Physical Review Letters}}
  \textbf{\bibinfo{volume}{124}}, \bibinfo{pages}{200501}
  (\bibinfo{year}{2020}).

\bibitem{kaufman2014two}
\bibinfo{author}{Kaufman, A.} \emph{et~al.}
\newblock \bibinfo{title}{Two-particle quantum interference in tunnel-coupled
  optical tweezers}.
\newblock \emph{\bibinfo{journal}{Science}} \textbf{\bibinfo{volume}{345}},
  \bibinfo{pages}{306--309} (\bibinfo{year}{2014}).

\bibitem{preiss2015strongly}
\bibinfo{author}{Preiss, P.~M.} \emph{et~al.}
\newblock \bibinfo{title}{Strongly correlated quantum walks in optical
  lattices}.
\newblock \emph{\bibinfo{journal}{Science}} \textbf{\bibinfo{volume}{347}},
  \bibinfo{pages}{1229--1233} (\bibinfo{year}{2015}).

\bibitem{lopes2015atomic}
\bibinfo{author}{Lopes, R.} \emph{et~al.}
\newblock \bibinfo{title}{Atomic Hong--Ou--Mandel experiment}.
\newblock \emph{\bibinfo{journal}{Nature}} \textbf{\bibinfo{volume}{520}},
  \bibinfo{pages}{66} (\bibinfo{year}{2015}).

\bibitem{lewis2014proposal}
\bibinfo{author}{Lewis-Swan, R.} \& \bibinfo{author}{Kheruntsyan, K.}
\newblock \bibinfo{title}{Proposal for demonstrating the Hong-Ou-Mandel
  effect with matter waves}.
\newblock \emph{\bibinfo{journal}{Nature Communications}}
  \textbf{\bibinfo{volume}{5}}, \bibinfo{pages}{3752} (\bibinfo{year}{2014}).

\bibitem{dussarrat2017two}
\bibinfo{author}{Dussarrat, P.} \emph{et~al.}
\newblock \bibinfo{title}{Two-particle four-mode interferometer for atoms}.
\newblock \emph{\bibinfo{journal}{Physical Review Letters}}
  \textbf{\bibinfo{volume}{119}}, \bibinfo{pages}{173202}
  (\bibinfo{year}{2017}).

\bibitem{kaufman2018hong}
\bibinfo{author}{Kaufman, A.~M.}, \bibinfo{author}{Tichy, M.~C.},
  \bibinfo{author}{Mintert, F.}, \bibinfo{author}{Rey, A.~M.} \&
  \bibinfo{author}{Regal, C.~A.}
\newblock \bibinfo{title}{The hong--ou--mandel effect with atoms}.
\newblock In \emph{\bibinfo{booktitle}{Advances In Atomic, Molecular, and
  Optical Physics}}, vol.~\bibinfo{volume}{67}, \bibinfo{pages}{377--427}
  (\bibinfo{publisher}{Elsevier}, \bibinfo{year}{2018}).

\bibitem{liu1998quantum}
\bibinfo{author}{Liu, R.}, \bibinfo{author}{Odom, B.},
  \bibinfo{author}{Yamamoto, Y.} \& \bibinfo{author}{Tarucha, S.}
\newblock \bibinfo{title}{Quantum interference in electron collision}.
\newblock \emph{\bibinfo{journal}{Nature}} \textbf{\bibinfo{volume}{391}},
  \bibinfo{pages}{263} (\bibinfo{year}{1998}).

\bibitem{bocquillon2012electron}
\bibinfo{author}{Bocquillon, E.} \emph{et~al.}
\newblock \bibinfo{title}{Electron quantum optics: partitioning electrons one
  by one}.
\newblock \emph{\bibinfo{journal}{Physical Review Letters}}
  \textbf{\bibinfo{volume}{108}}, \bibinfo{pages}{196803}
  (\bibinfo{year}{2012}).

\bibitem{bocquillon2013coherence}
\bibinfo{author}{Bocquillon, E.} \emph{et~al.}
\newblock \bibinfo{title}{Coherence and indistinguishability of single
  electrons emitted by independent sources}.
\newblock \emph{\bibinfo{journal}{Science}} \bibinfo{pages}{1232572}
  (\bibinfo{year}{2013}).

\bibitem{feve2007demand}
\bibinfo{author}{F{\`e}ve, G.} \emph{et~al.}
\newblock \bibinfo{title}{An on-demand coherent single-electron source}.
\newblock \emph{\bibinfo{journal}{Science}} \textbf{\bibinfo{volume}{316}},
  \bibinfo{pages}{1169--1172} (\bibinfo{year}{2007}).

\bibitem{freulon2015hong}
\bibinfo{author}{Freulon, V.} \emph{et~al.}
\newblock \bibinfo{title}{Hong-ou-mandel experiment for temporal investigation
  of single-electron fractionalization}.
\newblock \emph{\bibinfo{journal}{Nature Communications}}
  \textbf{\bibinfo{volume}{6}}, \bibinfo{pages}{6854} (\bibinfo{year}{2015}).

\bibitem{dubois2013minimal}
\bibinfo{author}{Dubois, J.} \emph{et~al.}
\newblock \bibinfo{title}{Minimal-excitation states for electron quantum optics
  using levitons}.
\newblock \emph{\bibinfo{journal}{Nature}} \textbf{\bibinfo{volume}{502}},
  \bibinfo{pages}{659} (\bibinfo{year}{2013}).

\bibitem{levitov1996electron}
\bibinfo{author}{Levitov, L.~S.}, \bibinfo{author}{Lee, H.} \&
  \bibinfo{author}{Lesovik, G.~B.}
\newblock \bibinfo{title}{Electron counting statistics and coherent states of
  electric current}.
\newblock \emph{\bibinfo{journal}{Journal of Mathematical Physics}}
  \textbf{\bibinfo{volume}{37}}, \bibinfo{pages}{4845--4866}
  (\bibinfo{year}{1996}).

\bibitem{keeling2006minimal}
\bibinfo{author}{Keeling, J.}, \bibinfo{author}{Klich, I.} \&
  \bibinfo{author}{Levitov, L.}
\newblock \bibinfo{title}{Minimal excitation states of electrons in
  one-dimensional wires}.
\newblock \emph{\bibinfo{journal}{Physical Review Letters}}
  \textbf{\bibinfo{volume}{97}}, \bibinfo{pages}{116403}
  (\bibinfo{year}{2006}).

\bibitem{glattli2017levitons}
\bibinfo{author}{Glattli, D.~C.} \& \bibinfo{author}{Roulleau, P.~S.}
\newblock \bibinfo{title}{Levitons for electron quantum optics}.
\newblock \emph{\bibinfo{journal}{Physica Status Solidi (b)}}
  \textbf{\bibinfo{volume}{254}} (\bibinfo{year}{2017}).

\bibitem{glattli2016hanbury}
\bibinfo{author}{Glattli, D.} \& \bibinfo{author}{Roulleau, P.}
\newblock \bibinfo{title}{Hanbury-brown twiss noise correlation with time
  controlled quasi-particles in ballistic quantum conductors}.
\newblock \emph{\bibinfo{journal}{Physica E: Low-dimensional Systems and
  Nanostructures}} \textbf{\bibinfo{volume}{76}}, \bibinfo{pages}{216--222}
  (\bibinfo{year}{2016}).

\bibitem{jullien2014quantum}
\bibinfo{author}{Jullien, T.} \emph{et~al.}
\newblock \bibinfo{title}{Quantum tomography of an electron}.
\newblock \emph{\bibinfo{journal}{Nature}} \textbf{\bibinfo{volume}{514}},
  \bibinfo{pages}{603} (\bibinfo{year}{2014}).

\bibitem{ahmed2017guided}
\bibinfo{author}{Ahmed, M.~H.}, \bibinfo{author}{Jeske, J.} \&
  \bibinfo{author}{Greentree, A.~D.}
\newblock \bibinfo{title}{Guided magnonic michelson interferometer}.
\newblock \emph{\bibinfo{journal}{Scientific Reports}}
  \textbf{\bibinfo{volume}{7}}, \bibinfo{pages}{41472} (\bibinfo{year}{2017}).

\bibitem{ferraro2015nonlocal}
\bibinfo{author}{Ferraro, D.}, \bibinfo{author}{Rech, J.},
  \bibinfo{author}{Jonckheere, T.} \& \bibinfo{author}{Martin, T.}
\newblock \bibinfo{title}{Nonlocal interference and Hong-Ou-Mandel collisions
  of single bogoliubov quasiparticles}.
\newblock \emph{\bibinfo{journal}{Physical Review B}}
  \textbf{\bibinfo{volume}{91}}, \bibinfo{pages}{075406}
  (\bibinfo{year}{2015}).

\bibitem{khan2014two}
\bibinfo{author}{Khan, M.} \& \bibinfo{author}{Leuenberger, M.~N.}
\newblock \bibinfo{title}{Two-dimensional fermionic hong-ou-mandel interference
  with massless dirac fermions}.
\newblock \emph{\bibinfo{journal}{Physical Review B}}
  \textbf{\bibinfo{volume}{90}}, \bibinfo{pages}{075439}
  (\bibinfo{year}{2014}).

\bibitem{sun2014mean}
\bibinfo{author}{Sun, Z.-Y.}, \bibinfo{author}{Kevrekidis, P.~G.} \&
  \bibinfo{author}{Kr{\"u}ger, P.}
\newblock \bibinfo{title}{Mean-field analog of the hong-ou-mandel experiment
  with bright solitons}.
\newblock \emph{\bibinfo{journal}{Physical Review A}}
  \textbf{\bibinfo{volume}{90}}, \bibinfo{pages}{063612}
  (\bibinfo{year}{2014}).

\end{thebibliography}

\end{document}